\documentclass[
  aps,pra,twocolumn,amsmath,
  tightenlines,superscriptaddress,
  11pt,a4paper,longbibliography,
  floatfix
]{revtex4-2}
\usepackage[utf8]{inputenc}
\usepackage[T1]{fontenc}
\usepackage[english]{babel} 
\usepackage{amsfonts}
\usepackage{textcomp}
\usepackage{silence}
\usepackage{graphicx}
\usepackage{svg}
\usepackage{rotating}
\usepackage{longtable}
\usepackage{booktabs}
\usepackage{verbatim}
\usepackage{makecell}
\usepackage{multirow}
\usepackage{empheq}
\usepackage{siunitx}
\DeclareSIUnit\bar{bar}
\newcommand{\betaL}{\beta_{\mathrm L}}
\newcommand{\Ic}{I_{\mathrm{c}}}
\newcommand{\IcOne}{I_{\mathrm{c}1}}
\newcommand{\IcTwo}{I_{\mathrm{c}2}}
\DeclareSIUnit\fluxquantum{\ensuremath{\Phi_0}}
\newcommand{\eref}[1]{Eq.~(\ref{#1})}
\newcommand{\fref}[1]{Fig.~\ref{#1}}
\newcommand{\tref}[1]{Table~\ref{#1}}

\newcommand{\aref}[1]{App.~\ref{#1}}
\newcommand{\frefs}[1]{Figs.~\ref{#1}}

\usepackage[
  colorlinks=true,
  allcolors=blue,
  bookmarksnumbered=true
]{hyperref}
\begin{document}
\raggedbottom

\title{Efficient flip-chip and on-chip-based modulation of flux-tunable superconducting resonators}

\author{Achintya Paradkar}
\affiliation{Department of Microtechnology and Nanoscience (MC2), Chalmers University of Technology, Kemivägen 9, Gothenburg, SE-41296, Sweden}
\author{Paul Nicaise}
\affiliation{Department of Microtechnology and Nanoscience (MC2), Chalmers University of Technology, Kemivägen 9, Gothenburg, SE-41296, Sweden}
\author{Karim Dakroury}
\affiliation{Department of Microtechnology and Nanoscience (MC2), Chalmers University of Technology, Kemivägen 9, Gothenburg, SE-41296, Sweden}
\author{Fabian Resare}
\affiliation{Department of Microtechnology and Nanoscience (MC2), Chalmers University of Technology, Kemivägen 9, Gothenburg, SE-41296, Sweden}
\author{Christian Dejaco}
\affiliation{Institute for Quantum Optics and Quantum Information, Austrian Academy of Sciences, 6020 Innsbruck, Austria}
\affiliation{University of Innsbruck, Institute for Experimental Physics, 6020 Innsbruck Austria}
\author{Lukas Deeg}
\affiliation{Institute for Quantum Optics and Quantum Information, Austrian Academy of Sciences, 6020 Innsbruck, Austria}
\affiliation{University of Innsbruck, Institute for Experimental Physics, 6020 Innsbruck Austria}
\author{Gerhard Kirchmair}
\affiliation{Institute for Quantum Optics and Quantum Information, Austrian Academy of Sciences, 6020 Innsbruck, Austria}
\affiliation{University of Innsbruck, Institute for Experimental Physics, 6020 Innsbruck Austria}
\author{Witlef Wieczorek}
\email{witlef.wieczorek@chalmers.se}
\affiliation{Department of Microtechnology and Nanoscience (MC2), Chalmers University of Technology, Kemivägen 9, Gothenburg, SE-41296, Sweden}

\begin{abstract} 
We demonstrate the efficient modulation of flux-tunable superconducting resonators (FTRs) using flip-chip or on-chip-based input coils. The FTRs we use are aluminum-based quarter-wave coplanar waveguide resonators terminated with \SI{100}{\micro\meter} or \SI{200}{\micro\meter}-wide square loop dc superconducting quantum interference devices (SQUIDs) employing \SI{1}{\micro\meter}-sized Josephson junctions. We employ SQUIDs with a geometric loop inductance of up to \SI{0.7}{\nano\henry} to increase the flux transfer efficiency. The geometric inductance of the SQUID results in a non-zero screening parameter $\betaL$, whose branch switching effect is mitigated by using asymmetric junctions. We achieve flux modulation of the FTRs by more than one \si{\giga\hertz} and flux responsivities of up to tens of \si{\giga\hertz/\fluxquantum} with \si{\micro\ampere}-scale on-chip currents. We compare flip-chip with on-chip input-coil-based flux modulation, where the former is realized through galvanically connected and closely spaced chips, while the latter is achieved through superconducting air-bridge connections. We achieve a flux-transfer efficiency from the input coil to the SQUID loop of up to 20\%. Our work paves the way for efficient low current flux modulation of FTRs and sensitive measurement of flux signals.
\end{abstract}
\maketitle

\section{Introduction}

Superconducting quantum circuits are widely employed in quantum technologies due to their high coherence, controllability, coupling strength, and potential for scalability \cite{Devoret_2013, Wendin_2017, Krantz_2019, Blais_2021}. Flux-tunable superconducting elements are crucial components in superconducting quantum circuits, enabling flux-tunable transmon-like qubits \cite{Hutchings_2017, Mergenthaler_2021, Garcia_2022}, flux-tunable couplers \cite{Pierre_2014, Kafri_2017, Menke_2022}, Josephson parametric amplifiers \cite{Yamamoto_2008, Roch_2012, Simoen_2015, Pogorzalek_2017}, or flux-tunable resonators (FTRs) \cite{Palacios_2008, Sandberg_2008, Krantz_2013, Kennedy_2019, Uhl2023}. FTRs are typically implemented as coplanar waveguide (CPW) or lumped element resonators that are terminated by a dc superconducting quantum interference device (SQUID) \cite{Palacios_2008}. The flux modulation of the FTR exhibits a tunable response that can be as large as tens of \si{\giga\hertz/\fluxquantum}, making the FTR a suitable device for sensing of magnetic flux.

FTRs have been employed, for example, for flux-based detection of the motion of micromechanical resonators \cite{Nation_2008, Etaki_2008, Poot_2010, Rodrigues2019, Zoepfl_2020, Schmidt_2020, Luschmann2022, Zoepfl_2023, Schmidt2024}, for detecting the electronic state of spin ensembles \cite{Kubo_2010}, for photon–pressure coupling of superconducting resonators \cite{Eichler_2018, Bothner_2020}, for intra-cavity photon generation \cite{Wilson_2010}, qubit read-out \cite{Krantz_2016}, or for dark-matter searches \cite{Zhao_2025}. For flux sensing using FTRs, it is crucial to transport the flux signal efficiently from the source to the SQUID loop of the FTR. This is challenging when the SQUID loop cannot be placed near the source for readout, for example, due to limits imposed by the critical field of the superconducting material employed in the FTR. In such cases, the flux can be transported to the FTR via a flux transformer \cite{Duret_1984, Knuutila_1988, Yi_2000, Granata_2011, Chukharkin_2012}, in which a pick-up coil is located at the source and connected to an input coil located at the FTR. It is crucial that the input coil transports the flux efficiently into the SQUID loop without affecting the coherence of the FTR. For certain applications, it may be necessary to increase the geometric inductance of the SQUID loop for optimal flux-transfer efficiency. This can then lead to a non-zero screening parameter $\betaL$, which can result in branch switching behavior of the resonance frequency of the FTR \cite{Tesche_1977, LefevreSeguin1992, Bhupathi2016, Pogorzalek_2017, Kennedy_2019, Uhl2023}.

\begin{figure*}[t!hbp]
    \centering
    \includegraphics[width=\textwidth, keepaspectratio]{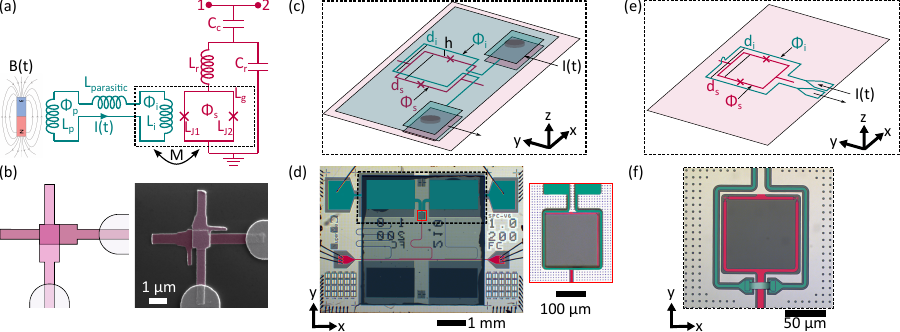}
    \caption{Chip-based modulation of a flux-tunable resonator (FTR) using integrated input coils. 
    (a) Equivalent lumped element circuit model: the FTR (red) is inductively coupled via a flux transformer (green) to an external magnetic source. 
    (b) Left: design of a \SI{1}{\micro\meter}-wide Josephson junction. Right: false-colored SEM image of the fabricated junction where the dark electrode is patterned first; the thin appendages on either side are artifacts of the shadow evaporation technique.
    (c) Schematic of a flip-chip input coil (green) concentric with the SQUID (red) and galvanically connected to the bottom chip via superconducting indium bumps \cite{paradkar_2025}. 
    (d) Device based on schematic shown in (c): \SI{200}{\micro\meter}-wide SQUID-loop FTR on a silicon bottom chip with a NbN input coil on a sapphire top chip (false-colored). 
    (e) Schematic of an on-chip input coil (green) patterned concentrically around the SQUID loop (red). 
    (f) Device based on schematic shown in (e): optical image (false-colored) of a \SI{100}{\micro\meter}-wide SQUID-loop FTR; the input coil is routed over the CPW via an aluminum air bridge \cite{Osman_2021, Osman_2024}.}
    \label{fig:Fig_1}
\end{figure*}

In our work, we demonstrate two efficient methods to modulate FTRs using chip-based input coils. The first method involves positioning the input coil on a separate chip above the SQUID loop and galvanically connecting it through superconducting interconnects in a simple flip-chip assembly \cite{paradkar_2025}. The second method involves placing the input coil concentrically with the SQUID loop on the same chip, made possible through air-bridge interconnects \cite{Osman_2021,Osman_2024}. Our FTRs use large-loop dc SQUIDs with asymmetric Josephson junctions (JJs). The large loop increases flux-transfer efficiency, while the asymmetry mitigates the branch switching effect \cite{Shevchuk_2017, Frattini_2017, Frattini_2018, Lescanne_2020, Hillmann_2022, Lu_2023, Eriksson_2024}. We account for the large loop inductance in our FTR model through a finite screening parameter $\betaL$ and for the junction asymmetry through an asymmetry parameter $\alpha$ \cite{Tesche_1977, LefevreSeguin1992}. We analyze the efficiency of our flux-transfer methods and their effect on the intrinsic quality factor of the FTR. Finally, we employ a flux transformer to couple a source magnetic field to the FTR via the input coil while keeping the source and FTR physically separated.

\section{Results}

\fref{fig:Fig_1}(a) shows the lumped-element circuit representation of the FTR, which is a quarter-wave CPW terminated by a dc SQUID at the grounded end, i.e., at the current anti-node. The FTR is capacitively coupled to a transmission line for readout at the voltage anti-node. A magnetic field $B(t)$ is measured through a pick-up loop, whose flux $\Phi_p$ is transported to an input coil at the SQUID via a flux transformer. The resonance frequency of the FTR depends on the flux $\Phi_s$ through the SQUID approximately as \cite{Wallquist_2006, Palacios_2008, Wustmann2013, Pogorzalek_2017, Schmidt_2020} (a detailed derivation can be found in \aref{SI:FTRtheory})
\begin{equation}
\omega_r(\Phi_s) \simeq \frac{\omega_0}{1 + \gamma(\Phi_s)},\label{eq:FTRmod}
\end{equation}
with $\omega_0 = A\sqrt{1/(L_rC_r)}$ the resonance frequency of the bare CPW (with inductance $L_r$, capacitance $C_r$, and a scaling factor $A$) and the inductance participation ratio $\gamma$, which quantifies the nonlinearity of the FTR via
\begin{equation}
\gamma(\Phi_s) = \frac{L_S(\Phi_s)}{lL_l},
\end{equation}
i.e., the ratio of total SQUID inductance $L_S(\Phi_s)$, which includes contributions from the Josephson inductances ($L_{Ji}(\Phi_s)$) of each JJ and the loop's geometric inductance ($L_g$), to the linear inductance per unit length ($L_l$) and total CPW length ($l$); we provide the full equation in \aref{subsec:dcSQUID}. 

The FTRs we employ have a large geometric inductance of up to $L_g=\SI{750}{\pico\henry}$, which gives rise to a non-negligible screening current. As a result, the flux $\Phi_s$ threading the SQUID loop differs from the externally applied flux $\Phi_e$ set by the coil. Following Ref.~\cite{Tesche_1977}, these quantities are related by (see \aref{subsec:dcSQUID})
\begin{equation}
\label{eq:beta_L}
    \Phi_e = \Phi_s
    \pm \Phi_0\frac{\betaL}{2}\frac{(1-\alpha^2)\sin\!\left(\pi\frac{\Phi_s}{\Phi_0}\right)}{\sqrt{1 + \alpha^2\tan^2\!\left(\pi\frac{\Phi_s}{\Phi_0}\right)}},
\end{equation}
where $\betaL = 2L_g I_0/\Phi_0$ is the screening parameter, $I_0\sim \SI{375}{\nano\ampere}$ is the average critical current of our Josephson junctions, and $\alpha$ is the junction asymmetry defined as
\begin{equation}
    \alpha = (\IcOne - \IcTwo)/(\IcOne + \IcTwo),
    \label{eq:alpha}
\end{equation}
with $I_{{\text{c}}i}$ being the critical current of the $i$th junction. In our devices, we expect $\betaL\sim 0.3$ for the \SI{200}{\micro\meter}-wide dc SQUID. We use the simplifying assumption that the SQUID perturbs the FTR boundary only locally at its grounded end. However, as our SQUID loops are rather large, i.e., up to 12\% of the CPW length, we expect a deviation from this assumption and introduce a phenomenological scaling factor $A$ in \eref{eq:FTRmod}.

We fabricate the wiring layer and ground plane of the FTR by patterning a \SI{150}{\nano\meter} thick Al film on a high-resistivity $\langle111\rangle$ silicon substrate. The JJs are Manhattan-style \cite{Potts_2001} Al/AlO$_x$/Al junctions deposited using the shadow evaporation technique \cite{Osman_2024}. We realized \SI{1}{\micro\meter}-sized JJs, as shown in \fref{fig:Fig_1}(b) (for fabrication details, see \aref{SI:experiment}), to reduce the nonlinearity of the FTR, i.e., its Kerr coefficient $\mathcal{K}$, which, among other factors, is reduced for larger junction sizes. Our Manhattan-style junction design, combined with the shadow evaporation process, allows us to pattern one junction smaller than the other due to resist shadowing effects (see \aref{SI:device}). We determine the critical current of the JJs from the normal-state resistance ($R_N$) above the critical temperature using the Ambegaokar–Baratoff relation $\Ic = \pi\Delta/(2\mathrm{e}R_N)$ \cite{Tinkham_2004a}, where $\Delta$ is the superconducting gap with $\Delta\sim\SI{180}{\micro\electronvolt}$ for Al. $R_N$ is estimated by measuring the junctions at room temperature and accounting for a 10\% increase at \SI{4}{\kelvin} \cite{Zheng_2023}. We obtain $\IcOne\approx\SI{500}{\nano\ampere}$ and $\IcTwo\approx\SI{250}{\nano\ampere}$, with corresponding inductances of $L_{J1}\approx\SI{650}{\pico\henry}$ and $L_{J2}\approx\SI{1300}{\pico\henry}$, respectively, leading to $\alpha\sim0.3$. 

\fref{fig:Fig_1}(c) shows a schematic of the flip-chip-based input coil on the FTR, and \fref{fig:Fig_1}(d) shows its implementation. The top chip of the flip-chip assembly contains a \SI{215}{\micro\meter}-wide input coil, patterned on a sapphire substrate, and placed above a \SI{200}{\micro\meter}-wide SQUID, both of square geometry. We galvanically connect the top and bottom chips using indium bumps and achieve a chip separation of approximately \SI{50}{\micro\meter}; for details, see Ref. \cite{paradkar_2025}. Note that the separation can, in principle, be made smaller by using smaller indium bumps. Our second integrated modulation technique relies on a \SI{125}{\micro\meter}-wide on-chip input coil patterned around a \SI{100}{\micro\meter}-wide square SQUID loop (\fref{fig:Fig_1}(e,f)). The input coil needs to bridge the FTR central line, which is achieved through the use of a superconducting air bridge; for fabrication details, see Refs. \cite{Osman_2021, Osman_2024}. Note that the on-chip input coil could, in principle, also be located within the SQUID loop using two air bridge connections.

\begin{figure}[t!hbp]
    \centering
    \includegraphics[width=\columnwidth, keepaspectratio]{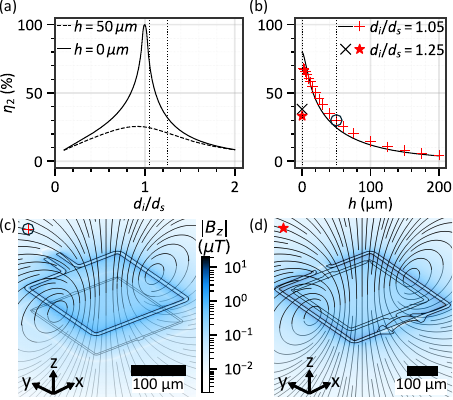}
    \caption{Flux transfer efficiency $\eta_2$ between the input coil and the SQUID loop. Efficiency as a function of (a) the ratio of input coil loop width to SQUID loop width $d_i/d_s$ and (b) axial loop separation $h$. The black lines represent analytic results assuming one-dimensional square loops, while the red markers are results from FEM simulations of realistic geometries of the on-chip (star) and flip-chip (cross) configurations. The dotted vertical lines in (a) are at $d_i/d_s$ of $1.05$ and $1.25$ and in (b) at $h$ of $\SI{0}{\micro\meter}$ and $\SI{50}{\micro\meter}$. FEM-simulated magnetic field distribution for (c) flip-chip and (d) on-chip configurations assuming an input coil current of \SI{100}{\micro\ampere}.}
    \label{fig:Fig_2}
\end{figure}

\fref{fig:Fig_2}(a) shows the expected flux transfer efficiency $\eta_2=\Phi_e/\Phi_i$ from the input coil of width $d_i$ to the SQUID loop of width $d_s$, for the flip-chip case (axial coil separation $h = \SI{50}{\micro\meter}$) and for the on-chip case ($h = \SI{0}{\micro\meter}$), with $\Phi_i$ being the flux in the input coil; for details, see \aref{app:eff_flux}. Note that one can obtain the mutual inductance $M$ between the SQUID loop and the input coil via $M = L_i\cdot\eta_2$, where $L_i$ is the inductance of the input coil. The simulated input coil inductance for the flip-chip coil loop is $\SI{567}{\pico\henry}$ and $\SI{348}{\pico\henry}$ for the on-chip coil loop. As expected, $\eta_2$ is maximum when the loop widths are the same, i.e., $d_i/d_s = 1$. Our implementation achieves loop ratios close to this optimal value, i.e., $d_i/d_s = 1.05$ for the flip-chip case, whereas for the on-chip case, we have to compromise to $d_i/d_s = 1.25$, as the input coil must be slightly larger than the SQUID loop it surrounds. We place the input coil outside the SQUID loop. Placing it inside instead would require an additional air-bridge, which would increase the parasitic coupling loss of the FTR. \fref{fig:Fig_2}(b) shows $\eta_2$ as a function of loop distance $h$. We compare analytic results for $\eta_2$, assuming an idealized geometry of one-dimensional square loops, to FEM simulations of the real geometries of the flip-chip and on-chip cases; see \fref{fig:Fig_2}(c,d). The simulated results follow the trend of the analytical calculation but are slightly lower at smaller separations; see \fref{fig:Fig_2}(b). The reason is that the analytical model assumes a one-dimensional geometry with no coil openings, while the simulations are performed for the actual geometry. The simulation results of both the flip-chip and on-chip methods yield similar efficiencies of about $30\%$. In the flip-chip method, the efficiency is mainly limited by the axial loop separation ($h = \SI{50}{\micro\meter}$), while in the on-chip method, it is limited by the constraint $d_i > d_s$ imposed by routing the input coil around the SQUID loop.

\begin{figure}[t!bhp]
    \centering
    \includegraphics[width=\columnwidth, keepaspectratio]{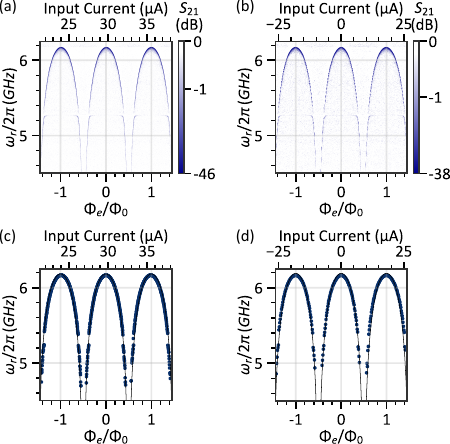}
    \caption{Multiple $\Phi_0$ modulations of an FTR with a \SI{200}{\micro\meter}-wide SQUID modulated at 0.33 intra-cavity photons via (a,c) an external bias coil and (b,d) via a flip-chip-based input coil. (a,b) show the background-subtracted FTR modulation in dependence of coil current and applied flux. (c,d) show the extracted resonance frequency of the FTR (dots), the solid line is the model accounting for finite screening and junction asymmetry.}
    \label{fig:Fig_3}
\end{figure}

We turn now to the characterization of the FTRs. \frefs{fig:Fig_3}(a) and (b) compare external-coil-based and input-coil-based modulation of an FTR with a \SI{200}{\micro\meter}-wide SQUID, respectively. We observe the expected periodic dependence of the resonator frequency over multiple $\Phi_0$ modulations achieved via the external multi-winding bias coil (\fref{fig:Fig_3}(a)) as well as with the flip-chip-based input coil (\fref{fig:Fig_3}(b)). We see no discernible difference between the two modulation types. We observe an avoided crossing close to \SI{5.3}{\giga\hertz}, which we attribute to the hybridization of the FTR with a parasitic slotline mode on the chip. Note that we obtain the primary $x$-axis by converting the applied current $I_\text{in}$ to the external flux seen by the SQUID $\Phi_e = \Phi_0\cdot (I_\text{in} - I_{\rm off})/I_{\Phi_0}$, where $I_{\rm off}$ is the current at which $\Phi_e/\Phi_0$ takes an integer value, and $I_{\Phi_0}$ is the current required for modulation by one \si{\fluxquantum} \cite{Tesche_1977}.

\begin{figure}[!htbp]
    \centering
    \includegraphics[width=\columnwidth, keepaspectratio]{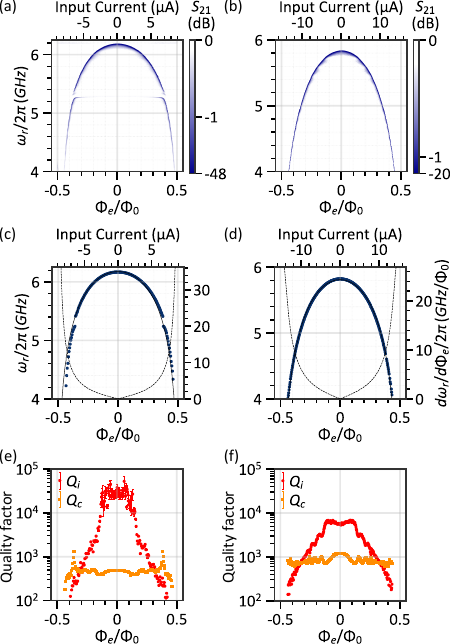}
    \caption{Modulation of FTRs via flip-chip-based (a,c,e) and on-chip-based (b,d,f) input coil measured at 0.33 and 0.83 intra-cavity photons, respectively. (a,b) show background-subtracted plots of the FTR modulation in dependence of input coil current and input flux. (c,d) show the resonance frequency (dots), extracted via a circle-fit routine, as a function of input flux. The solid line and the dotted line are the fit result for the model according to \eref{eq:FTRmod} and its derivative, respectively. (e,f) show the intrinsic and coupling quality factors of the FTRs as a function of input flux.}
    \label{fig:Fig_4}
\end{figure}

\frefs{fig:Fig_3}(c) and (d) show the resonance frequencies of the FTRs extracted as the minima of the scattering parameter $|S_{21}|$ near resonance, after background subtraction. The solid curve shows the model according to \eref{eq:FTRmod} with the circuit parameters extracted from high-resolution measurements shown in \fref{fig:Fig_4}; see also \tref{tab:ftr_fit_results}. We account for the finite screening current and junction asymmetry using \eref{eq:beta_L} and \eref{eq:alpha}, respectively. The model shows good agreement with the data. The FTR does not exhibit branch switching behavior due to the use of the asymmetric dc SQUID, similar to previous work using asymmetric SQUID-like elements \cite{Frattini_2017, Sivak_2019,Hutchings_2017,Garcia_2022}.

\frefs{fig:Fig_4}(a) and (b) show high-resolution measurements of the FTR modulation achieved through the flip-chip and the on-chip input coil, respectively. We use these measurements to extract the parameters of our devices. We observe that we require only tens of \si{\micro\ampere} of current via the chip-based coils to realize a modulation of one flux quantum. \frefs{fig:Fig_4}(c) and (d) show the resonance frequencies of the FTRs determined from a circle-fit routine \cite{Probst_2015, Zoepfl_2020, Zoepfl_2023} (see \aref{SI:data_analysis}). We use the expected device parameters as input to the model fit according to \eref{eq:FTRmod} and obtain fitted parameters that are close to our design values (see \tref{tab:ftr_fit_results}) with a fit result that shows good agreement with the data. Independently, we modulated the FTRs using the external bias coil only (see \frefs{fig:Fig_S3}(a) and (b)) and fitted their tuning behavior (see \frefs{fig:Fig_S3}(c) and (d)). The corresponding fitted parameters (see \tref{tab:ftr_fit_results}) are in good agreement with those determined from the chip-based input coil modulations. Hence, both the chip-based and external-coil-based modulation methods tune the FTR in a similar way. The FTRs reach a flux responsivity $\partial\omega_r/\partial\Phi$ of up to $2\pi\cdot\SI{20}{\giga\hertz\per{\Phi_0}}$, making them suitable devices as sensitive flux detectors.

We use the observed modulation by one flux quantum to determine the flux transfer efficiency $\eta_2$ from its relation to the mutual inductance $\eta_2 = M/L_i$; see \aref{app:eff_flux}. From \frefs{fig:Fig_4}(a) and (b), we extract the input coil current $I_{\Phi_0}$ required for one period of resonance modulation, corresponding to a change of $\Phi_e$ by one flux quantum. With $M = \Phi_0/I_{\Phi_0}$ and using the previously mentioned simulated input coil inductances, we obtain experimental efficiencies of 21\% and 19\%, respectively. These values are lower than the simulated efficiencies of 30\% and 33\%, respectively, which do not account for flux screening in the SQUID loop. The finite SQUID loop inductance causes partial screening of the applied flux, reducing the experimentally determined flux transfer efficiency.

Next, we study the quality factors of the FTRs as a function of chip-based flux biasing. \frefs{fig:Fig_4}(e) and (f) show the intrinsic ($Q_i$) and coupling ($Q_c$) quality factors as a function of input flux. Note that our FTRs are overcoupled near zero flux bias, which may lead to a systematic uncertainty when determining $Q_i$ \cite{Rieger_2023, Baitly_2024}. We observe a maximum $Q_i$ of about $3\cdot10^4$ and $0.7\cdot10^4$ for flip-chip and on-chip devices, respectively, near the zero flux bias point, comparable to previous works \cite{Palacios_2008, Zoepfl_2023} (see \tref{tab:stateoftheart}). These values are similar to those obtained using external coil modulation only (\fref{fig:Fig_S3}), i.e., $3\cdot10^4$ and $0.8\cdot10^4$. Hence, we observe that driving current on-chip is not detrimental to the intrinsic quality factor of the FTRs. Note that we attribute the slightly lower $Q_i$ of the on-chip device to parasitic capacitive coupling between the resonator and the nearby input coil. This effect is suppressed in the flip-chip geometry due to the vertical separation of \SI{50}{\micro\meter} between the input coil and the FTR.

We observe a decrease in the effective $Q_i$ with increasing flux bias, which is consistent with previous works \cite{Palacios_2008, Sandberg_2008, Schmidt_2020, Zoepfl_2023, Schmidt2024}. We attribute this primarily to flux noise-induced linewidth broadening due to the increased flux responsivity away from the zero-flux bias point. The coupling quality factors $Q_c$ at the zero-flux bias point are 490 and 1170 for the flip-chip and on-chip devices, which are close to their design values of 470 and 1250, respectively. The slight variations in $Q_c$ with flux bias arise from flux-dependent changes in the characteristic impedance and resonance frequency of the FTR.

\begin{figure}[!htbp]
    \centering
    \includegraphics[width=\columnwidth, keepaspectratio]{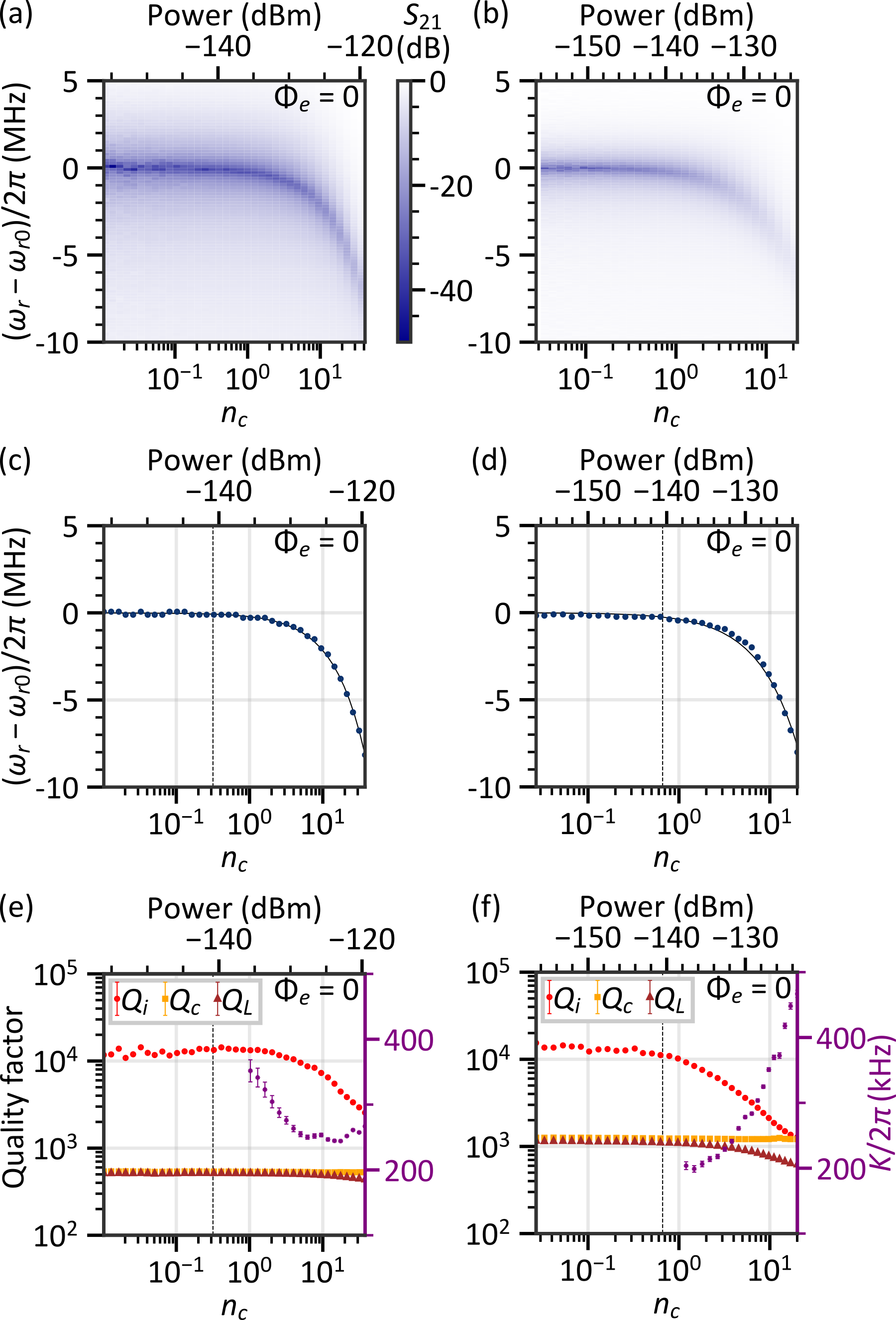}
    \caption{Kerr nonlinearity of FTRs. (a) and (b) show the lowering of the resonance frequency as a function of intra-cavity photon number due to the Kerr coefficient for the flip-chip and on-chip modulated FTRs, respectively. (c,d) Circle fit results obtained from power sweep measurements at the zero-flux-bias point of the FTRs with flip-chip-based input coil (c) and on-chip-based input coil (d). The data is fit using \eref{eq:Kerr} to obtain the Kerr coefficients. (e) and (f) show the intrinsic, coupling, and loaded quality factors ($Q_L^{-1} = Q_i^{-1} + Q_c^{-1}$), and the Kerr coefficient as a function of intra-cavity photons. The dashed lines in both plots show the power at which the flux modulation measurements, shown in \fref{fig:Fig_4}, were taken for the respective devices.}
    \label{fig:Fig_5}
\end{figure}

We also briefly study the nonlinear behavior of the FTR, which is inherited from the nonlinearity of the JJs \cite{Palacios_2008}. To this end, we performed power sweeps at the zero-flux bias point. \frefs{fig:Fig_5}(a) and (b) show the resonance frequency of the FTR for increasing intra-cavity photon numbers. The frequency pulling to lower frequencies, observed at higher powers, is due to the negative Kerr nonlinearity of the FTR. We can model this behavior as \cite{Nation_2008, Diaz-Naufal_2025}
\begin{equation}
    \label{eq:Kerr}
    \omega_r(n_c) = \omega_r(0) - \mathcal{K}n_c,
\end{equation}
where $\omega_r(n_c)$ is the resonance frequency for $n_c$ intra-cavity photons with the flux-dependent Kerr coefficient $\mathcal{K}(\Phi_s)$. From our data shown in \frefs{fig:Fig_5}(c) and (d), we determine $\mathcal{K}/2\pi=\SI{216}{\kilo\hertz\per n_c}$ and $\mathcal{K}/2\pi=\SI{381}{\kilo\hertz\per n_c}$ for the flip-chip device and the on-chip device, respectively. These values are similar to their respective predicted values of \SI{184}{\kilo\hertz\per n_c} and \SI{371}{\kilo\hertz\per n_c}, calculated at zero flux bias. 

\frefs{fig:Fig_5}(e) and (f) show the quality factors for varying intra-cavity photon numbers. While $Q_c$ remains constant, as expected, $Q_i$ decreases with larger drive power, similar to Ref.~\cite{Watanabe_2009}.

Finally, we employed an FTR in a proof-of-principle experiment as a flux detector. To this end, we used an FTR with a \SI{200}{\micro\meter} on-chip input coil to detect a magnetic source via the scheme depicted in \fref{fig:Fig_1}(a). The magnetic source field was generated on an independent chip, which contained a multi-winding coil for creating a magnetic field as well as a pick-up coil; for details of a similar device, see Refs.~\cite{martiIEEE2022,martiPRA2023}. By injecting a current into the multi-winding coil, a magnetic field was generated, which was detected through the pick-up coil with a current-to-flux transduction of \SI{0.27}{\fluxquantum \per \micro\ampere}. The pick-up coil was connected through Al wire bonds to off-chip \SI{5}{\milli\meter} long superconducting Al leads, whose ends were wire-bonded to the input coil connections on the FTR chip. A dc current sweep of \SI{230}{\micro\ampere} through the multi-winding coil resulted in a modulation by one flux quantum of the FTR; see \aref{SI:data_analysis}. This yields a total flux transfer efficiency of 1.6\%, close to the estimate of 1.5\% based on the inductances involved in the circuit. Our efficiency is an order of magnitude larger than that achieved in Ref. \cite{Schmidt2024}, which employed a similar flux transfer architecture.

\section{Conclusion}

We have demonstrated on-chip and flip-chip-based modulation of flux-tunable superconducting resonators. We achieved flux tuning by one flux quantum with tens of \si{\micro\ampere} of current and flux transfer efficiencies from the input coil to the SQUID loop of up to 20\%. This flux transfer efficiency can be further improved by using washer-type SQUIDs \cite{Ketchen_1981, Ketchen_1982, Gross_1990}, which allow for a decrease in loop geometric inductance while keeping the effective area similar \cite{Drung_2007, Granata_2016, Xie_2017}. Furthermore, the use of multiwinding input coils \cite{Dantsker_1997, Xie_2017} would decrease the contribution of parasitic inductance to the total flux transfer efficiency. Some applications may require a reduction in the nonlinearity of the FTR, which could be realized by operating the nonlinear resonator at a Kerr-free point via the use of superconducting nonlinear asymmetric inductive elements (SNAILs) \cite{Frattini_2017, Frattini_2018, Lu_2023, Eriksson_2024} or asymmetrically threaded SQUIDs \cite{Lescanne_2020, Hillmann_2022}. Efficient on-chip modulation of flux-tunable superconducting resonators is key for measuring small magnetic flux signals and for precise and low-current tuning of resonators, as required in quantum sensing and computing applications.

\begin{acknowledgments}
We gratefully acknowledge support from Amr Osman, Anita Fadavi, and Jonas Bylander in device microfabrication, and Irshad Ahmad in SEM imaging. We acknowledge fruitful discussions with the entire team of the SuperMeQ EU project. This work was supported in part by the Horizon Europe 2021-2027 framework program of the European Union under Grant Agreement No. 101080143 (SuperMeQ), the European Research Council under Grant No. 101087847 (ERC Consolidator SuperQLev), the Knut and Alice Wallenberg (KAW) Foundation through a Wallenberg Academy Fellowship and Scholar (WW), and the Wallenberg Center for Quantum Technology (WACQT). Samples were fabricated at the Chalmers Myfab Nanofabrication Laboratory.
\end{acknowledgments}

\section*{Data availability}
Data underlying the results presented in this paper are available in the open-access Zenodo database: \href{https://doi.org/10.5281/zenodo.18079253}{10.5281/zenodo.18079253}.

\clearpage
\appendix
\section{Asymmetric dc SQUID}\label{subsec:dcSQUID}

\begin{figure}[t!bhp]
    \centering
    \includegraphics[width=\columnwidth, keepaspectratio]{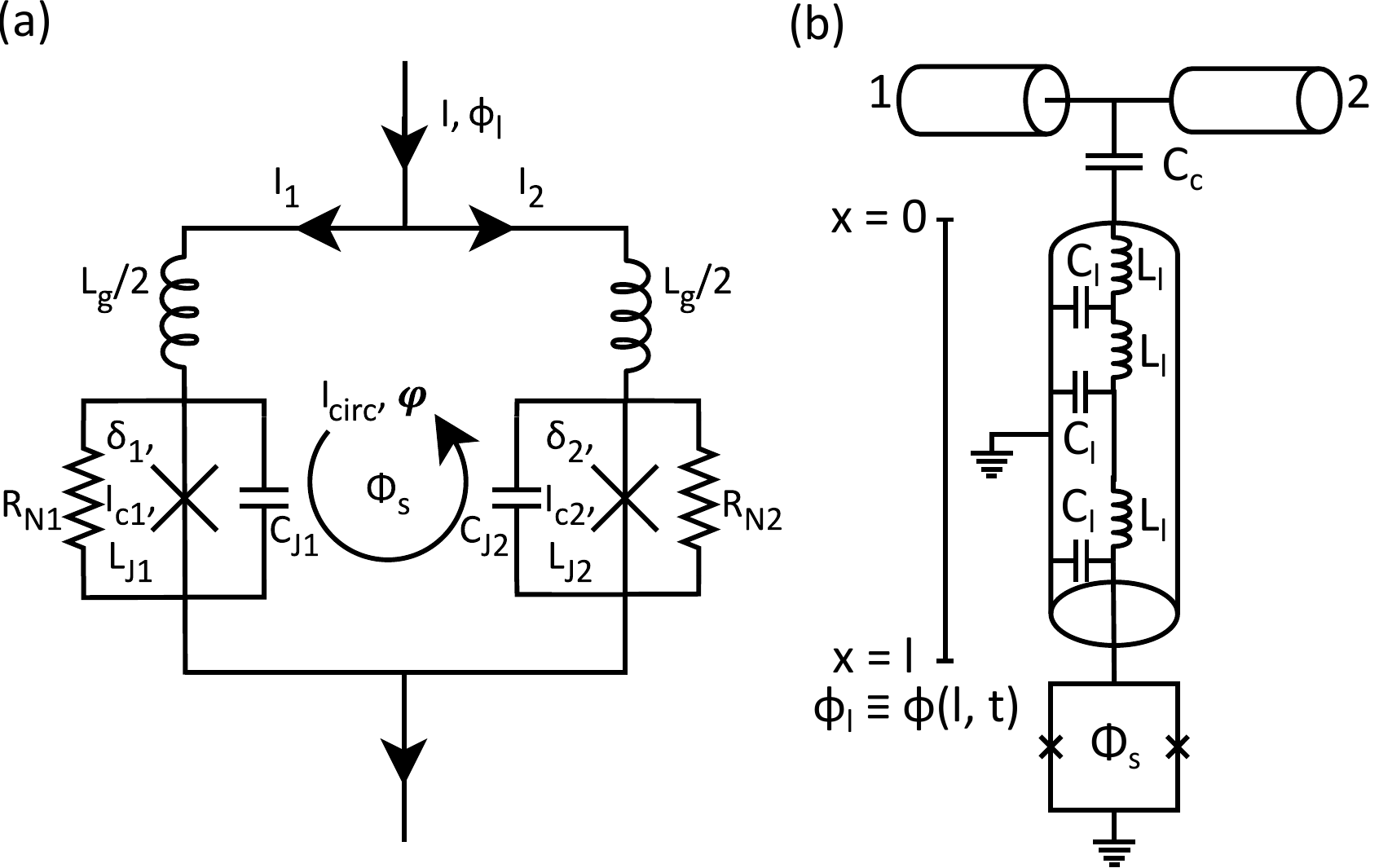}
    \caption{(a) Schematic of a dc SQUID with asymmetric Josephson junctions. (b) Transmission line model of a $\lambda$/4 flux-tunable resonator (FTR) capacitively coupled to a transmission feedline in notch-type configuration.}
    \label{fig:Fig_S0}
\end{figure}

We treat the dc SQUID core as a lumped, lossless, quasi-static termination at the CPW end $x=l$ with asymmetric JJs, similar to Ref.~\cite{Tesche_1977}; see \fref{fig:Fig_S0}. The junction capacitance and shunt are neglected for the static relations, and we assume small-signal operation so that the SQUID behaves as an inductive load. The loop geometry is symmetric, with total geometric inductance $L_g$ split equally between the two arms, so that each arm has geometric inductance $L_{\rm arm}=L_g/2$. A general nonzero transport current $I\equiv I_1+I_2$ can be supplied by the CPW ground termination.

\subsection{Transport and circulating current, screening relation}

\paragraph{Definitions.}

We use the terminal phase (common mode) $\phi_l(t)\equiv \phi(l,t)$ and the loop phase (differential mode) $\varphi(t)$, expressed via the gauge-invariant phase drops across the two Josephson junctions (JJs)
\begin{equation}\label{eq:squid_phases_short_asym}
\delta_1=\phi_l+\varphi,\qquad \delta_2=\phi_l-\varphi .
\end{equation}
The current across each JJ is given by the Josephson relation
\begin{align}
    I_k&=I_{\text{c}k}\sin\delta_k,
\end{align}
with $k=1,2$. We introduce an asymmetry in the critical current of the JJs with the parameter $\alpha$ while keeping the loop geometry symmetric:
\begin{align}\label{eq:asym_def}
\alpha &= (\IcOne-\IcTwo)/(\IcOne+\IcTwo)
\end{align}
and define the mean critical current $I_0$ as
\begin{equation}
\qquad
I_0=(\IcOne+\IcTwo)/2.
\end{equation}
From $I_0$ and $\alpha$, it follows that
\begin{equation}\label{eq:Ic_from_I0_alpha}
\IcOne = I_0(1+\alpha), \qquad
\IcTwo = I_0(1-\alpha).
\end{equation}
Finally, we use the screening parameter $\betaL$ and the reduced external flux $\phi_e$ defined as
\begin{equation}\label{eq:betaSigma}
\betaL\equiv\frac{2 L_g I_0}{\Phi_{0}}, \qquad
\phi_e\equiv\frac{\Phi_e}{\Phi_{0}}.
\end{equation}

\paragraph{Josephson junction RCSJ model.}

We are interested in the small-signal input inductance of the dc SQUID as seen by the CPW node. We treat the SQUID as a linear termination with a small ac perturbation around a chosen static bias
\begin{equation*}
    \delta_{1,2}(t)=\delta_{1,2}+\delta^{\rm ac}_{1,2}(t),\qquad |\delta^{\rm ac}_{1,2}|\ll 1.
\end{equation*}
In the resistively and capacitively shunted junction (RCSJ) model, the small-signal admittance is
\begin{equation*}
    Y_J(\omega)=\frac{1}{i\omega L_J}+i\omega C_J+\frac{1}{R},
\end{equation*}
with $Z_J(\omega)=Y_J(\omega)^{-1}$. In the quasistatic limit with negligible $C_J$ and $R\to\infty$, $Z_J(\omega)\approx i\omega L_J$.

For each junction, we use the small-signal Josephson inductance at the chosen static bias as
\begin{equation}\label{eq:LJk_def}
L_{Jk}(\delta_k) = \frac{\Phi_{0}}{2\pi\,I_{ck} \cos\delta_k},
\qquad k=1,2,
\end{equation}
evaluated at the operating point $(\bar\phi_l,\bar\varphi)$ determined by the external flux and bias current.

\paragraph{Input inductance.}

The inductance of each SQUID arm is given as
\begin{equation}\label{eq:L_series_exact}
L_{\rm arm, 1} \equiv L_{J1}+\frac{L_g}{2},\qquad
L_{\rm arm, 2} \equiv L_{J2}+\frac{L_g}{2}.
\end{equation}

The total SQUID inductance seen by the CPW node is then
\begin{align}\label{eq:LS_exact_parallel_series}
\nonumber L_S&=\frac{L_{\rm arm, 1} L_{\rm arm, 2}}{L_{\rm arm, 1}+L_{\rm arm, 2}}\\
&=\frac{\big(L_{J1}+L_g/2\big)\big(L_{J2}+L_g/2\big)}
{L_{J1}+L_{J2}+L_g}.
\end{align}
The small-signal input inductance $L_S(\bar\varphi,\alpha)$ is then evaluated at the operating point $(\bar\phi_l,\bar\varphi)$ of the SQUID.

\paragraph{Transport current.}

The transport current through the SQUID is given as
\begin{align}\label{eq:transport_general}
I &= I_1+I_2\\\nonumber
&=\IcOne\sin(\phi_l+\varphi)+\IcTwo\sin(\phi_l-\varphi)\\\nonumber
&= 2I_0\cos\varphi\,\sin\phi_l \;+\; 2\alpha I_0\sin\varphi\,\cos\phi_l.
\end{align}
We can rewrite this equation by defining 
\begin{align}\label{eq:Rpsi0_defs}
R(\varphi)&=2I_0\sqrt{\cos^2\varphi+\alpha^2\sin^2\varphi},\\
\psi_0(\varphi)&=\operatorname{atan2}\!\big(\alpha\sin\varphi,\ \cos\varphi\big),
\end{align}
so that we obtain
\begin{equation}\label{eq:I_R_form}
I \;=\; R(\varphi)\,\sin\!\big(\phi_l+\psi_0(\varphi)\big),\, |I|\le R(\varphi).
\end{equation}
This equation can be regarded as the Josephson relation of an effective JJ, whose critical current $R(\varphi)$ is phase dependent. Resolving this equation for $\phi_l$ yields
\begin{align}\label{eq:phi_l_branches}
\phi_l &= -\psi_0(\varphi) + (-1)^n \arcsin\!\Big(\frac{I}{R(\varphi)}\Big) + n\pi,\\
\nonumber& \text{with}\,\arcsin\Big(\frac{I}{R(\varphi)}\Big)\in[-\pi/2,\pi/2],\\
\nonumber& n\in\{0,1\} \text{ and }\,|I|\le R(\varphi).
\end{align}
These two solutions for $\phi_l$ result in two screening branches, as we will see later.

\paragraph{Circulating current.}
The circulating current is given as
\begin{align}\label{eq:I_circ_branch_general}
\nonumber I_{\rm circ}&=\frac{I_1-I_2}{2} \\
&= I_0\left(\alpha\sin\phi_l\cos\varphi+\cos\phi_l\sin\varphi\right).
\end{align}
Rewriting this equation by using \eref{eq:phi_l_branches} for $\phi_l$ and defining
$D(\varphi)= R(\varphi)/(2I_0)$, we get
\begin{widetext}
\begin{equation}\label{eq:Icirc_asym_general_final}
I_{\rm circ}(\varphi) = \pm I_0\frac{(1-\alpha^2)\,\cos\varphi\,\sin\varphi}{D(\varphi)}\,\sqrt{1-\Big(\frac{I}{2I_0 D(\varphi)}\Big)^2}
\;+\; \frac{\alpha}{2}\,\frac{I}{D(\varphi)^2}.
\end{equation}
\end{widetext}
For $I = 0$, this reduces to
\begin{equation}\label{eq:Icirc_I0_main}
I_{\mathrm{circ}}(\varphi) = \pm\, I_0 (1 - \alpha^2) 
\frac{\sin\varphi}{\sqrt{1 + \alpha^2 \tan^2\varphi}}.
\end{equation}

\paragraph{Fluxoid constraint.}

Applying the gauge-invariant fluxoid quantization around the SQUID loop, we obtain
\begin{equation}\label{eq:fluxoid_general}
\oint\!\Big(\nabla\phi - \frac{2\pi}{\Phi_{0}}\mathbf A\Big)\!\cdot d\boldsymbol\ell \;=\; 2\pi m,\, m\in\mathbb Z.
\end{equation}
The phase jumps across the junctions are $\delta_1$ and $-\delta_2$, and the arm contributions provide the self-flux from the circulating current. With external flux $\Phi_e$ and symmetric arms of total inductance $L_g$ carrying a circulating current $I_{\rm circ}$, one obtains \cite{Tesche_1977, LefevreSeguin1992}
\begin{align}\label{eq:fluxoid_asym}
\nonumber\delta_1-\delta_2+\frac{2\pi}{\Phi_{0}}L_g I_{\rm circ}(\varphi)
&= 2\pi\frac{\Phi_e}{\Phi_{0}}+2\pi m\\
\Longleftrightarrow2\varphi+\pi\,\betaL\frac{I_{\rm circ}(\varphi)}{I_0}&=2\pi(\phi_e+m),
\end{align}
where $\delta_{1,2}=\phi_l\pm\varphi$. In the limit $L_g\to 0$ (that is, $\betaL\to 0$), this reduces to $\varphi=\pi(\phi_e+m)$.

Using
\begin{equation}
    \varphi=\pi\Big(\frac{\Phi_s}{\Phi_{0}}+m\Big),
\end{equation}
with the screened flux $\Phi_s$, we can write 
\begin{equation}\label{eq:Phi_s_def_asym_final}
    \Phi_s \equiv \Phi_e - L_g I_{\rm circ}(\varphi).
\end{equation}
\eref{eq:Icirc_asym_general_final} and \eref{eq:Phi_s_def_asym_final} give the implicit $\Phi_s(\Phi_e)$ at finite $I$.

\begin{figure*}[t!bhp]
    \centering
    \includegraphics[width=\textwidth, keepaspectratio]{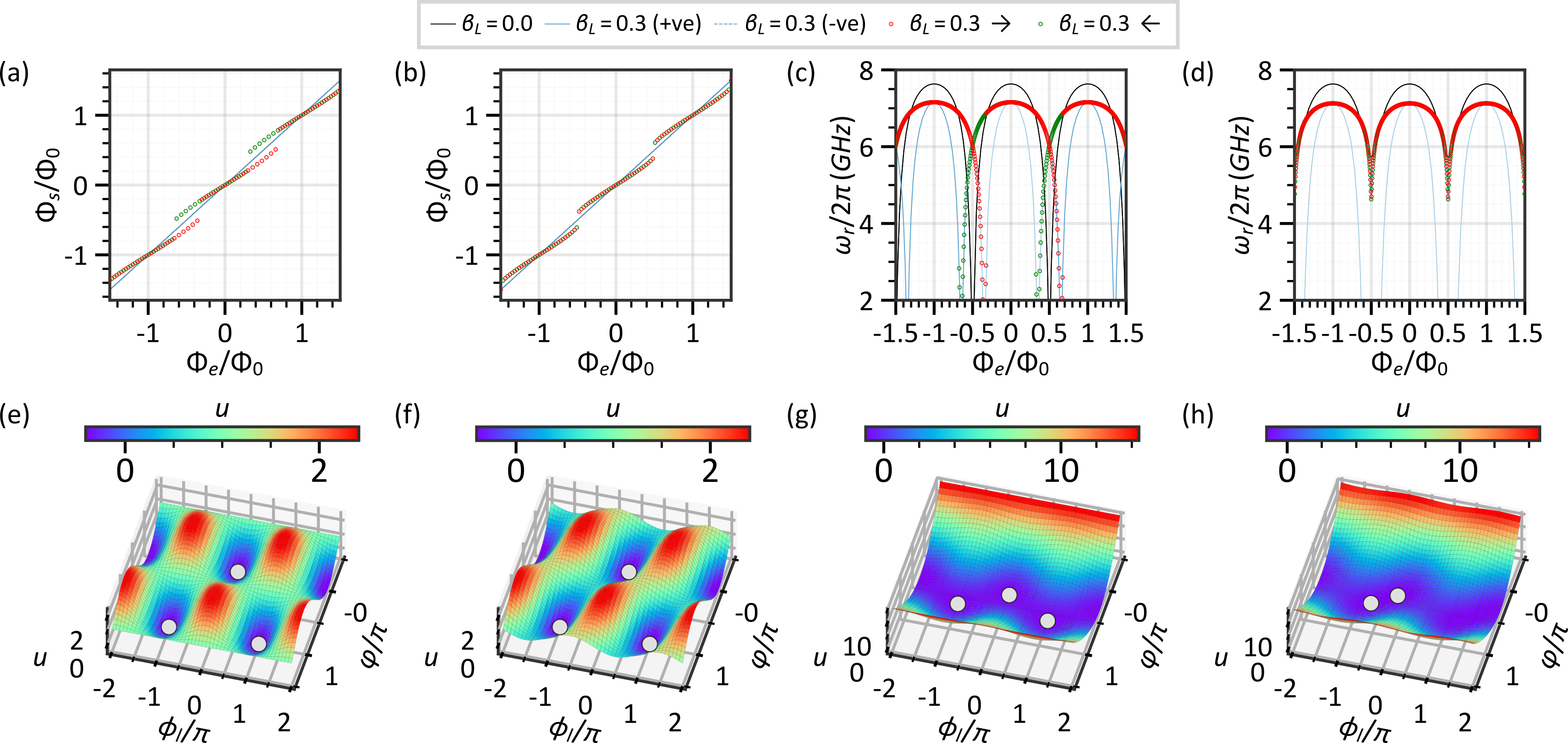}
    \caption{Symmetric vs.~asymmetric dc SQUID. (a,b) Loop flux vs.~applied flux for $\betaL = 0$ and $\betaL = 0.3$ with symmetric ($\alpha=0$) and asymmetric ($\alpha=0.3$) junctions. (c,d) Corresponding FTR resonance frequency vs.~applied flux. (e-h) Normalized dc SQUID potential $U(\phi_l,\varphi)/E_0$ for $\betaL \rightarrow 0$ (e,f) and $0.3$ (g,h), symmetric ($\alpha=0$) and asymmetric ($\alpha=0.3$), at external flux bias $\Phi_e = 0.5\Phi_{0}$.}
    \label{fig:Fig_S1}
\end{figure*}

For $I=0$, representative screening characteristics $\Phi_s(\Phi_e)$ for symmetric ($\alpha=0$) and asymmetric ($\alpha=0.3$) SQUIDs at $\betaL = 0$ and $\betaL = 0.3$ are plotted in \frefs{fig:Fig_S1}(a) and (b).

For sufficiently large screening, \eref{eq:fluxoid_asym} becomes multivalued in $\varphi(\phi_e)$ over finite intervals, leading to a switching between the two screening branches. In the symmetric and zero-transport limit, the onset occurs at $\betaL\ge 2/\pi$.

\subsection{Potential energy}
\label{sec:SQUID_potential}

For static operation (zero voltage), the potential energy of the dc SQUID biased by external flux $\Phi_e$ and transport current $I$ is \cite{Tesche_1977, LefevreSeguin1992, Pogorzalek_2017}
\begin{widetext}
\begin{equation}\label{eq:U_dimensional}
U(\phi_l,\varphi)
= -\frac{\Phi_{0}}{2\pi}\!\left[\IcOne\cos(\phi_l+\varphi)+\IcTwo\cos(\phi_l-\varphi)\right]
\;+\;\frac{1}{2L_g}\!\left[\Phi_{0}\cdot\!\left(\frac{\varphi}{\pi}-m\right)-\Phi_e\right]^{\!2}
\;-\;\frac{\Phi_{0}}{2\pi}\,I\,\phi_l.
\end{equation}
\end{widetext}
The three terms are, respectively, the Josephson energy of the two junctions, the magnetic energy of the loop, and the tilt of the potential washboard due to current bias at the node. The magnetic-energy term in \eref{eq:U_dimensional} accounts for screening through the circulating current (differential mode). Using $\Phi_s=\Phi_{0}\cdot(\varphi/\pi-m)$ and $E_0=\Phi_{0} I_0/2\pi$, $i= I/I_0$, we can rewrite \eref{eq:U_dimensional} as
\begin{widetext}
\begin{equation}\label{eq:u_dimensionless}
u(\phi_l,\varphi)\equiv\frac{U}{E_0}
= -\big(1+\alpha\big)\cos(\phi_l+\varphi)
- \big(1-\alpha\big)\cos(\phi_l-\varphi)
+ \frac{2\pi}{\betaL}\!\left(\frac{\varphi}{\pi}-m-\phi_e\right)^{\!2}
- i\,\phi_l.
\end{equation}
\end{widetext}
Representative landscapes of $u(\phi_l,\varphi)$ for $\betaL = 0$ and $\betaL = 0.3$, and for symmetric ($\alpha=0$) and asymmetric ($\alpha=0.3$) junctions at $\Phi_e = 0.5\Phi_{0}$, are shown in \fref{fig:Fig_S1}(e)-(h).

\paragraph{Branch switching criterion.}
Differentiating with respect to $\phi_l$ and $\varphi$ gives
\begin{align}
\label{eq:U_phi_stationary}\frac{\partial U}{\partial\phi_l}&=\frac{\Phi_{0}}{2\pi}\!\left[\IcOne\sin(\phi_l+\varphi)+\IcTwo\sin(\phi_l-\varphi)-I\right],\\
\label{eq:U_varphi_stationary}
\frac{\partial U}{\partial\varphi}
\nonumber&=\frac{\Phi_{0}}{2\pi}\!\left[\IcOne\sin(\phi_l+\varphi)-\IcTwo\sin(\phi_l-\varphi)\right]\\
&+\frac{\Phi_{0}}{\pi L_g}\!\left[\Phi_{0}\cdot\!\left(\frac{\varphi}{\pi}-m\right)-\Phi_e\right].
\end{align}
When setting each derivative to zero, \eref{eq:U_phi_stationary} gives the transport-current relation for $I$, while \eref{eq:U_varphi_stationary} reduces to the fluxoid constraint \eref{eq:Phi_s_def_asym_final}. A given state $(\phi_l,\varphi)$ is a stable operating point if the Hessian matrix $H(\phi_l,\varphi)$ of $U$ is positive definite (i.e., $\det H>0$).
\begin{widetext}
\begin{align}
\nonumber H(\phi_l,\varphi)&=
\begin{pmatrix}
U_{\phi\phi} & U_{\phi\varphi}\\[4pt]
U_{\phi\varphi} & U_{\varphi\varphi}
\end{pmatrix}\\
&= \frac{\Phi_{0}}{2\pi}
\begin{pmatrix}
\!\IcOne\cos(\phi_l+\varphi)+\IcTwo\cos(\phi_l-\varphi) &
\!\IcOne\cos(\phi_l+\varphi)-\IcTwo\cos(\phi_l-\varphi) \\
\!\IcOne\cos(\phi_l+\varphi)-\IcTwo\cos(\phi_l-\varphi) &
\!\IcOne\cos(\phi_l+\varphi)+\IcTwo\cos(\phi_l-\varphi)+\frac{2\Phi_{0}}{\pi L_g}
\end{pmatrix}.
\end{align}
\end{widetext}

Here, the derivatives are evaluated at the stationary point $(\bar\phi_l,\bar\varphi)$ that satisfies \eref{eq:U_phi_stationary} and \eref{eq:U_varphi_stationary} when set to zero. Branch switching occurs when
\begin{equation}\label{eq:detH_zero_general}
\det H \;=\; U_{\phi\phi}U_{\varphi\varphi}-U_{\phi\varphi}^2 \;=\; 0,
\end{equation}
This simplifies to 
\begin{align}\label{eq:switching_condition_explicit}
 \IcOne\cos(\phi_l+\varphi)+\IcTwo\cos(\phi_l-\varphi)&\\
\nonumber + \pi\betaL I_0\,(1-\alpha^2)\,\cos(\phi_l+\varphi)\cos(\phi_l-\varphi)&=0.
\end{align}

For example, for $\alpha=0$ and $I=0$, the transport equation forces $\sin\phi_l=0$, so $\phi_l=0$. Then \eref{eq:switching_condition_explicit} reduces to
\begin{align*}
\cos\varphi\cdot\big(2+\pi\betaL\cos\varphi\big)&=0.
\end{align*}
A nontrivial turning point is obtained when $\cos\varphi=-2/(\pi\betaL)$, which exists when $\betaL\ge 2/\pi$. In \fref{fig:Fig_S1}(e)-(h), this onset of hysteresis corresponds to the appearance of multiple minima and saddle points in the potential landscape as $\betaL$ is increased from 0 to 0.3 and asymmetry is introduced.

\section{Theory of a flux-tunable resonator}
\label{SI:FTRtheory}

Our FTR consists of a $\lambda/4$ CPW of length $l$ with distributed capacitance $C_l$ and inductance $L_l$ per unit length, and the CPW is terminated with a dc SQUID at the grounded end; see \fref{fig:Fig_S0}(b).
We represent the SQUID termination by a total junction capacitance $C_S$ and by a small-signal termination inductance: in the quasistatic dc limit this is the one-port input inductance $L_S(\bar\varphi,\alpha)$. Shunt resistances and parasitic capacitances are ignored. We now derive how this SQUID inductance modifies the resonator modes and obtain a transcendental equation for the eigenfrequencies, see also Ref.~\cite{Wallquist_2006,Wustmann2013}. We fix coordinates such that the open end of the CPW is at $x=0$ and the grounded end is at $x=l$, where the dc SQUID provides a lumped shunt termination to the ground. We denote the terminal phase at the SQUID end by $\phi_l(t)\equiv\phi(l,t)$.

\subsection{Resonance frequency}

\paragraph{Lagrangian.}

The Lagrangian of a CPW of length $l$ with distributed capacitance $C_l$ and inductance $L_l$ per unit length is \cite{Wallquist_2006,Wustmann2013}
\begin{equation}
\mathcal{L}_{\rm cpw}
=\int_0^l \left[
\frac{C_l}{2}\,\dot{\Phi}^{\,2}
-\frac{1}{2L_l}\left(\partial_x \Phi\right)^{\!2}
\right] dx.
\end{equation}
Substituting the node flux $\Phi(x,t)=\left(\frac{\Phi_0}{2\pi}\right)\phi(x,t)$, where $\phi(x,t)$ is the dimensionless phase, we obtain
\begin{equation}\label{L_cav}
\mathcal{L}_{\rm cpw}[\phi] = \left(\frac{\Phi_0}{2\pi}\right)^2\int_0^l \frac{C_l}{2} \left[\dot{\phi}^2 - v^2 (\phi')^2\right]\,dx\,,
\end{equation}
with the wave speed $v=1/\sqrt{L_l C_l}$. At the grounded end $x=l$, the dc SQUID is treated as a lumped, flux-dependent termination obtained by choosing a static operating point $(\bar\phi_l,\bar\varphi)$ set by the external flux and bias current, and then considering only small ac perturbations around this point.

In this small-signal regime, the internal loop phase ($\varphi$) adjusts quasi-statically to the terminal phase ($\phi_l$), so that the SQUID can be represented, as seen from the CPW node, by a single effective capacitance $C_S$ and a small-signal input inductance $L_S(\bar\varphi,\alpha)$. For the asymmetric SQUID defined in \aref{subsec:dcSQUID}, these parameters are
\begin{align}
\nonumber C_S &= C_{J1} + C_{J2},\\
L_S(\bar\varphi,\alpha) &= \frac{L_{\rm arm, 1} L_{\rm arm, 2}}{L_{\rm arm, 1}+L_{\rm arm, 2}},
\label{eq:junction_params}
\end{align}
with $L_{\rm arm,1}$ and $L_{\rm arm,2}$, evaluated at the operating point $(\bar\phi_l,\bar\varphi)$, as given in \eref{eq:L_series_exact}. The small-signal SQUID Lagrangian, written solely in terms of $\phi_l(t)$, is
\begin{equation}\label{L_SQUID}
\mathcal{L}_S[\phi_l]
= \left(\frac{\Phi_0}{2\pi}\right)^2\left[\frac{C_S}{2}\,\dot{\phi}_l^{\,2}
- \frac{\phi_l^{\,2}}{2L_S(\bar\varphi,\alpha)}\right].
\end{equation}
This is a lumped shunt termination at $x=l$ which modifies only the boundary conditions.
The total Lagrangian is then
\begin{equation}\label{L_total}
\mathcal{L}_{\rm FTR}[\phi,\phi_l] = \mathcal{L}_{\rm cpw}[\phi] + \mathcal{L}_S[\phi_l].
\end{equation}
Varying the field in the interior of the line yields the one-dimensional wave equation,
\begin{align}\label{eq:1D_wave_eqn}
\ddot{\phi}(x,t)-v^2\,\phi''(x,t)&=0.
\end{align}
Since the SQUID acts as a lumped termination at $x=l$, it only modifies the endpoint condition.

In the perfect short-termination limit at $x=l$, the bare $\lambda/4$ fundamental mode has a mode shape
\begin{align*}
u_0(x)&=\cos\!\left(kx\right), 
\end{align*}
with wave vector $k=\pi/2l$. We write the distributed phase as consisting of a time-dependent part multiplied by the spatial profile $u_0(x)$: 
\[
\phi(x,t)=\phi_0(t)\,u_0(x).
\]
Using $u_0'(x)=-k\sin(kx)$ and 
\[
\int_0^l u_0(x)^2\,dx=\frac{l}{2}, \,
\int_0^l [u_0'(x)]^2\,dx=\frac{k^2 l}{2}=\frac{\pi^2}{8l},
\]
yields then for \eref{L_cav}
\[
\mathcal L_{\rm cpw}[\phi]=\Bigl(\frac{\Phi_0}{2\pi}\Bigr)^{\!2}
\left[\frac{C_l}{2}\,\frac{l}{2}\,\dot\phi_0^{\,2}
-\frac{1}{2L_l}\,\frac{\pi^2}{8l}\,\phi_0^{\,2}\right].
\]
We compare this result to a single-degree-of-freedom LC Lagrangian,
\[
\mathcal L_{\rm mode}=\Bigl(\frac{\Phi_0}{2\pi}\Bigr)^{\!2}\left[\frac{C_r}{2}\,\dot\phi_0^{\,2}-\frac{\phi_0^{\,2}}{2L_r}\right],
\]
and read off the modal lumped parameters
\[
\,C_r=\frac{1}{2}\,C_l\,l\,, \qquad
\,L_r=\frac{8}{\pi^2}\,L_l\,l\,.
\]

\paragraph{Boundary condition.}

We now derive the boundary condition for $\phi(x,t)$ at the time-dependent SQUID termination. Using \eref{L_cav} and keeping the endpoint terms, the boundary variation gives the CPW contribution:
\begin{equation}
\left.\frac{\partial\mathcal{L}_{\rm cpw}}{\partial \phi'}\right|_{x}
= -\left(\frac{\Phi_0}{2\pi}\right)^2\frac{1}{L_l}\,\phi'(x,t).
\end{equation}
At the open end $x=0$, the current must vanish. Hence,
\begin{equation}\label{BC_open}
\phi'(0,t)=0.
\end{equation}
At the SQUID end $x=l$, the variation of \eref{L_SQUID} gives
\begin{equation}
\frac{\partial\mathcal{L}_S}{\partial\phi_l}
= -\left(\frac{\Phi_0}{2\pi}\right)^2\frac{\phi_l}{L_{S}},\,
\frac{\partial\mathcal{L}_S}{\partial\dot\phi_l}
= \left(\frac{\Phi_0}{2\pi}\right)^2 C_S\,\dot\phi_l.
\end{equation}
The endpoint condition reads as follows:
\begin{equation}
\left.\frac{\partial\mathcal{L}_{\rm cpw}}{\partial \phi'}\right|_{x=l}
+\frac{\partial\mathcal{L}_S}{\partial\phi_l}
- \frac{d}{dt}\!\left(\frac{\partial\mathcal{L}_S}{\partial\dot\phi_l}\right)=0,
\end{equation}
and yields the lumped load boundary
\begin{equation}\label{BC_l_full}
\phi'(l,t)= -\,L_l\!\left[\frac{\phi_l(t)}{L_{S}} + C_S\,\ddot\phi_l(t)\right].
\end{equation}

In the quasistatic-load limit $C_S\to 0$, this reduces to the purely inductive condition
\begin{equation}\label{BC_final}
\phi'(l,t)= -\,\frac{L_l}{L_{S}}\,\phi_l(t).
\end{equation}
This inductive boundary condition assumes $\omega \ll 1/\sqrt{L_{S}\,C_S}$ evaluated near $\omega\approx\omega_r$ (typically $\sim 2\pi\cdot\SI{26}{\giga\hertz}$ in our devices with $L_{S}\sim\SI{600}{\pico\henry}$ and $C_S\sim\SI{60}{\femto\farad}$). This expression provides the flux-dependent boundary condition of the FTR.

\paragraph{Transcendental equation of an FTR.}

Assuming a harmonic solution $\phi(x,t)=\psi(x)\,e^{-i\omega t}$ with $k=\omega\sqrt{L_lC_l}$ and inserting this ansatz into \eref{eq:1D_wave_eqn}, the spatial amplitude satisfies
\begin{equation}
\psi''(x)+k^2\psi(x)=0. 
\end{equation}
Applying the open-end boundary \eref{BC_open} to the ansatz $\psi(x)=A\cos(kx)+B\sin(kx)$ imposes $\psi'(0)=0$, which gives $B=0$ and thus $\psi(x)=A\cos(kx)$. At $x=l$,
\[
\psi(l)=A\cos(kl),\qquad \psi'(l)=-Ak\sin(kl).
\]
Using \eref{BC_l_full} in the frequency domain $(\ddot\phi_l\to -\omega^2\phi_l)$, we obtain

\begin{equation}\label{tangent_eq_cap}
\tan(kl)=\frac{L_l}{kL_{S}}\bigl(1-\omega^2 C_S L_{S}\bigr).
\end{equation}
Solutions exist for frequencies well below the junction plasma frequency, $\omega \ll \omega_p$, with $\omega_p = 1/\sqrt{L_J(\bar{\varphi})\,C_J} \approx 2\pi\cdot\SI{32}{\giga\hertz}$ for $L_J \approx \SI{800}{\pico\henry}$ and $C_J \approx \SI{30}{\femto\farad}$, and below the series LC resonance associated with the shunt capacitance, $\omega \ll 1/\sqrt{L_{S}\,C_S}$ evaluated near $\omega\approx\omega_r$. In the purely inductive limit, i.e., $C_S\to 0$, this reduces to
\begin{equation}\label{tangent_eq}
\tan(kl)=\frac{L_l}{kL_{S}}.
\end{equation}

For later normalization, we consider the bare quarter-wave limit of the FTR boundary. Taking $L_{S}\to 0$ in \eref{tangent_eq} forces $\tan(kl)\to\infty$; hence,
\[
\cos(kl)=0  \Rightarrow kl=\frac{(2n+1)\pi}{2}, \, n=0,1,2,\ldots
\]
With $k=\omega\sqrt{L_lC_l}$, this gives the bare quarter-wave modes
\begin{equation}\label{eq:bare_quarter_wave}
\omega_n=\frac{(2n+1)\pi}{2l\sqrt{L_lC_l}},
\end{equation}
and for $n=0$:
\begin{equation}\label{eq:bare_omega}
\omega_0=\frac{\pi}{2l\sqrt{L_lC_l}}.
\end{equation}

\paragraph{Solution of the transcendental equation.}
\label{subsec:solution_transcendental}
We now solve Eq.~\eqref{tangent_eq} approximately. We define $\theta$ as
\[
\theta=kl=\omega\,l\sqrt{L_lC_l}=\frac{\pi}{2}\frac{\omega}{\omega_0},
\] and introduce
\begin{equation}\label{eq:participation_ratio}
\gamma(\omega)\equiv\frac{L_{S}}{lL_l},
\end{equation}
which is the SQUID termination inductance participation ratio. With this, we can write the transcendental equation as
\begin{equation}\label{eq:tan-theta}
\tan\theta=\frac{1}{\gamma\,\theta}.
\end{equation}

For the fundamental mode, $\theta$ lies close to $\pi/2$. We set $\theta=\pi/2-\epsilon$ with $|\epsilon|\ll 1$:
\[
\tan\theta=\cot\epsilon=\frac{1}{\epsilon}+O(\epsilon).
\]
Thus, we can write
\[
\theta\tan\theta=\frac{1}{\gamma}
\;\Rightarrow\;
\Bigl(\frac{\pi}{2}-\epsilon\Bigr)\frac{1}{\epsilon}+O(\epsilon)=\frac{1}{\gamma}.
\]
To leading order, this yields
\[
\frac{\pi}{2\epsilon}-1\approx\frac{1}{\gamma}
\;\Rightarrow\;
\epsilon\approx\frac{\pi}{2}\,\frac{\gamma}{1+\gamma}.
\]
Hence
\[
\theta=\frac{\pi}{2}-\epsilon\approx\frac{\pi}{2}\,\frac{1}{1+\gamma}
\quad\Rightarrow\quad
\frac{\omega}{\omega_0}\approx\frac{1}{1+\gamma},
\]
so that
\begin{align}\label{eq:omega-shift}
\omega \approx \frac{\omega_0}{1+\gamma(\omega)},
\qquad
\gamma(\omega)=\frac{L_{S}}{lL_l}.
\end{align}
 
The resulting flux dependence of the fundamental FTR resonance frequency for symmetric ($\alpha=0$) and asymmetric ($\alpha=0.3$) SQUID parameters is shown in \frefs{fig:Fig_S1}(c) and (d), respectively.

\subsection{Evaluation of FTR resonance frequency}
\label{subsec:recipe_analytic_curve}

To obtain $\omega_r(\Phi)$ for non-zero JJ asymmetry $\alpha$ and a finite screening parameter $\betaL$, we use the following steps.

We determine the static dc SQUID operating point: For the eigenmode analysis, we assume a transport current $I=0$. We solve the fluxoid constraint \eref{eq:fluxoid_asym} together with the circulating-current expression \eref{eq:Icirc_asym_general_final} to obtain $\bar\varphi(\Phi_e)$ on the desired screening branch, and obtain $\phi_l(\Phi_e)$ from \eref{eq:phi_l_branches}. Then, we compute the screened flux from \eref{eq:Phi_s_def_asym_final}.

We determine the flux-dependent SQUID inductance: We compute the junction phase drops via \eref{eq:squid_phases_short_asym} and the Josephson inductances via \eref{eq:LJk_def}. We insert these into \eref{eq:LS_exact_parallel_series} to obtain $L_S(\Phi_s)$.

We determine the participation ratio and resonance frequency: We evaluate $\gamma(\Phi_s)$ using \eref{eq:participation_ratio}. Then, we compute $\omega_r(\Phi_s)$ using \eref{eq:omega-shift} with $\omega_0$ as defined in \eref{eq:bare_omega}.

\section{Flux transfer efficiency}
\label{app:eff_flux}

When a current $I_{in}$ flows through the closed loop contour $C_i$ of the input coil, it generates a magnetic vector potential $\mathbf{A}$ at any point in space $\mathbf{r}$. In the Coulomb gauge, this vector potential is given by the Biot–Savart law:
\begin{equation}
\mathbf{A}(\mathbf{r}) = \frac{\mu_0 I_{in}}{4\pi} \oint_{C_i} \frac{d\boldsymbol{\ell}'}{|\mathbf{r} - \mathbf{r}'|}\,,
\label{eq:A_coil}
\end{equation}
where $|\mathbf{r} - \mathbf{r}'|$ is the distance between the source point $\mathbf{r}'$ on the input coil and $\mathbf{r}$.

Ignoring the screening flux due to finite loop inductance, the external flux coupled into the SQUID loop as $\Phi_e$ is defined as the line integral of $\mathbf{A}$ along the SQUID loop contour $C_s$:
\begin{equation}
\Phi_e = \oint_{C_s} \mathbf{A}(\mathbf{r}) \cdot d\boldsymbol{\ell}\,,
\label{eq:flux_via_A}
\end{equation}

Substituting \eref{eq:A_coil} into \eref{eq:flux_via_A} yields:
\begin{equation}
\Phi_e = \frac{\mu_0 I_{in}}{4\pi} \oint_{C_s}\oint_{C_i} \frac{d\boldsymbol{\ell}_s \cdot d\boldsymbol{\ell}_i}{|\mathbf{r}_s - \mathbf{r}_i|}\,.
\label{eq:mutual_inductance}
\end{equation}

This double line integral represents the Neumann formula for mutual inductance between the SQUID loop and the input coil \cite{rosa1908self}. The mutual inductance $M$ relates the current in the input coil to the flux in the SQUID loop through $\Phi_e = M \cdot I_{in}$. From \eref{eq:mutual_inductance}, we can express $M$ as:
\begin{equation}
M = \frac{\mu_0}{4\pi} \oint_{C_s}\oint_{C_i} \frac{d\boldsymbol{\ell}_s \cdot d\boldsymbol{\ell}_i}{|\mathbf{r}_s - \mathbf{r}_i|}\,.
\label{eq:M_explicit}
\end{equation}

The self-inductance of the input coil follows a similar formula, but with both path integrals taken over the same contour $C_i$:
\begin{equation}
L_i = \frac{\mu_0}{4\pi} \oint_{C_i}\oint_{C_i} \frac{d\boldsymbol{\ell}_i \cdot d\boldsymbol{\ell}'_i}{|\mathbf{r}_i - \mathbf{r}'_i|}\,.
\label{eq:L_Self}
\end{equation}

The self-flux in the input coil is then $\Phi_i = L_i \cdot I_{in}$, and the flux efficiency is re-written as:
\begin{equation}
    \eta_2 = \Phi_e/\Phi_i = M/L_i.
\end{equation}

\section{Experimental methods}\label{SI:experiment}

\begin{figure*}[t!bhp]
\centering
{\includegraphics[width=\textwidth, keepaspectratio]{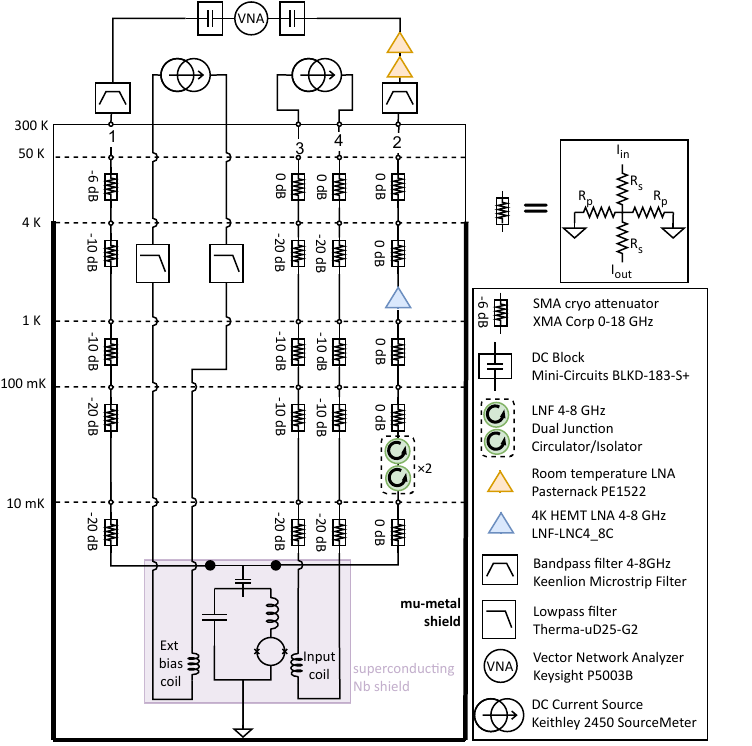}}
\caption{Wiring diagram illustrating the electrical connections and signal paths in the experimental setup.}
\label{fig:Fig_S7}
\end{figure*}

This section describes the microfabrication of devices, as well as the measurement setup and characterization techniques utilized.

\subsection{Device fabrication and assembly}
\label{SI:device}

The flux-tunable resonators are patterned on a $\langle 111\rangle$ silicon chip that is glued to an OFHC copper sample holder using BF-6 glue. The chip contains a bare CPW resonator, an FTR with a \SI{10}{\micro\meter}-wide SQUID loop, and an FTR with a \SI{100}{\micro\meter} or \SI{200}{\micro\meter}-wide SQUID loop that is coupled to an integrated input coil placed either on-chip or via flip chip. The ground plane, junction electrodes, patch layer, and air-bridges are all made of aluminum. To galvanically connect to the flip-chip via indium microspheres, we require an under-bump metallization layer on the connecting pads, which prevents the diffusion of In from the microsphere into the Al layer, forming a non-superconducting intermetallic compound \cite{Wade1973}; for details of fabrication and thin film properties, see Ref.~\cite{paradkar_2025}. This UBM layer is \SI{50}{\nano\meter} thick NbN. The UBM layer is only patterned on the four flip-chip bonding pads; see \fref{fig:Fig_1}(d). The JJ oxidation parameters are \SI{2}{\milli\bar} pressure in a \SI{100}{sccm} static MFC environment for 20\,minutes, which is expected to yield a thin oxide layer of approximately \SI{2}{\nano\meter}. The first and second aluminum junction electrodes are \SI{50}{\nano\meter} and \SI{100}{\nano\meter} thick, respectively, while the patch layer is \SI{200}{\nano\meter} thick.

The width of our Manhattan-style JJ fabricated via the shadow evaporation process cannot reliably exceed \SI{750}{\nano\meter}, as it is limited by the total resist height and the length of the undercut region. Therefore, we fabricated our \SI{1}{\micro\meter} JJs by designing the electrodes to be \SI{500}{\nano\meter}-wide and making only the junction area \SI{1}{\micro\meter}-wide, with an additional \SI{1}{\micro\meter} offset on each electrode to compensate for the resist shadow (see inset of \fref{fig:Fig_1}(b)). This increases the effective overlapping area of the junction, despite some appendages created as an artifact of the shadow evaporation and resist undercut. In our design, one junction was patterned smaller than the other junction, resulting in an asymmetric dc SQUID. 

The sample holder has a machined recess below the chip that leaves an air gap between the substrate and the metallic base. This prevents the chip from forming a large parallel plate capacitor with the ground of the sample holder and reduces parasitic coupling to box modes. The chip ground plane is tied to the sample holder in a controlled manner using aluminum wire bonds distributed along all edges of the chip to provide uniform grounding. The CPW feedline is wire bonded to the exposed, milled center pin of an SMA connector. The dc flux bias lines for the on-chip input coil are wire bonded to a PTFE insulated beryllium copper pin that is soldered to a twisted pair of NbTi wires. An additional external flux bias coil, made from Niobium Titanium (NbTi) wire and wound with more than 1000 turns on a \SI{10}{\milli\meter} bobbin glued to the top of the sample holder, is connected to the cryostat dc wiring via pin headers.

\subsection{Experimental setup}
\label{SI:cryosetup}

An overview of the setup is shown in \fref{fig:Fig_S7}. The sample is mounted to the mixing chamber plate of a BlueFors LD250 dilution refrigerator. It is placed inside a Niobium shield that is wrapped with a mu-metal (annealed soft ferromagnet) sheet. Additional shielding is provided by a tin-plated copper shield fixed to the mixing chamber plate and a mu-metal shield in the vacuum can. The microwave feedlines and the input coil lines are driven via standard microwave wiring of the fridge, with a total of \SI{66}{\decibel} and \SI{60}{\decibel}, respectively. \fref{fig:Fig_S7} also shows the circuit of the attenuators, which reduce the current applied to the chip-based input coils. All measurements in this work were obtained at a base temperature of ~\SI{20}{\milli\kelvin}. The FTR feedline is driven using a Keysight P5003B Vector Network Analyzer (VNA) in transmission (to measure the scattering parameter $S_{21}$), while both the input coil and the external biasing coil are driven using a Keithley 2450 sourcemeter current source.

\subsection{Circle fitting technique}
\label{SI:circlefit}

We analyze resonator traces using the circle-fitting routine from Ref.~\cite{Deeg_2025}. We use the model of a single notch-type microwave resonator in transmission. 

\subsubsection{Linear circle fitting}

In the linear regime, we model the complex transmission of a single-mode notch resonator as
\begin{equation}
S_{21}(\omega)
= 1 - \frac{Q_L}{Q_c^{\mathrm{complex}}}\,
\frac{1}{1+2iQ_L\Big(\frac{\omega-\omega_r}{\omega_r}\Big)},
\label{eq:S21-Qform}
\end{equation}
with a resonance frequency $\omega_r$, a loaded quality factor $Q_L$, and a complex coupling
$Q_c^{\mathrm{complex}}\equiv Q_c e^{i\phi}$ that accounts for small impedance mismatches. The quality factors are related to the decay rates via
\begin{equation}
\frac{1}{Q_L}=\frac{1}{Q_i}+\frac{1}{Q_c},\qquad
Q_L=\frac{\omega_r}{\kappa},\quad Q_c=\frac{\omega_r}{\kappa_c},
\end{equation}
where $\kappa=\kappa_i+\kappa_c$ is the total linewidth. From the complex fit, we extract the effective real coupling
\begin{equation}
\frac{1}{Q_c^{\mathrm{eff}}}=\Re\!\left(\frac{1}{Q_c^{\mathrm{complex}}}\right),
\qquad
\frac{1}{Q_L}=\frac{1}{Q_i}+\frac{1}{Q_c^{\mathrm{eff}}}.
\label{eq:Qeff}
\end{equation}

Before fitting, the raw VNA data are corrected for a smooth complex background: we remove a linear phase ramp and a weak linear tilt in $|S_{21}|$ taken from off-resonant data, which is equivalent to dividing by a complex baseline $S_{\mathrm{bg}}(\omega)$ over the narrow bandwidth of each resonance.

In the complex plane, the background-corrected response traces an approximately circular arc as the probe frequency is swept. For each trace, we fit a circle to the complex data points $S_{21}(f_i)$ using a standard algebraic least-squares procedure to obtain the circle center $z_c$ and radius $r_0$. We then map the off-resonant point on this circle to $1+0i$ by a global complex normalization, and finally remove the effect of the complex coupling $Q_c^{\mathrm{complex}} = Q_c^{\mathrm{abs}} e^{i\phi}$ by rotating the normalized circle around $1+0i$ by the angle $\phi$ and scaling radial distances from this point by $\cos\phi$. This yields the matched-circle representation, in which the circle center lies on the real axis and the radius $r_0^{(\mathrm{matched})}$ obeys
\begin{equation}
  r_0^{(\mathrm{matched})}
  = \frac{Q_L}{2\,Q_c^{\mathrm{eff}}}
  = \frac{\kappa_c}{2\,\kappa}.
  \label{eq:radius-identity}
\end{equation}
In the fit, this identity is included as a soft constraint in addition to the complex residuals of \eref{eq:S21-Qform}, ensuring that the extracted circle parameters remain consistent with the underlying resonator model.

\subsubsection{Nonlinear circle fitting}

For power-dependent traces, we apply the same preprocessing and circle-fitting routine to each trace but interpret the fitted resonance frequency as a Kerr-shifted effective resonance $\omega_r - \mathcal{K} n$ with an intracavity photon number $n$.

At higher drive, the mode behaves as a driven Duffing oscillator with a Kerr coefficient $\mathcal{K}$. The steady state satisfies
\begin{equation}
a=\frac{\sqrt{\kappa_c}\,s_{\mathrm{in}}}{\frac{\kappa}{2}-i\big[\Delta-\mathcal{K} n\big]},\qquad
n=|a|^2,
\label{eq:ss-a}
\end{equation}
with detuning $\Delta=\omega-\omega_r$ and input amplitude $s_{\mathrm{in}}$, such that $|s_{\mathrm{in}}|^2$ is the incident photon flux. We obtain a real cubic equation for $n$ \cite{Deeg_2025},
\begin{equation}
\mathcal{K}^2 n^3-2\Delta \mathcal{K} n^2+\Big(\Delta^2+\frac{\kappa^2}{4}\Big)n-\kappa_c\,|s_{\mathrm{in}}|^2=0.
\label{eq:cubic}
\end{equation}
The power at the device $P_g$ determines $|s_{\mathrm{in}}|^2$ via $|s_{\mathrm{in}}|^2=P_g/(\hbar\omega)$. In the bistable regime, \eref{eq:cubic} has three real roots; the smallest and largest dynamically stable roots define the low and high photon number branches.

The nonlinear transmission follows from evaluating the linear response at the Kerr-shifted resonance,
\begin{equation}
S_{21}^{(\mathrm{nl})}(\omega,n)=1-\frac{\kappa_c}{\frac{\kappa}{2}-i\big[\omega-(\omega_r-\mathcal{K} n)\big]}.
\label{eq:S21-nl}
\end{equation}
This can be written in terms of quality factors and in the matched-circle representation, where $Q_c^{\mathrm{eff}}$ is real, as
\begin{equation}
S_{21}^{(\mathrm{nl})}(f,n)=1-\frac{Q_L}{Q_c^{\mathrm{eff}}}\,
\frac{1}{1+2iQ_L\Big(\frac{f-(f_r-\frac{\mathcal{K}}{2\pi} n)}{f_r}\Big)},
\end{equation}
with $f=\omega/2\pi$. Thus, for each power, the circle fit provides $\omega_r(P)$ and $Q_L(P)$, while the power dependence of $\omega_r(P)$ is evaluated through the Kerr model as a function of the intracavity photon number.

\section{Supplementary data}\label{SI:data_analysis}

\subsection{Bare CPW resonator}
\label{SI:bareCPW}

\begin{figure}[t!bhp]
    \centering
    \includegraphics[width=\columnwidth, keepaspectratio]{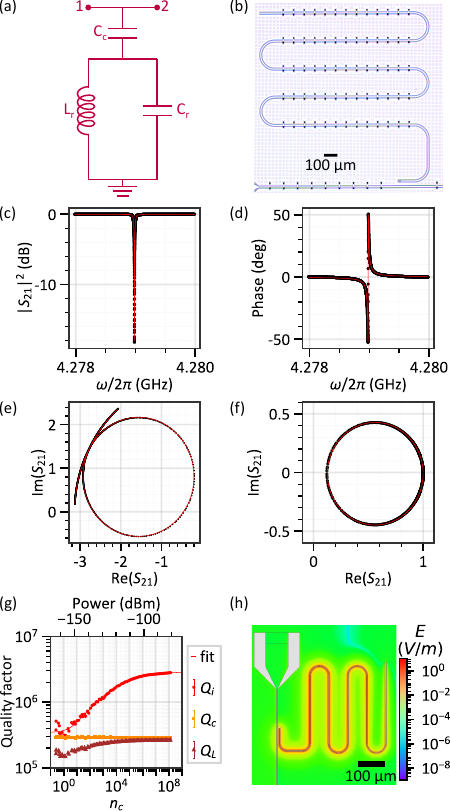}
    \caption{Bare CPW resonator. (a) Lumped-element circuit of a $\lambda/4$ CPW resonator capacitively coupled to a transmission line in notch-type configuration. (b) Optical micrograph of the aluminum CPW with air bridges across the gaps that suppress odd (slotline-like) modes and equalize the electric potential of the ground planes on either side of the center conductor. (c) Magnitude and (d) phase of the transmission coefficient $S_{21}$ around the fundamental resonance. (e) Complex $S_{21}$ data at \SI{-114}{dBm} input power and (f) corresponding normalized circle used for parameter extraction. (g) Internal ($Q_i$), coupling ($Q_c$), and loaded ($Q_L$) quality factors versus intra-cavity photon number $n_{\mathrm{c}}$, together with a two-level system (TLS) saturation fit for $Q_i(n_{\mathrm{c}})$. (h) HFSS eigenmode simulation showing the electric field distribution of the CPW resonator.}
    \label{fig:Fig_S2}
\end{figure}

The CPW resonator is \SI{10}{\micro\meter}-wide with a \SI{2}{\micro\meter} gap from the ground on either side. One end is shorted to ground, while the other end is left open and capacitively coupled to the transmission line for readout, realizing a $\lambda/4$ notch-type resonator with an electric-field antinode at the coupling capacitor. The corresponding lumped-element circuit in \fref{fig:Fig_S2}(a) shows that the resonance is probed as a dip in transmission, and the internal and coupling losses are obtained from the complex $S_{21}$ response. The optical micrograph in \fref{fig:Fig_S2}(b) shows the fabricated aluminum CPW with air bridges across the gaps, which suppress odd (slotline-like) modes and equalize the electric field on both sides of the center conductor, thereby allowing a well-defined quasi-TEM fundamental mode \cite{Chung-Yi_1995, Chen_2014}. The simulated eigenmode in \fref{fig:Fig_S2}(h) confirms that the electric field is concentrated in the CPW gaps and closely follows the expected $\lambda/4$ profile, validating the use of a single-mode description in \aref{SI:FTRtheory}.

We characterize the resonator by measuring the complex transmission parameter $S_{21}$ around the fundamental mode. The magnitude and phase traces in \frefs{fig:Fig_S2}(c) and (d) exhibit the characteristic notch and dispersive phase swing of an overcoupled CPW resonator, indicating that the coupling loss dominates over the internal loss above single-photon powers. To extract reliable quality factors in the presence of cable delays and impedance mismatches, we apply the linear circle-fit procedure described in \aref{SI:experiment} to the complex $S_{21}$ data. \frefs{fig:Fig_S2}(e) and (f) illustrate how the raw complex response at \SI{-40}{dBm} is mapped onto a normalized circle, from which we obtain the internal, coupling, and loaded quality factors $Q_i$, $Q_c$, and $Q_L$, respectively. The power dependence of these quantities, shown in \fref{fig:Fig_S2}(g), reveals that $Q_i$ increases with the intra-cavity photon number while $Q_c$ remains essentially constant.

The observed behavior of $Q_i$ is well described by a standard two-level system (TLS) saturation model \cite{Burnett_2018} of the form
\[
Q_i^{-1}(n_{\mathrm{c}}) = \delta_0 + \frac{\delta_{\mathrm{TLS}}}{1 + \bigl(n_{\mathrm{c}}/n_{\mathrm{c}}^\ast\bigr)^{\beta'}},
\]
where $\delta_0$ is a residual loss term unrelated to TLS, $\delta_{\mathrm{TLS}}$ is the low-power TLS loss contribution, $\beta'$ is the saturation exponent, and $n_{\mathrm{c}}^\ast$ is the characteristic photon number at which the TLS contribution begins to saturate. For the bare CPW resonator, we obtain
\begin{align}
\nonumber \delta_0 &= 3.4\times 10^{-7},\\
\nonumber \delta_{\mathrm{TLS}} &= 2.6\times 10^{-6},\\
\nonumber \beta' &= 0.295,\\
\nonumber n_{\mathrm{c}}^\ast &= 3.30,
\end{align}
which correspond to a low-power quality factor $Q_{i,0} = 1/(\delta_0+\delta_{\mathrm{TLS}}) \approx 3.5\times 10^{5}$ and a high-power saturation value $Q_{i,\infty} = 1/\delta_0 \approx 2.9\times 10^{6}$. The extracted parameters are consistent with loss dominated by TLS \cite{Burnett_2018}. An $n_{\mathrm{c}}^\ast$ of order a few photons, together with $\beta' \simeq 0.3$, indicates that the TLS bath begins to saturate close to the single-photon regime, consistent with observations on aluminum CPW resonators fabricated using a similar process \cite{Biznarova_2024}. 

\subsection{FTR parameters}

\tref{tab:ftr_fit_results} shows the FTR parameters determined from fitting our experimental data.

\begin{figure}[t!bhp]
    \centering
    \includegraphics[width=\columnwidth, keepaspectratio]{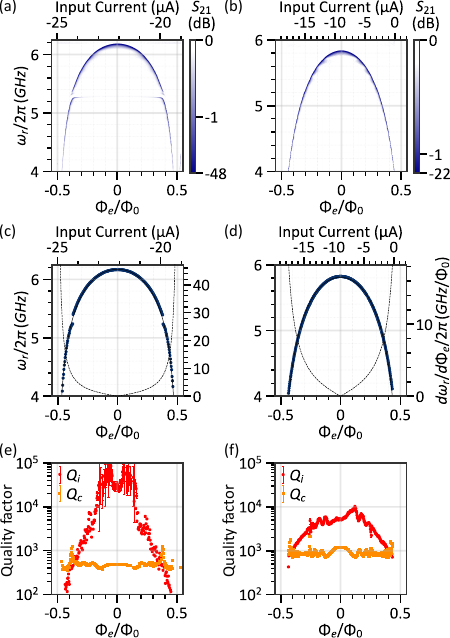}
    \caption{Flux modulation using external bias coil of the flip-chip-based (a,c,e) and on-chip-based (b,d,f) FTRs measured at 0.33 and 0.83 intra-cavity photons, respectively.}
    \label{fig:Fig_S3}
\end{figure}

\fref{fig:Fig_S3} shows the modulation using an external bias coil of the flip-chip and on-chip FTR devices.

\begin{table*}[t!bhp]
    \centering
    \caption{Fitted parameters for three FTRs all with \SI{1}{\micro\meter}-wide Josephson junction: one with a \SI{200}{\micro\meter}-wide SQUID loop modulated via a flip-chip input coil, one with a \SI{100}{\micro\meter}-wide SQUID loop modulated via an on-chip input coil, and a reference device with a \SI{10}{\micro\meter}-wide SQUID loop.}
    \label{tab:ftr_fit_results}
    \begin{ruledtabular}
    \begin{tabular}{l c c c c c c c c c c c}
        Parameter & Symbol & Unit
          & \multicolumn{3}{c}{\SI{200}{\micro\meter} loop FTR}
          & \multicolumn{3}{c}{\SI{100}{\micro\meter} loop FTR}
          & \multicolumn{3}{c}{\SI{10}{\micro\meter} loop FTR} \\
        \addlinespace[2pt]
        \colrule
        \multicolumn{12}{l}{\textbf{CPW parameters}} \\
        \addlinespace[2pt]
        \colrule
        CPW length & $l_{\mathrm{r}}$ & \si{\micro\meter}
          & \multicolumn{3}{c}{3259}
          & \multicolumn{3}{c}{3440}
          & \multicolumn{3}{c}{3320} \\
        CPW inductance & $L_{\mathrm{r}}$ & \si{\pico\henry}
          & \multicolumn{3}{c}{881.2}
          & \multicolumn{3}{c}{930.1}
          & \multicolumn{3}{c}{897.7} \\
        CPW capacitance & $C_{\mathrm{r}}$ & \si{\femto\farad}
          & \multicolumn{3}{c}{354.2}
          & \multicolumn{3}{c}{373.9}
          & \multicolumn{3}{c}{360.9} \\
        CPW frequency & $\omega_{\mathrm{0}}$ & \si{\giga\hertz}
          & \multicolumn{3}{c}{9.008}
          & \multicolumn{3}{c}{8.534}
          & \multicolumn{3}{c}{8.843} \\
        \addlinespace[2pt]
        \colrule
        \multirow{2}{*}{\textbf{SQUID parameters}} & & 
          & \multicolumn{3}{c}{Flip-chip device}
          & \multicolumn{3}{c}{On-chip device}
          & \multicolumn{3}{c}{Reference device} \\
        \cmidrule(lr){4-6}\cmidrule(lr){7-9}\cmidrule(lr){10-12}
        & & & Analytic & Input & Ext & Analytic & Input & Ext & Analytic & Ext \\
        \addlinespace[2pt]
        \colrule
        Current per $\Phi_{0}$          & $I_{\Phi_{0}}$    & \si{\micro\ampere\per\Phi_{0}}    & -  & 17.8  & 5.8   & -  & 30.8  & 20    & -  & 83.0 \\
        Scaling factor                  & $A$               & -                                 & -     & 1.1   & 1.06  & -     & 0.98  & 0.98  & -     & 0.98 \\
        Junction asymmetry              & $\alpha$          & -                                 & ~0.33 & 0.3   & 0.35  & ~0.33 & 0.35  & 0.33  & 0.33  & 0.35 \\
        Avg critical current        & $I_0$             & \si{\nano\ampere}                 & ~400  & 361   & 405   & ~400  & 394   & 400   & 452   & 444 \\
        Avg JJ inductance    & $L_{\mathrm{J}}$  & \si{\pico\henry}                  & 823   & 913   & 812   & 823   & 836   & 823   & 727   & 740 \\
        SQUID inductance          & $L_{\mathrm{S}}$  & \si{\pico\henry}                  & 600   & 663   & 605   & 492   & 509   & 502   & 364   & 380 \\
        Loop inductance       & $L_{\mathrm{g}}$  & \si{\pico\henry}                  & 697   & 776   & 723   & 331   & 331   & 332   & 10.8  & 13.0 \\
        Screening parameter             & $\beta_{L}$       & -                                 & 0.27  & 0.27  & 0.28  & 0.11  & 0.13  & 0.13  & 0.005 & 0.006 \\
        Participation ratio             & $\gamma$          & -                                 & 0.55  & 0.61  & 0.56  & 0.43  & 0.44  & 0.44  & 0.33 & 0.34 \\
    \end{tabular}
    \end{ruledtabular}
\end{table*}

\begin{table*}[t!bhp]
\centering
\caption{FTR parameters of similar devices.}
\label{tab:stateoftheart}
\begin{ruledtabular}
\begin{tabular}{lcccccccc}
Reference
& JJ material 
& $\omega_0/2\pi$\,(\si{\giga\hertz}) 
& $Q_{i}$\,($\Phi_{s}=0$)
& $L_g$\,(\si{\pico\henry})
& loop area\,(\si{\square\micro\metre})
& $L_J$\,(\si{\pico\henry})
& $\gamma$ 
& $\beta_L$ \\
\addlinespace[2pt]
\colrule
Chalmers I \cite{Sandberg_2008} 
& Al/AlOx/Al
& 4.77 
& \SI{10e3}{}
& - 
& \SI{30}{}
& \SI{274}{}
& - 
& - \\
Chalmers II \cite{Svensson_2018} 
& Al/AlOx/Al
& 5.46 
& \SI{400e3}{}
& - 
& \SI{100}{}
& \SI{350}{}
& 0.1 
& - \\
Innsbruck \cite{Zoepfl_2023} 
& Nb/Al/Nb
& 8.18 
& \SI{7.2e3}{}
& \SI{56}{}
& \SI{216}{}
& \SI{17}{} 
& 0.004 
& \SI{1.03}{} \\
TU Delft \cite{Rodrigues2019} 
& Al/AlOx/Al
& 5.22 
& \SI{19e3}{}
& \SI{150}{}
& \SI{90}{}
& \SI{13}{} 
& 0.02 
& \SI{3.7}{} \\
WMI \cite{Luschmann2022} 
& Al/AlOx/Al
& 7.96 
& \SI{6e3}{}
& \SI{19}{}
& \SI{45}{}
& \SI{360}{} 
& 0.23 
& \SI{0.013}{} \\
IQOQI \cite{Schmidt2024} 
& Al/AlOx/Al
& 4.44 
& \SI{0.9e3}{} 
& \SI{120}{}
&\SI{3000}{} 
& \SI{360}{}
& 0.19 
& \SI{0.06}{} \\
\addlinespace[2pt]
\colrule
This Work 
& Al/AlOx/Al
& 6.18 
& \SI{10e3}{}
& \SI{750}{}
& \SI{40000}{}
& \SI{800}{} 
& 0.6 
& \SI{0.28}{} \\
\end{tabular}
\end{ruledtabular}
\end{table*}

\subsection{10-\textmu m-wide loop FTR characterization}
\label{SI:uncoupledFTR}

Our chip design also contained a smaller-loop FTR with a \SI{10}{\micro\meter}-wide SQUID loop and \SI{1}{\micro\meter}-wide JJs. This device has no input coil and is modulated solely by the external bias coil. We use it as a fabrication reference. For input-coil-based modulation of the large-loop FTRs, the external bias coil detunes the \SI{10}{\micro\meter}-loop reference FTR, thereby avoiding mode hybridization (see \aref{SI:data_analysis}).

\begin{figure*}[t!bhp]
    \centering
    \includegraphics[keepaspectratio]{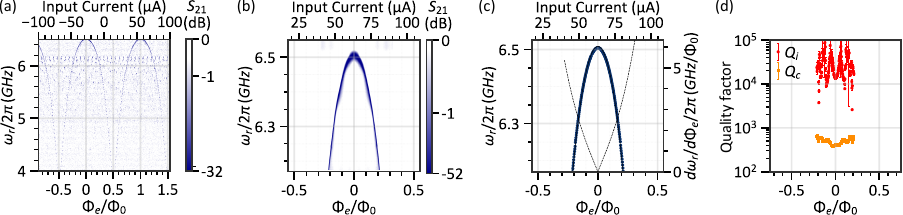}
    \caption{Flux modulation using external bias coil of an FTR with \SI{10}{\micro\meter}-wide SQUID loop and \SI{1}{\micro\meter}-wide Josephson junctions. (a) Multiple $\Phi_0$ modulations showing the \SI{10}{\micro\meter} loop FTR up to a frequency of \SI{6.5}{\giga\hertz} and the \SI{200}{\micro\meter} loop FTR up to a frequency of \SI{6.2}{\giga\hertz}. (b) High-resolution measurement of the \SI{10}{\micro\meter} loop FTR. (c) Resonance frequency and (d) quality factors as a function of flux, extracted via circle fitting, taken at 0.76 intra-cavity photons.}
    \label{fig:Fig_S4}
\end{figure*}

\begin{figure*}[!htbp]
    \centering
    \includegraphics[keepaspectratio]{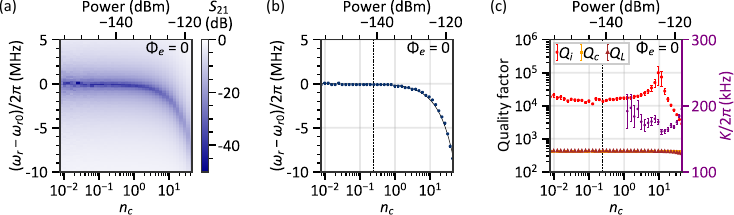}
    \caption{Power sweep of an uncoupled FTR with \SI{10}{\micro\meter}-wide SQUID loop and \SI{1}{\micro\meter}-wide Josephson junctions. We determine a Kerr coefficient of $\mathcal{K}/2\pi = \SI{187}{\kilo\hertz\per n_c}$ by fitting to \eref{eq:Kerr}.}
    \label{fig:Fig_S5}
\end{figure*}

\fref{fig:Fig_S4} shows the flux tuning behavior of the reference FTR. We first take a coarse measurement with multiple modulations to obtain the current required for one flux quantum. Then we take a high-resolution measurement to obtain the FTR's parameters through circle fitting as summarized in \tref{tab:ftr_fit_results}. \fref{fig:Fig_S5} shows a power sweep of the reference FTR. With circle-fit determined $Q_i$ limited by overcoupling, we obtain $Q_i$ close to $10^5$, comparable to state-of-the-art results~\cite{Svensson_2018}; see \tref{tab:stateoftheart}.

\subsection{Flux transfer efficiency}\label{SI:trapmodulation}

\begin{table}[t!bhp]
\centering
\caption{Comparison of flux transfer parameters for flip‐chip and on‐chip approaches.}
\label{tab:fluxeffi_comparison}
\begin{ruledtabular}
\begin{tabular}{l c ccc ccc}
 & & \multicolumn{3}{c}{Flip‐chip device} & \multicolumn{3}{c}{On‐chip device} \\
\addlinespace[2pt]
Parameter & Unit & Cal. & Sim. & Exp. & Cal. & Sim. & Exp. \\
\addlinespace[2pt]
\colrule
$L_{i}$               & pH          & 565  & 567  & -     & 315   & 348   & -    \\
$M=\Phi_0/I_{\Phi_0}$ & pH          & 153  & 169  & 116   & 121   & 116   & 67   \\
$\eta_2$              & \%          & 29.3 & 29.8 & 20.8  & 38.4  & 33.1  & 19.3 \\
\end{tabular}
\end{ruledtabular}
\end{table}

\tref{tab:fluxeffi_comparison} shows the collected values for the flux transfer efficiency between the input coil and the SQUID loop. We determine the inductance for the square coil geometry of length $l$ and wire width $w$ via \cite{Paul_2011}
\begin{align}
L_{\text{coil2}}(l,w) &= \frac{2\mu_0\,l}{\pi}\left[\sqrt{2} - 2 + \ln\!\left(\frac{4l}{w\,(1+\sqrt{2})}\right)\right].
\end{align}

\begin{figure}[t!bhp]
\centering
{\includegraphics[width=\columnwidth, keepaspectratio]{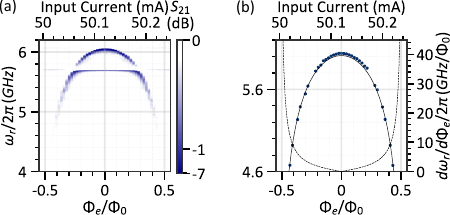}}
\caption{Modulation of a \SI{200}{\micro\meter}-SQUID FTR coupled to an on-chip input coil by running current through a magnetic coil placed concentrically around a pickup coil. Measurements performed at 1.45 intra-cavity photons.}
\label{fig:Fig_S6}
\end{figure}

To obtain the total flux transfer efficiency, we use a pickup coil concentrically placed around an on-chip magnetic field generating coil \cite{martiIEEE2022, martiPRA2023}. From a FEM simulation of the coils, we determine the current-to-flux transduction to be \SI{0.27}{\fluxquantum \per \micro\ampere}. In our experiment (see \fref{fig:Fig_S6}), a current of \SI{230}{\micro\ampere} provides modulation of the FTR by one \si{\fluxquantum}. This implies $\Phi_p = \SI{62}{\Phi_{0}}$ and $\Phi_e = \SI{1}{\Phi_{0}}$, which yield $\eta = \Phi_e/\Phi_p = 1.6\%$. We can compare this value to the expected efficiency determined from the involved inductances as
\begin{equation}
    \label{eq:total_eta}
    \eta = \frac{\Phi_e}{\Phi_p} = \frac{M}{L_p + L_{\rm wire} + L_i} =1.5\%,
\end{equation}
with a mutual inductance of $M = \SI{469}{\pico\henry}$ between the \SI{210}{\micro\meter}-wide input coil and the \SI{200}{\micro\meter}-wide SQUID loop, a simulated pickup coil inductance of $L_p = \SI{2.02}{\nano\henry}$, a simulated input coil inductance of $L_i =\SI{485}{\pico\henry}$, and an analytically estimated parasitic inductance of $L_{p} \approx \SI{28.7}{\nano\henry}$. This parasitic inductance is given by two dc pins made of superconducting aluminum that connect the pickup coil to the input coil via bond wires. The single bond wire inductance is $L_{\rm bondwire} =\SI{1.65}{\nano\henry}$ (in total, four have been used), and the single dc pin inductance is $L_{\rm DCpin} = \SI{11.0}{\nano\henry}$, both estimated for a cylindrical geometry of length $l$ and radius $r$ using \cite{Qi_2001}
\begin{equation}
    L = \frac{\mu_0l}{2\pi}\biggl(\log\frac{2l}{r} - 0.75\biggr).
\end{equation}

\clearpage
\section{Symbols and parameters}

\begin{longtable*}{lll}
\caption{Symbols and parameters used.\label{tab:symbols}}\\
\toprule
Symbol & Description & Unit \\
\midrule
\endfirsthead

\multicolumn{3}{c}{{\tablename\ \thetable{} -- continued from previous page}}\\
\toprule
Symbol & Description & Unit \\
\midrule
\endhead

\bottomrule
\endfoot

\bottomrule
\endlastfoot

\multicolumn{3}{l}{\textbf{Geometry and material parameters}}\\
\addlinespace[2pt]
$l$                      & length of coplanar waveguide                                  & \si{\meter} \\
$x\in[0,l]$              & Longitudinal coordinate; $x=0$ open, $x=l$ SQUID end          & \si{\meter} \\
$d_s$                    & SQUID loop width (linear size)                                & \si{\meter} \\
$d_i$                    & Input coil loop width                                         & \si{\meter} \\
$h$                      & Axial separation between input coil and SQUID loop            & \si{\meter} \\
$C_l$                    & CPW capacitance per unit length                               & \si{\farad\per\meter} \\
$L_l$                    & CPW inductance per unit length                                & \si{\henry\per\meter} \\
$C_r$                    & Effective quarter-wave modal capacitance                      & \si{\farad} \\
$L_r$                    & Effective quarter-wave modal inductance                       & \si{\henry} \\
$L_g$                    & SQUID loop geometric inductance                               & \si{\henry} \\
$L_i$                    & Self inductance of the input coil                             & \si{\henry} \\
$M$                      & Mutual inductance (input coil to SQUID loop)                  & \si{\henry} \\

\midrule
\multicolumn{3}{l}{\textbf{Fields, fluxes, and phases}}\\
\addlinespace[2pt]
$\Phi(x,t)$              & Node flux along the CPW                                       & \si{\weber} \\
$\phi(x,t)$              & Dimensionless phase, $\Phi=(\Phi_0/2\pi)\phi$                 & - \\
$\Phi_l(t)$              & Node flux at the SQUID termination ($x=l$)                    & \si{\weber} \\
$\phi_l(t)$              & Terminal phase at $x=l$, $\phi_l\equiv\phi(l,t)$              & - \\
$\varphi(t)$             & Loop (differential) SQUID phase                               & - \\
$\bar{\varphi}$          & Static bias value of the loop phase                           & - \\
$I_1,\,I_2$              & Currents through junctions 1 and 2                            & \si{\ampere} \\
$I$                      & Transport current supplied by the CPW                         & \si{\ampere} \\
$I_{\rm circ}$           & Circulating current in the SQUID loop                         & \si{\ampere} \\
$\Phi_e$                 & External flux threading the SQUID loop                        & \si{\weber} \\
$\Phi_s$                 & Screened (internal) SQUID loop flux                           & \si{\weber} \\
$\Phi_p$                 & Flux in the pickup coil                                       & \si{\weber} \\
$\Phi_i$                 & Self flux of the input coil                                   & \si{\weber} \\
$\phi_{\mathrm e}$       & Reduced external flux, $\phi_{\mathrm e}=\Phi_e/\Phi_0$       & - \\
$\Phi_0$                 & Flux quantum, $\Phi_0=h/2e$                                   & \si{\weber} \\
$\mathbf{A}(\mathbf{r})$ & Magnetic vector potential at position $\mathbf{r}$            & \si{\weber\per\meter} \\

\midrule
\multicolumn{3}{l}{\textbf{Spectral and mode quantities}}\\
\addlinespace[2pt]
$v$                      & Wave speed on the CPW, $v=1/\sqrt{L_l C_l}$                   & \si{\meter\per\second} \\
$k$                      & Wavenumber, $k=\omega\sqrt{L_l C_l}=\omega/v$                 & \si{\per\meter} \\
$\psi(x)$                & Mode profile in $\phi(x,t)=\psi(x)e^{-i\omega t}$             & - \\
$u_0(x)$                 & Bare quarter-wave mode shape, $u_0(x)=\cos(kx)$               & - \\
$\omega$                 & Angular drive frequency                                       & \si{\radian\per\second} \\
$\omega_r$               & Angular resonance frequency of the FTR                        & \si{\radian\per\second} \\
$\omega_r(\Phi_s)$       & Flux dependent resonance frequency                            & \si{\radian\per\second} \\
$\omega_0$               & Bare quarter-wave fundamental CPW frequency                   & \si{\radian\per\second} \\
$\theta$                 & Electrical length, $\theta=kl=(\pi/2)(\omega/\omega_0)$       & - \\
$\gamma$                 & SQUID inductance participation ratio $\gamma = L_S/(lL_l)$    & - \\

\midrule
\multicolumn{3}{l}{\textbf{Lagrangians and effective circuit parameters}}\\
\addlinespace[2pt]
$\mathcal{L}_{\rm cpw}$      & Lagrangian of the distributed CPW                         & \si{\joule} \\
$\mathcal{L}_S$              & Effective SQUID Lagrangian                                & \si{\joule} \\
$\mathcal{L}_{\rm FTR}$      & Total Lagrangian (CPW plus SQUID termination)             & \si{\joule} \\
$C_S$                        & Effective SQUID capacitance                               & \si{\farad} \\
$L_J(\bar{\varphi})$         & Average Josephson inductance at loop bias $\bar{\varphi}$ & \si{\henry} \\
$L_S(\bar{\varphi})$         & Effective SQUID inductance seen by the CPW                & \si{\henry} \\
$L_{\rm arm,1},L_{\rm arm,2}$& Series inductances of the two SQUID arms                  & \si{\henry} \\

\midrule
\multicolumn{3}{l}{\textbf{dc SQUID parameters and screening}}\\
\addlinespace[2pt]
$\delta_1,\delta_2$       & Junction phase drops across junctions 1 and 2                & - \\
$I_{ck}$                 & Critical current of a single $k^{\rm th}$ junction             & \si{\ampere} \\
$\IcOne,\IcTwo$           & Critical currents of junctions 1 and 2                       & \si{\ampere} \\
$I_0$                     & Average junction critical current $(\IcOne+\IcTwo)/2$        & \si{\ampere} \\
$\alpha$                  & Junction asymmetry $(\IcOne-\IcTwo)/(\IcOne+\IcTwo)$         & - \\
$\betaL$                  & Screening parameter, $\betaL=2L_g I_0/\Phi_0$                & - \\
$L_{J1},L_{J2}$           & Josephson inductances of junctions 1 and 2                   & \si{\henry} \\
$C_J$                     & Junction capacitance                                         & \si{\farad} \\
$L_S(\bar\varphi,\alpha)$ & Small-signal input inductance of the SQUID at bias $(\bar\varphi,\alpha)$ & \si{\henry} \\
$\omega_p$                & Junction plasma frequency, $\omega_p = 1/\sqrt{L_J(\bar\varphi)\,C_J}$ & \si{\radian\per\second} \\
$R(\varphi)$              & Amplitude in $I=R(\varphi)\sin(\phi_l+\psi_0)$               & \si{\ampere} \\
$\psi_0(\varphi)$         & Phase offset in transport relation                           & - \\
$U(\phi_l,\varphi)$       & SQUID potential energy                                       & \si{\joule} \\
$u(\phi_l,\varphi)$       & Dimensionless potential, $u=U/E_0$                           & - \\

\midrule
\multicolumn{3}{l}{\textbf{Mode quantization and nonlinearity}}\\
\addlinespace[2pt]
$\Phi_r$                 & Generalized mode flux at the open end                         & \si{\weber} \\
$Q_r$                    & Conjugate charge of the mode                                  & \si{\coulomb} \\
$\Phi_{\rm zpf}$         & Zero-point flux fluctuation of the mode                       & \si{\weber} \\
$\lambda(\Phi_s)$        & Termination factor, $\lambda = \phi_l/\phi_0 = \cos\theta(\Phi_s)$ & - \\

\midrule
\multicolumn{3}{l}{\textbf{Flux transfer and transformer parameters}}\\
\addlinespace[2pt]
$\eta_2$                 & Flux transfer efficiency, $\eta_2=\Phi_e/\Phi_i=M/L_i$        & - \\
$\eta$                   & Total pickup to SQUID flux efficiency, $\eta=\Phi_e/\Phi_p$   & - \\
$I_{in}$                 & Input coil current                                            & \si{\ampere} \\
$I_{off}$                & Coil current corresponding to $\Phi_e=0$                      & \si{\ampere} \\
$I_{\Phi_0}$             & Coil current needed to change $\Phi_e$ by one $\Phi_0$        & \si{\ampere} \\[4pt]

\midrule
\multicolumn{3}{l}{\textbf{Quality factors and Kerr nonlinearity}}\\
\addlinespace[2pt]
$Q_i$                    & Internal quality factor                                       & - \\
$Q_c$                    & Coupling quality factor                                       & - \\
$Q_L$                 & Loaded quality factor, $1/Q_L=1/Q_i+1/Q_c$                 & - \\
$n_c$                    & Intra cavity photon number                                    & - \\
$\mathcal{K}$            & Kerr coefficient in angular frequency units                   & \si{\radian\per\second} \\

\midrule
\multicolumn{3}{l}{\textbf{Input-output and circle fit}}\\
\addlinespace[2pt]
$a$                      & Intracavity field amplitude                                   & - \\
$\kappa$                 & Total decay rate, $\kappa=\kappa_i+\kappa_c$                  & \si{\radian\per\second} \\
$\Delta$                 & Detuning, $\Delta=\omega-\omega_r$                            & \si{\radian\per\second} \\
$S_{21}$                 & Complex transmission coefficient                              & - \\
$\phi$                   & Phase offset from impedance mismatch                          & - \\
$\tau$                   & Electrical delay of the microwave line                        & \si{\second} \\[4pt]

\bottomrule
\end{longtable*}

\clearpage
\bibliographystyle{apsrev4-2}
\bibliography{references}

\begin{thebibliography}{82}%
\makeatletter
\providecommand \@ifxundefined [1]{%
 \@ifx{#1\undefined}
}%
\providecommand \@ifnum [1]{%
 \ifnum #1\expandafter \@firstoftwo
 \else \expandafter \@secondoftwo
 \fi
}%
\providecommand \@ifx [1]{%
 \ifx #1\expandafter \@firstoftwo
 \else \expandafter \@secondoftwo
 \fi
}%
\providecommand \natexlab [1]{#1}%
\providecommand \enquote  [1]{``#1''}%
\providecommand \bibnamefont  [1]{#1}%
\providecommand \bibfnamefont [1]{#1}%
\providecommand \citenamefont [1]{#1}%
\providecommand \href@noop [0]{\@secondoftwo}%
\providecommand \href [0]{\begingroup \@sanitize@url \@href}%
\providecommand \@href[1]{\@@startlink{#1}\@@href}%
\providecommand \@@href[1]{\endgroup#1\@@endlink}%
\providecommand \@sanitize@url [0]{\catcode `\\12\catcode `\$12\catcode
  `\&12\catcode `\#12\catcode `\^12\catcode `\_12\catcode `\%12\relax}%
\providecommand \@@startlink[1]{}%
\providecommand \@@endlink[0]{}%
\providecommand \url  [0]{\begingroup\@sanitize@url \@url }%
\providecommand \@url [1]{\endgroup\@href {#1}{\urlprefix }}%
\providecommand \urlprefix  [0]{URL }%
\providecommand \Eprint [0]{\href }%
\providecommand \doibase [0]{https://doi.org/}%
\providecommand \selectlanguage [0]{\@gobble}%
\providecommand \bibinfo  [0]{\@secondoftwo}%
\providecommand \bibfield  [0]{\@secondoftwo}%
\providecommand \translation [1]{[#1]}%
\providecommand \BibitemOpen [0]{}%
\providecommand \bibitemStop [0]{}%
\providecommand \bibitemNoStop [0]{.\EOS\space}%
\providecommand \EOS [0]{\spacefactor3000\relax}%
\providecommand \BibitemShut  [1]{\csname bibitem#1\endcsname}%
\let\auto@bib@innerbib\@empty
\bibitem [{\citenamefont {Devoret}\ and\ \citenamefont
  {Schoelkopf}(2013)}]{Devoret_2013}%
  \BibitemOpen
  \bibfield  {author} {\bibinfo {author} {\bibfnamefont {M.~H.}\ \bibnamefont
  {Devoret}}\ and\ \bibinfo {author} {\bibfnamefont {R.~J.}\ \bibnamefont
  {Schoelkopf}},\ }\href {https://doi.org/10.1126/science.1231930} {\bibfield
  {journal} {\bibinfo  {journal} {Science}\ }\textbf {\bibinfo {volume}
  {339}},\ \bibinfo {pages} {1169–1174} (\bibinfo {year} {2013})}\BibitemShut
  {NoStop}%
\bibitem [{\citenamefont {Wendin}(2017)}]{Wendin_2017}%
  \BibitemOpen
  \bibfield  {author} {\bibinfo {author} {\bibfnamefont {G.}~\bibnamefont
  {Wendin}},\ }\href {https://doi.org/10.1088/1361-6633/aa7e1a} {\bibfield
  {journal} {\bibinfo  {journal} {Reports on Progress in Physics}\ }\textbf
  {\bibinfo {volume} {80}},\ \bibinfo {pages} {106001} (\bibinfo {year}
  {2017})}\BibitemShut {NoStop}%
\bibitem [{\citenamefont {Krantz}\ \emph {et~al.}(2019)\citenamefont {Krantz},
  \citenamefont {Kjaergaard}, \citenamefont {Yan}, \citenamefont {Orlando},
  \citenamefont {Gustavsson},\ and\ \citenamefont {Oliver}}]{Krantz_2019}%
  \BibitemOpen
  \bibfield  {author} {\bibinfo {author} {\bibfnamefont {P.}~\bibnamefont
  {Krantz}}, \bibinfo {author} {\bibfnamefont {M.}~\bibnamefont {Kjaergaard}},
  \bibinfo {author} {\bibfnamefont {F.}~\bibnamefont {Yan}}, \bibinfo {author}
  {\bibfnamefont {T.~P.}\ \bibnamefont {Orlando}}, \bibinfo {author}
  {\bibfnamefont {S.}~\bibnamefont {Gustavsson}},\ and\ \bibinfo {author}
  {\bibfnamefont {W.~D.}\ \bibnamefont {Oliver}},\ }\href
  {https://doi.org/10.1063/1.5089550} {\bibfield  {journal} {\bibinfo
  {journal} {Applied Physics Reviews}\ }\textbf {\bibinfo {volume} {6}},\
  \bibinfo {pages} {021318} (\bibinfo {year} {2019})}\BibitemShut {NoStop}%
\bibitem [{\citenamefont {Blais}\ \emph {et~al.}(2021)\citenamefont {Blais},
  \citenamefont {Grimsmo}, \citenamefont {Girvin},\ and\ \citenamefont
  {Wallraff}}]{Blais_2021}%
  \BibitemOpen
  \bibfield  {author} {\bibinfo {author} {\bibfnamefont {A.}~\bibnamefont
  {Blais}}, \bibinfo {author} {\bibfnamefont {A.~L.}\ \bibnamefont {Grimsmo}},
  \bibinfo {author} {\bibfnamefont {S.}~\bibnamefont {Girvin}},\ and\ \bibinfo
  {author} {\bibfnamefont {A.}~\bibnamefont {Wallraff}},\ }\href
  {https://doi.org/10.1103/revmodphys.93.025005} {\bibfield  {journal}
  {\bibinfo  {journal} {Reviews of Modern Physics}\ }\textbf {\bibinfo {volume}
  {93}},\ \bibinfo {pages} {025005} (\bibinfo {year} {2021})}\BibitemShut
  {NoStop}%
\bibitem [{\citenamefont {Hutchings}\ \emph {et~al.}(2017)\citenamefont
  {Hutchings}, \citenamefont {Hertzberg}, \citenamefont {Liu}, \citenamefont
  {Bronn}, \citenamefont {Keefe}, \citenamefont {Brink}, \citenamefont {Chow},\
  and\ \citenamefont {Plourde}}]{Hutchings_2017}%
  \BibitemOpen
  \bibfield  {author} {\bibinfo {author} {\bibfnamefont {M.~D.}\ \bibnamefont
  {Hutchings}}, \bibinfo {author} {\bibfnamefont {J.~B.}\ \bibnamefont
  {Hertzberg}}, \bibinfo {author} {\bibfnamefont {Y.}~\bibnamefont {Liu}},
  \bibinfo {author} {\bibfnamefont {N.~T.}\ \bibnamefont {Bronn}}, \bibinfo
  {author} {\bibfnamefont {G.~A.}\ \bibnamefont {Keefe}}, \bibinfo {author}
  {\bibfnamefont {M.}~\bibnamefont {Brink}}, \bibinfo {author} {\bibfnamefont
  {J.~M.}\ \bibnamefont {Chow}},\ and\ \bibinfo {author} {\bibfnamefont
  {B.~L.~T.}\ \bibnamefont {Plourde}},\ }\href
  {https://doi.org/10.1103/PhysRevApplied.8.044003} {\bibfield  {journal}
  {\bibinfo  {journal} {Phys. Rev. Appl.}\ }\textbf {\bibinfo {volume} {8}},\
  \bibinfo {pages} {044003} (\bibinfo {year} {2017})}\BibitemShut {NoStop}%
\bibitem [{\citenamefont {Mergenthaler}\ \emph {et~al.}(2021)\citenamefont
  {Mergenthaler}, \citenamefont {Müller}, \citenamefont {Ganzhorn},
  \citenamefont {Paredes}, \citenamefont {Müller}, \citenamefont {Salis},
  \citenamefont {Adiga}, \citenamefont {Brink}, \citenamefont {Sandberg},
  \citenamefont {Hertzberg},\ and\ \citenamefont {et~al.}}]{Mergenthaler_2021}%
  \BibitemOpen
  \bibfield  {author} {\bibinfo {author} {\bibfnamefont {M.}~\bibnamefont
  {Mergenthaler}}, \bibinfo {author} {\bibfnamefont {C.}~\bibnamefont
  {Müller}}, \bibinfo {author} {\bibfnamefont {M.}~\bibnamefont {Ganzhorn}},
  \bibinfo {author} {\bibfnamefont {S.}~\bibnamefont {Paredes}}, \bibinfo
  {author} {\bibfnamefont {P.}~\bibnamefont {Müller}}, \bibinfo {author}
  {\bibfnamefont {G.}~\bibnamefont {Salis}}, \bibinfo {author} {\bibfnamefont
  {V.~P.}\ \bibnamefont {Adiga}}, \bibinfo {author} {\bibfnamefont
  {M.}~\bibnamefont {Brink}}, \bibinfo {author} {\bibfnamefont
  {M.}~\bibnamefont {Sandberg}}, \bibinfo {author} {\bibfnamefont {J.~B.}\
  \bibnamefont {Hertzberg}},\ and\ \bibinfo {author} {\bibnamefont {et~al.}},\
  }\href {https://doi.org/10.1038/s41534-021-00491-2} {\bibfield  {journal}
  {\bibinfo  {journal} {npj Quantum Information}\ }\textbf {\bibinfo {volume}
  {7}},\ \bibinfo {pages} {157} (\bibinfo {year} {2021})}\BibitemShut {NoStop}%
\bibitem [{\citenamefont {Ch\'avez-Garcia}\ \emph {et~al.}(2022)\citenamefont
  {Ch\'avez-Garcia}, \citenamefont {Solgun}, \citenamefont {Hertzberg},
  \citenamefont {Jinka}, \citenamefont {Brink},\ and\ \citenamefont
  {Abdo}}]{Garcia_2022}%
  \BibitemOpen
  \bibfield  {author} {\bibinfo {author} {\bibfnamefont {J.~M.}\ \bibnamefont
  {Ch\'avez-Garcia}}, \bibinfo {author} {\bibfnamefont {F.}~\bibnamefont
  {Solgun}}, \bibinfo {author} {\bibfnamefont {J.~B.}\ \bibnamefont
  {Hertzberg}}, \bibinfo {author} {\bibfnamefont {O.}~\bibnamefont {Jinka}},
  \bibinfo {author} {\bibfnamefont {M.}~\bibnamefont {Brink}},\ and\ \bibinfo
  {author} {\bibfnamefont {B.}~\bibnamefont {Abdo}},\ }\href
  {https://doi.org/10.1103/PhysRevApplied.18.034057} {\bibfield  {journal}
  {\bibinfo  {journal} {Phys. Rev. Appl.}\ }\textbf {\bibinfo {volume} {18}},\
  \bibinfo {pages} {034057} (\bibinfo {year} {2022})}\BibitemShut {NoStop}%
\bibitem [{\citenamefont {Pierre}\ \emph {et~al.}(2014)\citenamefont {Pierre},
  \citenamefont {Svensson}, \citenamefont {Raman~Sathyamoorthy}, \citenamefont
  {Johansson},\ and\ \citenamefont {Delsing}}]{Pierre_2014}%
  \BibitemOpen
  \bibfield  {author} {\bibinfo {author} {\bibfnamefont {M.}~\bibnamefont
  {Pierre}}, \bibinfo {author} {\bibfnamefont {I.-M.}\ \bibnamefont
  {Svensson}}, \bibinfo {author} {\bibfnamefont {S.}~\bibnamefont
  {Raman~Sathyamoorthy}}, \bibinfo {author} {\bibfnamefont {G.}~\bibnamefont
  {Johansson}},\ and\ \bibinfo {author} {\bibfnamefont {P.}~\bibnamefont
  {Delsing}},\ }\href {https://doi.org/10.1063/1.4882646} {\bibfield  {journal}
  {\bibinfo  {journal} {Applied Physics Letters}\ }\textbf {\bibinfo {volume}
  {104}},\ \bibinfo {pages} {232604} (\bibinfo {year} {2014})}\BibitemShut
  {NoStop}%
\bibitem [{\citenamefont {Kafri}\ \emph {et~al.}(2017)\citenamefont {Kafri},
  \citenamefont {Quintana}, \citenamefont {Chen}, \citenamefont {Shabani},
  \citenamefont {Martinis},\ and\ \citenamefont {Neven}}]{Kafri_2017}%
  \BibitemOpen
  \bibfield  {author} {\bibinfo {author} {\bibfnamefont {D.}~\bibnamefont
  {Kafri}}, \bibinfo {author} {\bibfnamefont {C.}~\bibnamefont {Quintana}},
  \bibinfo {author} {\bibfnamefont {Y.}~\bibnamefont {Chen}}, \bibinfo {author}
  {\bibfnamefont {A.}~\bibnamefont {Shabani}}, \bibinfo {author} {\bibfnamefont
  {J.~M.}\ \bibnamefont {Martinis}},\ and\ \bibinfo {author} {\bibfnamefont
  {H.}~\bibnamefont {Neven}},\ }\href
  {https://doi.org/10.1103/PhysRevA.95.052333} {\bibfield  {journal} {\bibinfo
  {journal} {Phys. Rev. A}\ }\textbf {\bibinfo {volume} {95}},\ \bibinfo
  {pages} {052333} (\bibinfo {year} {2017})}\BibitemShut {NoStop}%
\bibitem [{\citenamefont {Menke}\ \emph {et~al.}(2022)\citenamefont {Menke},
  \citenamefont {Banner}, \citenamefont {Bergamaschi}, \citenamefont
  {Di~Paolo}, \citenamefont {Vepsäläinen}, \citenamefont {Weber},
  \citenamefont {Winik}, \citenamefont {Melville}, \citenamefont {Niedzielski},
  \citenamefont {Rosenberg},\ and\ \citenamefont {et~al.}}]{Menke_2022}%
  \BibitemOpen
  \bibfield  {author} {\bibinfo {author} {\bibfnamefont {T.}~\bibnamefont
  {Menke}}, \bibinfo {author} {\bibfnamefont {W.~P.}\ \bibnamefont {Banner}},
  \bibinfo {author} {\bibfnamefont {T.~R.}\ \bibnamefont {Bergamaschi}},
  \bibinfo {author} {\bibfnamefont {A.}~\bibnamefont {Di~Paolo}}, \bibinfo
  {author} {\bibfnamefont {A.}~\bibnamefont {Vepsäläinen}}, \bibinfo {author}
  {\bibfnamefont {S.~J.}\ \bibnamefont {Weber}}, \bibinfo {author}
  {\bibfnamefont {R.}~\bibnamefont {Winik}}, \bibinfo {author} {\bibfnamefont
  {A.}~\bibnamefont {Melville}}, \bibinfo {author} {\bibfnamefont {B.~M.}\
  \bibnamefont {Niedzielski}}, \bibinfo {author} {\bibfnamefont
  {D.}~\bibnamefont {Rosenberg}},\ and\ \bibinfo {author} {\bibnamefont
  {et~al.}},\ }\href {https://doi.org/10.1103/physrevlett.129.220501}
  {\bibfield  {journal} {\bibinfo  {journal} {Physical Review Letters}\
  }\textbf {\bibinfo {volume} {129}},\ \bibinfo {pages} {220501} (\bibinfo
  {year} {2022})}\BibitemShut {NoStop}%
\bibitem [{\citenamefont {Yamamoto}\ \emph {et~al.}(2008)\citenamefont
  {Yamamoto}, \citenamefont {Inomata}, \citenamefont {Watanabe}, \citenamefont
  {Matsuba}, \citenamefont {Miyazaki}, \citenamefont {Oliver}, \citenamefont
  {Nakamura},\ and\ \citenamefont {Tsai}}]{Yamamoto_2008}%
  \BibitemOpen
  \bibfield  {author} {\bibinfo {author} {\bibfnamefont {T.}~\bibnamefont
  {Yamamoto}}, \bibinfo {author} {\bibfnamefont {K.}~\bibnamefont {Inomata}},
  \bibinfo {author} {\bibfnamefont {M.}~\bibnamefont {Watanabe}}, \bibinfo
  {author} {\bibfnamefont {K.}~\bibnamefont {Matsuba}}, \bibinfo {author}
  {\bibfnamefont {T.}~\bibnamefont {Miyazaki}}, \bibinfo {author}
  {\bibfnamefont {W.~D.}\ \bibnamefont {Oliver}}, \bibinfo {author}
  {\bibfnamefont {Y.}~\bibnamefont {Nakamura}},\ and\ \bibinfo {author}
  {\bibfnamefont {J.~S.}\ \bibnamefont {Tsai}},\ }\href
  {https://doi.org/10.1063/1.2964182} {\bibfield  {journal} {\bibinfo
  {journal} {Applied Physics Letters}\ }\textbf {\bibinfo {volume} {93}},\
  \bibinfo {pages} {042510} (\bibinfo {year} {2008})}\BibitemShut {NoStop}%
\bibitem [{\citenamefont {Roch}\ \emph {et~al.}(2012)\citenamefont {Roch},
  \citenamefont {Flurin}, \citenamefont {Nguyen}, \citenamefont {Morfin},
  \citenamefont {Campagne-Ibarcq}, \citenamefont {Devoret},\ and\ \citenamefont
  {Huard}}]{Roch_2012}%
  \BibitemOpen
  \bibfield  {author} {\bibinfo {author} {\bibfnamefont {N.}~\bibnamefont
  {Roch}}, \bibinfo {author} {\bibfnamefont {E.}~\bibnamefont {Flurin}},
  \bibinfo {author} {\bibfnamefont {F.}~\bibnamefont {Nguyen}}, \bibinfo
  {author} {\bibfnamefont {P.}~\bibnamefont {Morfin}}, \bibinfo {author}
  {\bibfnamefont {P.}~\bibnamefont {Campagne-Ibarcq}}, \bibinfo {author}
  {\bibfnamefont {M.~H.}\ \bibnamefont {Devoret}},\ and\ \bibinfo {author}
  {\bibfnamefont {B.}~\bibnamefont {Huard}},\ }\href
  {https://doi.org/10.1103/physrevlett.108.147701} {\bibfield  {journal}
  {\bibinfo  {journal} {Physical Review Letters}\ }\textbf {\bibinfo {volume}
  {108}},\ \bibinfo {pages} {147701} (\bibinfo {year} {2012})}\BibitemShut
  {NoStop}%
\bibitem [{\citenamefont {Simoen}\ \emph {et~al.}(2015)\citenamefont {Simoen},
  \citenamefont {Chang}, \citenamefont {Krantz}, \citenamefont {Bylander},
  \citenamefont {Wustmann}, \citenamefont {Shumeiko}, \citenamefont {Delsing},\
  and\ \citenamefont {Wilson}}]{Simoen_2015}%
  \BibitemOpen
  \bibfield  {author} {\bibinfo {author} {\bibfnamefont {M.}~\bibnamefont
  {Simoen}}, \bibinfo {author} {\bibfnamefont {C.~W.}\ \bibnamefont {Chang}},
  \bibinfo {author} {\bibfnamefont {P.}~\bibnamefont {Krantz}}, \bibinfo
  {author} {\bibfnamefont {J.}~\bibnamefont {Bylander}}, \bibinfo {author}
  {\bibfnamefont {W.}~\bibnamefont {Wustmann}}, \bibinfo {author}
  {\bibfnamefont {V.}~\bibnamefont {Shumeiko}}, \bibinfo {author}
  {\bibfnamefont {P.}~\bibnamefont {Delsing}},\ and\ \bibinfo {author}
  {\bibfnamefont {C.~M.}\ \bibnamefont {Wilson}},\ }\href
  {https://doi.org/10.1063/1.4933265} {\bibfield  {journal} {\bibinfo
  {journal} {Journal of Applied Physics}\ }\textbf {\bibinfo {volume} {118}},\
  \bibinfo {pages} {154501} (\bibinfo {year} {2015})}\BibitemShut {NoStop}%
\bibitem [{\citenamefont {Pogorzalek}\ \emph {et~al.}(2017)\citenamefont
  {Pogorzalek}, \citenamefont {Fedorov}, \citenamefont {Zhong}, \citenamefont
  {Goetz}, \citenamefont {Wulschner}, \citenamefont {Fischer}, \citenamefont
  {Eder}, \citenamefont {Xie}, \citenamefont {Inomata}, \citenamefont
  {Yamamoto},\ and\ \citenamefont {et~al.}}]{Pogorzalek_2017}%
  \BibitemOpen
  \bibfield  {author} {\bibinfo {author} {\bibfnamefont {S.}~\bibnamefont
  {Pogorzalek}}, \bibinfo {author} {\bibfnamefont {K.~G.}\ \bibnamefont
  {Fedorov}}, \bibinfo {author} {\bibfnamefont {L.}~\bibnamefont {Zhong}},
  \bibinfo {author} {\bibfnamefont {J.}~\bibnamefont {Goetz}}, \bibinfo
  {author} {\bibfnamefont {F.}~\bibnamefont {Wulschner}}, \bibinfo {author}
  {\bibfnamefont {M.}~\bibnamefont {Fischer}}, \bibinfo {author} {\bibfnamefont
  {P.}~\bibnamefont {Eder}}, \bibinfo {author} {\bibfnamefont {E.}~\bibnamefont
  {Xie}}, \bibinfo {author} {\bibfnamefont {K.}~\bibnamefont {Inomata}},
  \bibinfo {author} {\bibfnamefont {T.}~\bibnamefont {Yamamoto}},\ and\
  \bibinfo {author} {\bibnamefont {et~al.}},\ }\href
  {https://doi.org/10.1103/physrevapplied.8.024012} {\bibfield  {journal}
  {\bibinfo  {journal} {Physical Review Applied}\ }\textbf {\bibinfo {volume}
  {8}},\ \bibinfo {pages} {024012} (\bibinfo {year} {2017})}\BibitemShut
  {NoStop}%
\bibitem [{\citenamefont {Palacios-Laloy}\ \emph {et~al.}(2008)\citenamefont
  {Palacios-Laloy}, \citenamefont {Nguyen}, \citenamefont {Mallet},
  \citenamefont {Bertet}, \citenamefont {Vion},\ and\ \citenamefont
  {Esteve}}]{Palacios_2008}%
  \BibitemOpen
  \bibfield  {author} {\bibinfo {author} {\bibfnamefont {A.}~\bibnamefont
  {Palacios-Laloy}}, \bibinfo {author} {\bibfnamefont {F.}~\bibnamefont
  {Nguyen}}, \bibinfo {author} {\bibfnamefont {F.}~\bibnamefont {Mallet}},
  \bibinfo {author} {\bibfnamefont {P.}~\bibnamefont {Bertet}}, \bibinfo
  {author} {\bibfnamefont {D.}~\bibnamefont {Vion}},\ and\ \bibinfo {author}
  {\bibfnamefont {D.}~\bibnamefont {Esteve}},\ }\href
  {https://doi.org/10.1007/s10909-008-9774-x} {\bibfield  {journal} {\bibinfo
  {journal} {Journal of Low Temperature Physics}\ }\textbf {\bibinfo {volume}
  {151}},\ \bibinfo {pages} {1034–} (\bibinfo {year} {2008})}\BibitemShut
  {NoStop}%
\bibitem [{\citenamefont {Sandberg}\ \emph {et~al.}(2008)\citenamefont
  {Sandberg}, \citenamefont {Wilson}, \citenamefont {Persson}, \citenamefont
  {Bauch}, \citenamefont {Johansson}, \citenamefont {Shumeiko}, \citenamefont
  {Duty},\ and\ \citenamefont {Delsing}}]{Sandberg_2008}%
  \BibitemOpen
  \bibfield  {author} {\bibinfo {author} {\bibfnamefont {M.}~\bibnamefont
  {Sandberg}}, \bibinfo {author} {\bibfnamefont {C.~M.}\ \bibnamefont
  {Wilson}}, \bibinfo {author} {\bibfnamefont {F.}~\bibnamefont {Persson}},
  \bibinfo {author} {\bibfnamefont {T.}~\bibnamefont {Bauch}}, \bibinfo
  {author} {\bibfnamefont {G.}~\bibnamefont {Johansson}}, \bibinfo {author}
  {\bibfnamefont {V.}~\bibnamefont {Shumeiko}}, \bibinfo {author}
  {\bibfnamefont {T.}~\bibnamefont {Duty}},\ and\ \bibinfo {author}
  {\bibfnamefont {P.}~\bibnamefont {Delsing}},\ }\href
  {https://doi.org/10.1063/1.2929367} {\bibfield  {journal} {\bibinfo
  {journal} {Applied Physics Letters}\ }\textbf {\bibinfo {volume} {92}},\
  \bibinfo {pages} {203501} (\bibinfo {year} {2008})}\BibitemShut {NoStop}%
\bibitem [{\citenamefont {Krantz}\ \emph {et~al.}(2013)\citenamefont {Krantz},
  \citenamefont {Reshitnyk}, \citenamefont {Wustmann}, \citenamefont
  {Bylander}, \citenamefont {Gustavsson}, \citenamefont {Oliver}, \citenamefont
  {Duty}, \citenamefont {Shumeiko},\ and\ \citenamefont
  {Delsing}}]{Krantz_2013}%
  \BibitemOpen
  \bibfield  {author} {\bibinfo {author} {\bibfnamefont {P.}~\bibnamefont
  {Krantz}}, \bibinfo {author} {\bibfnamefont {Y.}~\bibnamefont {Reshitnyk}},
  \bibinfo {author} {\bibfnamefont {W.}~\bibnamefont {Wustmann}}, \bibinfo
  {author} {\bibfnamefont {J.}~\bibnamefont {Bylander}}, \bibinfo {author}
  {\bibfnamefont {S.}~\bibnamefont {Gustavsson}}, \bibinfo {author}
  {\bibfnamefont {W.~D.}\ \bibnamefont {Oliver}}, \bibinfo {author}
  {\bibfnamefont {T.}~\bibnamefont {Duty}}, \bibinfo {author} {\bibfnamefont
  {V.}~\bibnamefont {Shumeiko}},\ and\ \bibinfo {author} {\bibfnamefont
  {P.}~\bibnamefont {Delsing}},\ }\href
  {https://doi.org/10.1088/1367-2630/15/10/105002} {\bibfield  {journal}
  {\bibinfo  {journal} {New Journal of Physics}\ }\textbf {\bibinfo {volume}
  {15}},\ \bibinfo {pages} {105002} (\bibinfo {year} {2013})}\BibitemShut
  {NoStop}%
\bibitem [{\citenamefont {Kennedy}\ \emph {et~al.}(2019)\citenamefont
  {Kennedy}, \citenamefont {Burnett}, \citenamefont {Fenton}, \citenamefont
  {Constantino}, \citenamefont {Warburton}, \citenamefont {Morton},\ and\
  \citenamefont {Dupont-Ferrier}}]{Kennedy_2019}%
  \BibitemOpen
  \bibfield  {author} {\bibinfo {author} {\bibfnamefont {O.}~\bibnamefont
  {Kennedy}}, \bibinfo {author} {\bibfnamefont {J.}~\bibnamefont {Burnett}},
  \bibinfo {author} {\bibfnamefont {J.}~\bibnamefont {Fenton}}, \bibinfo
  {author} {\bibfnamefont {N.}~\bibnamefont {Constantino}}, \bibinfo {author}
  {\bibfnamefont {P.}~\bibnamefont {Warburton}}, \bibinfo {author}
  {\bibfnamefont {J.}~\bibnamefont {Morton}},\ and\ \bibinfo {author}
  {\bibfnamefont {E.}~\bibnamefont {Dupont-Ferrier}},\ }\href
  {https://doi.org/10.1103/PhysRevApplied.11.014006} {\bibfield  {journal}
  {\bibinfo  {journal} {Phys. Rev. Appl.}\ }\textbf {\bibinfo {volume} {11}},\
  \bibinfo {pages} {014006} (\bibinfo {year} {2019})}\BibitemShut {NoStop}%
\bibitem [{\citenamefont {Uhl}\ \emph {et~al.}(2023)\citenamefont {Uhl},
  \citenamefont {Hackenbeck}, \citenamefont {F\"{u}ger}, \citenamefont
  {Kleiner}, \citenamefont {Koelle},\ and\ \citenamefont {Bothner}}]{Uhl2023}%
  \BibitemOpen
  \bibfield  {author} {\bibinfo {author} {\bibfnamefont {K.}~\bibnamefont
  {Uhl}}, \bibinfo {author} {\bibfnamefont {D.}~\bibnamefont {Hackenbeck}},
  \bibinfo {author} {\bibfnamefont {C.}~\bibnamefont {F\"{u}ger}}, \bibinfo
  {author} {\bibfnamefont {R.}~\bibnamefont {Kleiner}}, \bibinfo {author}
  {\bibfnamefont {D.}~\bibnamefont {Koelle}},\ and\ \bibinfo {author}
  {\bibfnamefont {D.}~\bibnamefont {Bothner}},\ }\href
  {https://doi.org/10.1063/5.0146524} {\bibfield  {journal} {\bibinfo
  {journal} {Applied Physics Letters}\ }\textbf {\bibinfo {volume} {122}},\
  \bibinfo {pages} {182603} (\bibinfo {year} {2023})}\BibitemShut {NoStop}%
\bibitem [{\citenamefont {Nation}\ \emph {et~al.}(2008)\citenamefont {Nation},
  \citenamefont {Blencowe},\ and\ \citenamefont {Buks}}]{Nation_2008}%
  \BibitemOpen
  \bibfield  {author} {\bibinfo {author} {\bibfnamefont {P.~D.}\ \bibnamefont
  {Nation}}, \bibinfo {author} {\bibfnamefont {M.~P.}\ \bibnamefont
  {Blencowe}},\ and\ \bibinfo {author} {\bibfnamefont {E.}~\bibnamefont
  {Buks}},\ }\href {https://doi.org/10.1103/PhysRevB.78.104516} {\bibfield
  {journal} {\bibinfo  {journal} {Phys. Rev. B}\ }\textbf {\bibinfo {volume}
  {78}},\ \bibinfo {pages} {104516} (\bibinfo {year} {2008})}\BibitemShut
  {NoStop}%
\bibitem [{\citenamefont {Etaki}\ \emph {et~al.}(2008)\citenamefont {Etaki},
  \citenamefont {Poot}, \citenamefont {Mahboob}, \citenamefont {Onomitsu},
  \citenamefont {Yamaguchi},\ and\ \citenamefont {van~der Zant}}]{Etaki_2008}%
  \BibitemOpen
  \bibfield  {author} {\bibinfo {author} {\bibfnamefont {S.}~\bibnamefont
  {Etaki}}, \bibinfo {author} {\bibfnamefont {M.}~\bibnamefont {Poot}},
  \bibinfo {author} {\bibfnamefont {I.}~\bibnamefont {Mahboob}}, \bibinfo
  {author} {\bibfnamefont {K.}~\bibnamefont {Onomitsu}}, \bibinfo {author}
  {\bibfnamefont {H.}~\bibnamefont {Yamaguchi}},\ and\ \bibinfo {author}
  {\bibfnamefont {H.~S.}\ \bibnamefont {van~der Zant}},\ }\href
  {https://doi.org/10.1038/nphys1057} {\bibfield  {journal} {\bibinfo
  {journal} {Nature Physics}\ }\textbf {\bibinfo {volume} {4}},\ \bibinfo
  {pages} {785–788} (\bibinfo {year} {2008})}\BibitemShut {NoStop}%
\bibitem [{\citenamefont {Poot}\ \emph {et~al.}(2010)\citenamefont {Poot},
  \citenamefont {Etaki}, \citenamefont {Mahboob}, \citenamefont {Onomitsu},
  \citenamefont {Yamaguchi}, \citenamefont {Blanter},\ and\ \citenamefont
  {van~der Zant}}]{Poot_2010}%
  \BibitemOpen
  \bibfield  {author} {\bibinfo {author} {\bibfnamefont {M.}~\bibnamefont
  {Poot}}, \bibinfo {author} {\bibfnamefont {S.}~\bibnamefont {Etaki}},
  \bibinfo {author} {\bibfnamefont {I.}~\bibnamefont {Mahboob}}, \bibinfo
  {author} {\bibfnamefont {K.}~\bibnamefont {Onomitsu}}, \bibinfo {author}
  {\bibfnamefont {H.}~\bibnamefont {Yamaguchi}}, \bibinfo {author}
  {\bibfnamefont {Y.~M.}\ \bibnamefont {Blanter}},\ and\ \bibinfo {author}
  {\bibfnamefont {H.~S.~J.}\ \bibnamefont {van~der Zant}},\ }\href
  {https://doi.org/10.1103/PhysRevLett.105.207203} {\bibfield  {journal}
  {\bibinfo  {journal} {Phys. Rev. Lett.}\ }\textbf {\bibinfo {volume} {105}},\
  \bibinfo {pages} {207203} (\bibinfo {year} {2010})}\BibitemShut {NoStop}%
\bibitem [{\citenamefont {Rodrigues}\ \emph {et~al.}(2019)\citenamefont
  {Rodrigues}, \citenamefont {Bothner},\ and\ \citenamefont
  {Steele}}]{Rodrigues2019}%
  \BibitemOpen
  \bibfield  {author} {\bibinfo {author} {\bibfnamefont {I.~C.}\ \bibnamefont
  {Rodrigues}}, \bibinfo {author} {\bibfnamefont {D.}~\bibnamefont {Bothner}},\
  and\ \bibinfo {author} {\bibfnamefont {G.~A.}\ \bibnamefont {Steele}},\
  }\href {https://doi.org/10.1038/s41467-019-12964-2} {\bibfield  {journal}
  {\bibinfo  {journal} {Nat. Commun.}\ }\textbf {\bibinfo {volume} {10}},\
  \bibinfo {pages} {1} (\bibinfo {year} {2019})}\BibitemShut {NoStop}%
\bibitem [{\citenamefont {Zoepfl}\ \emph {et~al.}(2020)\citenamefont {Zoepfl},
  \citenamefont {Juan}, \citenamefont {Schneider},\ and\ \citenamefont
  {Kirchmair}}]{Zoepfl_2020}%
  \BibitemOpen
  \bibfield  {author} {\bibinfo {author} {\bibfnamefont {D.}~\bibnamefont
  {Zoepfl}}, \bibinfo {author} {\bibfnamefont {M.~L.}\ \bibnamefont {Juan}},
  \bibinfo {author} {\bibfnamefont {C.~M.}\ \bibnamefont {Schneider}},\ and\
  \bibinfo {author} {\bibfnamefont {G.}~\bibnamefont {Kirchmair}},\ }\href
  {https://doi.org/10.1103/PHYSREVLETT.125.023601} {\bibfield  {journal}
  {\bibinfo  {journal} {Phys. Rev. Lett.}\ }\textbf {\bibinfo {volume} {125}},\
  \bibinfo {pages} {023601} (\bibinfo {year} {2020})}\BibitemShut {NoStop}%
\bibitem [{\citenamefont {Schmidt}\ \emph {et~al.}(2020)\citenamefont
  {Schmidt}, \citenamefont {T.~Amawi}, \citenamefont {Pogorzalek},
  \citenamefont {Deppe}, \citenamefont {Marx}, \citenamefont {Gross},\ and\
  \citenamefont {Huebl}}]{Schmidt_2020}%
  \BibitemOpen
  \bibfield  {author} {\bibinfo {author} {\bibfnamefont {P.}~\bibnamefont
  {Schmidt}}, \bibinfo {author} {\bibfnamefont {M.}~\bibnamefont {T.~Amawi}},
  \bibinfo {author} {\bibfnamefont {S.}~\bibnamefont {Pogorzalek}}, \bibinfo
  {author} {\bibfnamefont {F.}~\bibnamefont {Deppe}}, \bibinfo {author}
  {\bibfnamefont {A.}~\bibnamefont {Marx}}, \bibinfo {author} {\bibfnamefont
  {R.}~\bibnamefont {Gross}},\ and\ \bibinfo {author} {\bibfnamefont
  {H.}~\bibnamefont {Huebl}},\ }\href
  {https://doi.org/10.1038/s42005-020-00501-3} {\bibfield  {journal} {\bibinfo
  {journal} {Communications Physics}\ }\textbf {\bibinfo {volume} {3}},\
  \bibinfo {pages} {233} (\bibinfo {year} {2020})}\BibitemShut {NoStop}%
\bibitem [{\citenamefont {Luschmann}\ \emph {et~al.}(2022)\citenamefont
  {Luschmann}, \citenamefont {Schmidt}, \citenamefont {Deppe}, \citenamefont
  {Marx}, \citenamefont {Sanchez}, \citenamefont {Gross},\ and\ \citenamefont
  {Huebl}}]{Luschmann2022}%
  \BibitemOpen
  \bibfield  {author} {\bibinfo {author} {\bibfnamefont {T.}~\bibnamefont
  {Luschmann}}, \bibinfo {author} {\bibfnamefont {P.}~\bibnamefont {Schmidt}},
  \bibinfo {author} {\bibfnamefont {F.}~\bibnamefont {Deppe}}, \bibinfo
  {author} {\bibfnamefont {A.}~\bibnamefont {Marx}}, \bibinfo {author}
  {\bibfnamefont {A.}~\bibnamefont {Sanchez}}, \bibinfo {author} {\bibfnamefont
  {R.}~\bibnamefont {Gross}},\ and\ \bibinfo {author} {\bibfnamefont
  {H.}~\bibnamefont {Huebl}},\ }\href
  {https://doi.org/10.1038/s41598-022-05438-x} {\bibfield  {journal} {\bibinfo
  {journal} {Sci. Rep.}\ }\textbf {\bibinfo {volume} {12}},\ \bibinfo {pages}
  {1} (\bibinfo {year} {2022})}\BibitemShut {NoStop}%
\bibitem [{\citenamefont {Zoepfl}\ \emph {et~al.}(2023)\citenamefont {Zoepfl},
  \citenamefont {Juan}, \citenamefont {Diaz-Naufal}, \citenamefont {Schneider},
  \citenamefont {Deeg}, \citenamefont {Sharafiev}, \citenamefont {Metelmann},\
  and\ \citenamefont {Kirchmair}}]{Zoepfl_2023}%
  \BibitemOpen
  \bibfield  {author} {\bibinfo {author} {\bibfnamefont {D.}~\bibnamefont
  {Zoepfl}}, \bibinfo {author} {\bibfnamefont {M.~L.}\ \bibnamefont {Juan}},
  \bibinfo {author} {\bibfnamefont {N.}~\bibnamefont {Diaz-Naufal}}, \bibinfo
  {author} {\bibfnamefont {C.~M.~F.}\ \bibnamefont {Schneider}}, \bibinfo
  {author} {\bibfnamefont {L.~F.}\ \bibnamefont {Deeg}}, \bibinfo {author}
  {\bibfnamefont {A.}~\bibnamefont {Sharafiev}}, \bibinfo {author}
  {\bibfnamefont {A.}~\bibnamefont {Metelmann}},\ and\ \bibinfo {author}
  {\bibfnamefont {G.}~\bibnamefont {Kirchmair}},\ }\href
  {https://doi.org/10.1103/PhysRevLett.130.033601} {\bibfield  {journal}
  {\bibinfo  {journal} {Phys. Rev. Lett.}\ }\textbf {\bibinfo {volume} {130}},\
  \bibinfo {pages} {033601} (\bibinfo {year} {2023})}\BibitemShut {NoStop}%
\bibitem [{\citenamefont {Schmidt}\ \emph {et~al.}(2024)\citenamefont
  {Schmidt}, \citenamefont {Claessen}, \citenamefont {Higgins}, \citenamefont
  {Hofer}, \citenamefont {Hansen}, \citenamefont {Asenbaum}, \citenamefont
  {Zemlicka}, \citenamefont {Uhl}, \citenamefont {Kleiner}, \citenamefont
  {Gross}, \citenamefont {Huebl}, \citenamefont {Trupke},\ and\ \citenamefont
  {Aspelmeyer}}]{Schmidt2024}%
  \BibitemOpen
  \bibfield  {author} {\bibinfo {author} {\bibfnamefont {P.}~\bibnamefont
  {Schmidt}}, \bibinfo {author} {\bibfnamefont {R.}~\bibnamefont {Claessen}},
  \bibinfo {author} {\bibfnamefont {G.}~\bibnamefont {Higgins}}, \bibinfo
  {author} {\bibfnamefont {J.}~\bibnamefont {Hofer}}, \bibinfo {author}
  {\bibfnamefont {J.~J.}\ \bibnamefont {Hansen}}, \bibinfo {author}
  {\bibfnamefont {P.}~\bibnamefont {Asenbaum}}, \bibinfo {author}
  {\bibfnamefont {M.}~\bibnamefont {Zemlicka}}, \bibinfo {author}
  {\bibfnamefont {K.}~\bibnamefont {Uhl}}, \bibinfo {author} {\bibfnamefont
  {R.}~\bibnamefont {Kleiner}}, \bibinfo {author} {\bibfnamefont
  {R.}~\bibnamefont {Gross}}, \bibinfo {author} {\bibfnamefont
  {H.}~\bibnamefont {Huebl}}, \bibinfo {author} {\bibfnamefont
  {M.}~\bibnamefont {Trupke}},\ and\ \bibinfo {author} {\bibfnamefont
  {M.}~\bibnamefont {Aspelmeyer}},\ }\href
  {https://doi.org/10.1103/PhysRevApplied.22.014078} {\bibfield  {journal}
  {\bibinfo  {journal} {Phys. Rev. Appl.}\ }\textbf {\bibinfo {volume} {22}},\
  \bibinfo {pages} {014078} (\bibinfo {year} {2024})}\BibitemShut {NoStop}%
\bibitem [{\citenamefont {Kubo}\ \emph {et~al.}(2010)\citenamefont {Kubo},
  \citenamefont {Ong}, \citenamefont {Bertet}, \citenamefont {Vion},
  \citenamefont {Jacques}, \citenamefont {Zheng}, \citenamefont {Dréau},
  \citenamefont {Roch}, \citenamefont {Auffeves}, \citenamefont {Jelezko},\
  and\ \citenamefont {et~al.}}]{Kubo_2010}%
  \BibitemOpen
  \bibfield  {author} {\bibinfo {author} {\bibfnamefont {Y.}~\bibnamefont
  {Kubo}}, \bibinfo {author} {\bibfnamefont {F.~R.}\ \bibnamefont {Ong}},
  \bibinfo {author} {\bibfnamefont {P.}~\bibnamefont {Bertet}}, \bibinfo
  {author} {\bibfnamefont {D.}~\bibnamefont {Vion}}, \bibinfo {author}
  {\bibfnamefont {V.}~\bibnamefont {Jacques}}, \bibinfo {author} {\bibfnamefont
  {D.}~\bibnamefont {Zheng}}, \bibinfo {author} {\bibfnamefont
  {A.}~\bibnamefont {Dréau}}, \bibinfo {author} {\bibfnamefont {J.-F.}\
  \bibnamefont {Roch}}, \bibinfo {author} {\bibfnamefont {A.}~\bibnamefont
  {Auffeves}}, \bibinfo {author} {\bibfnamefont {F.}~\bibnamefont {Jelezko}},\
  and\ \bibinfo {author} {\bibnamefont {et~al.}},\ }\href
  {https://doi.org/10.1103/physrevlett.105.140502} {\bibfield  {journal}
  {\bibinfo  {journal} {Physical Review Letters}\ }\textbf {\bibinfo {volume}
  {105}},\ \bibinfo {pages} {140502} (\bibinfo {year} {2010})}\BibitemShut
  {NoStop}%
\bibitem [{\citenamefont {Eichler}\ and\ \citenamefont
  {Petta}(2018)}]{Eichler_2018}%
  \BibitemOpen
  \bibfield  {author} {\bibinfo {author} {\bibfnamefont {C.}~\bibnamefont
  {Eichler}}\ and\ \bibinfo {author} {\bibfnamefont {J.~R.}\ \bibnamefont
  {Petta}},\ }\href {https://doi.org/10.1103/PhysRevLett.120.227702} {\bibfield
   {journal} {\bibinfo  {journal} {Phys. Rev. Lett.}\ }\textbf {\bibinfo
  {volume} {120}},\ \bibinfo {pages} {227702} (\bibinfo {year}
  {2018})}\BibitemShut {NoStop}%
\bibitem [{\citenamefont {Bothner}\ \emph {et~al.}(2020)\citenamefont
  {Bothner}, \citenamefont {Rodrigues},\ and\ \citenamefont
  {Steele}}]{Bothner_2020}%
  \BibitemOpen
  \bibfield  {author} {\bibinfo {author} {\bibfnamefont {D.}~\bibnamefont
  {Bothner}}, \bibinfo {author} {\bibfnamefont {I.~C.}\ \bibnamefont
  {Rodrigues}},\ and\ \bibinfo {author} {\bibfnamefont {G.~A.}\ \bibnamefont
  {Steele}},\ }\href {https://doi.org/10.1038/s41567-020-0987-5} {\bibfield
  {journal} {\bibinfo  {journal} {Nature Physics}\ }\textbf {\bibinfo {volume}
  {17}},\ \bibinfo {pages} {85–91} (\bibinfo {year} {2020})}\BibitemShut
  {NoStop}%
\bibitem [{\citenamefont {Wilson}\ \emph {et~al.}(2010)\citenamefont {Wilson},
  \citenamefont {Duty}, \citenamefont {Sandberg}, \citenamefont {Persson},
  \citenamefont {Shumeiko},\ and\ \citenamefont {Delsing}}]{Wilson_2010}%
  \BibitemOpen
  \bibfield  {author} {\bibinfo {author} {\bibfnamefont {C.~M.}\ \bibnamefont
  {Wilson}}, \bibinfo {author} {\bibfnamefont {T.}~\bibnamefont {Duty}},
  \bibinfo {author} {\bibfnamefont {M.}~\bibnamefont {Sandberg}}, \bibinfo
  {author} {\bibfnamefont {F.}~\bibnamefont {Persson}}, \bibinfo {author}
  {\bibfnamefont {V.}~\bibnamefont {Shumeiko}},\ and\ \bibinfo {author}
  {\bibfnamefont {P.}~\bibnamefont {Delsing}},\ }\href
  {https://doi.org/10.1103/PhysRevLett.105.233907} {\bibfield  {journal}
  {\bibinfo  {journal} {Phys. Rev. Lett.}\ }\textbf {\bibinfo {volume} {105}},\
  \bibinfo {pages} {233907} (\bibinfo {year} {2010})}\BibitemShut {NoStop}%
\bibitem [{\citenamefont {Krantz}\ \emph {et~al.}(2016)\citenamefont {Krantz},
  \citenamefont {Bengtsson}, \citenamefont {Simoen}, \citenamefont
  {Gustavsson}, \citenamefont {Shumeiko}, \citenamefont {Oliver}, \citenamefont
  {Wilson}, \citenamefont {Delsing},\ and\ \citenamefont
  {Bylander}}]{Krantz_2016}%
  \BibitemOpen
  \bibfield  {author} {\bibinfo {author} {\bibfnamefont {P.}~\bibnamefont
  {Krantz}}, \bibinfo {author} {\bibfnamefont {A.}~\bibnamefont {Bengtsson}},
  \bibinfo {author} {\bibfnamefont {M.}~\bibnamefont {Simoen}}, \bibinfo
  {author} {\bibfnamefont {S.}~\bibnamefont {Gustavsson}}, \bibinfo {author}
  {\bibfnamefont {V.}~\bibnamefont {Shumeiko}}, \bibinfo {author}
  {\bibfnamefont {W.~D.}\ \bibnamefont {Oliver}}, \bibinfo {author}
  {\bibfnamefont {C.~M.}\ \bibnamefont {Wilson}}, \bibinfo {author}
  {\bibfnamefont {P.}~\bibnamefont {Delsing}},\ and\ \bibinfo {author}
  {\bibfnamefont {J.}~\bibnamefont {Bylander}},\ }\href
  {https://doi.org/10.1038/ncomms11417} {\bibfield  {journal} {\bibinfo
  {journal} {Nature Communications}\ }\textbf {\bibinfo {volume} {7}},\
  \bibinfo {pages} {11417} (\bibinfo {year} {2016})}\BibitemShut {NoStop}%
\bibitem [{\citenamefont {Zhao}\ \emph {et~al.}(2025)\citenamefont {Zhao},
  \citenamefont {Li}, \citenamefont {Dixit}, \citenamefont {Roy}, \citenamefont
  {Vrajitoarea}, \citenamefont {Banerjee}, \citenamefont {Anferov},
  \citenamefont {Lee}, \citenamefont {Schuster},\ and\ \citenamefont
  {Chou}}]{Zhao_2025}%
  \BibitemOpen
  \bibfield  {author} {\bibinfo {author} {\bibfnamefont {F.}~\bibnamefont
  {Zhao}}, \bibinfo {author} {\bibfnamefont {Z.}~\bibnamefont {Li}}, \bibinfo
  {author} {\bibfnamefont {A.~V.}\ \bibnamefont {Dixit}}, \bibinfo {author}
  {\bibfnamefont {T.}~\bibnamefont {Roy}}, \bibinfo {author} {\bibfnamefont
  {A.}~\bibnamefont {Vrajitoarea}}, \bibinfo {author} {\bibfnamefont
  {R.}~\bibnamefont {Banerjee}}, \bibinfo {author} {\bibfnamefont
  {A.}~\bibnamefont {Anferov}}, \bibinfo {author} {\bibfnamefont {K.-H.}\
  \bibnamefont {Lee}}, \bibinfo {author} {\bibfnamefont {D.~I.}\ \bibnamefont
  {Schuster}},\ and\ \bibinfo {author} {\bibfnamefont {A.}~\bibnamefont
  {Chou}},\ }\href {https://doi.org/10.1103/clp9-xc2n} {\bibfield  {journal}
  {\bibinfo  {journal} {Phys. Rev. Lett.}\ }\textbf {\bibinfo {volume} {135}},\
  \bibinfo {pages} {201002} (\bibinfo {year} {2025})}\BibitemShut {NoStop}%
\bibitem [{\citenamefont {Duret}\ and\ \citenamefont
  {Karp}(1984)}]{Duret_1984}%
  \BibitemOpen
  \bibfield  {author} {\bibinfo {author} {\bibfnamefont {D.}~\bibnamefont
  {Duret}}\ and\ \bibinfo {author} {\bibfnamefont {P.}~\bibnamefont {Karp}},\
  }\href {https://doi.org/10.1063/1.334182} {\bibfield  {journal} {\bibinfo
  {journal} {Journal of Applied Physics}\ }\textbf {\bibinfo {volume} {56}},\
  \bibinfo {pages} {1762–1768} (\bibinfo {year} {1984})}\BibitemShut
  {NoStop}%
\bibitem [{\citenamefont {Knuutila}\ \emph {et~al.}(1988)\citenamefont
  {Knuutila}, \citenamefont {Kajola}, \citenamefont {Seppä}, \citenamefont
  {Mutikainen},\ and\ \citenamefont {Salmi}}]{Knuutila_1988}%
  \BibitemOpen
  \bibfield  {author} {\bibinfo {author} {\bibfnamefont {J.}~\bibnamefont
  {Knuutila}}, \bibinfo {author} {\bibfnamefont {M.}~\bibnamefont {Kajola}},
  \bibinfo {author} {\bibfnamefont {H.}~\bibnamefont {Seppä}}, \bibinfo
  {author} {\bibfnamefont {R.}~\bibnamefont {Mutikainen}},\ and\ \bibinfo
  {author} {\bibfnamefont {J.}~\bibnamefont {Salmi}},\ }\href
  {https://doi.org/10.1007/bf00116869} {\bibfield  {journal} {\bibinfo
  {journal} {Journal of Low Temperature Physics}\ }\textbf {\bibinfo {volume}
  {71}},\ \bibinfo {pages} {369–392} (\bibinfo {year} {1988})}\BibitemShut
  {NoStop}%
\bibitem [{\citenamefont {Yi}\ \emph {et~al.}(2000)\citenamefont {Yi},
  \citenamefont {Zhang}, \citenamefont {Schubert}, \citenamefont {Zander},
  \citenamefont {Zeng},\ and\ \citenamefont {Klein}}]{Yi_2000}%
  \BibitemOpen
  \bibfield  {author} {\bibinfo {author} {\bibfnamefont {H.~R.}\ \bibnamefont
  {Yi}}, \bibinfo {author} {\bibfnamefont {Y.}~\bibnamefont {Zhang}}, \bibinfo
  {author} {\bibfnamefont {J.}~\bibnamefont {Schubert}}, \bibinfo {author}
  {\bibfnamefont {W.}~\bibnamefont {Zander}}, \bibinfo {author} {\bibfnamefont
  {X.~H.}\ \bibnamefont {Zeng}},\ and\ \bibinfo {author} {\bibfnamefont
  {N.}~\bibnamefont {Klein}},\ }\href {https://doi.org/10.1063/1.1322382}
  {\bibfield  {journal} {\bibinfo  {journal} {Journal of Applied Physics}\
  }\textbf {\bibinfo {volume} {88}},\ \bibinfo {pages} {5966–5974} (\bibinfo
  {year} {2000})}\BibitemShut {NoStop}%
\bibitem [{\citenamefont {Granata}\ \emph {et~al.}(2011)\citenamefont
  {Granata}, \citenamefont {Vettoliere},\ and\ \citenamefont
  {Russo}}]{Granata_2011}%
  \BibitemOpen
  \bibfield  {author} {\bibinfo {author} {\bibfnamefont {C.}~\bibnamefont
  {Granata}}, \bibinfo {author} {\bibfnamefont {A.}~\bibnamefont
  {Vettoliere}},\ and\ \bibinfo {author} {\bibfnamefont {M.}~\bibnamefont
  {Russo}},\ }\href {https://doi.org/10.1063/1.3521657} {\bibfield  {journal}
  {\bibinfo  {journal} {Review of Scientific Instruments}\ }\textbf {\bibinfo
  {volume} {82}},\ \bibinfo {pages} {013901} (\bibinfo {year}
  {2011})}\BibitemShut {NoStop}%
\bibitem [{\citenamefont {Chukharkin}\ \emph {et~al.}(2012)\citenamefont
  {Chukharkin}, \citenamefont {Kalabukhov}, \citenamefont {Schneiderman},
  \citenamefont {Öisjöen}, \citenamefont {Snigirev}, \citenamefont {Lai},\
  and\ \citenamefont {Winkler}}]{Chukharkin_2012}%
  \BibitemOpen
  \bibfield  {author} {\bibinfo {author} {\bibfnamefont {M.}~\bibnamefont
  {Chukharkin}}, \bibinfo {author} {\bibfnamefont {A.}~\bibnamefont
  {Kalabukhov}}, \bibinfo {author} {\bibfnamefont {J.~F.}\ \bibnamefont
  {Schneiderman}}, \bibinfo {author} {\bibfnamefont {F.}~\bibnamefont
  {Öisjöen}}, \bibinfo {author} {\bibfnamefont {O.}~\bibnamefont {Snigirev}},
  \bibinfo {author} {\bibfnamefont {Z.}~\bibnamefont {Lai}},\ and\ \bibinfo
  {author} {\bibfnamefont {D.}~\bibnamefont {Winkler}},\ }\href
  {https://doi.org/10.1063/1.4738782} {\bibfield  {journal} {\bibinfo
  {journal} {Applied Physics Letters}\ }\textbf {\bibinfo {volume} {101}},\
  \bibinfo {pages} {042602} (\bibinfo {year} {2012})}\BibitemShut {NoStop}%
\bibitem [{\citenamefont {Tesche}\ and\ \citenamefont
  {Clarke}(1977)}]{Tesche_1977}%
  \BibitemOpen
  \bibfield  {author} {\bibinfo {author} {\bibfnamefont {C.~D.}\ \bibnamefont
  {Tesche}}\ and\ \bibinfo {author} {\bibfnamefont {J.}~\bibnamefont
  {Clarke}},\ }\href {https://doi.org/10.1007/bf00655097} {\bibfield  {journal}
  {\bibinfo  {journal} {Journal of Low Temperature Physics}\ }\textbf {\bibinfo
  {volume} {29}},\ \bibinfo {pages} {301–331} (\bibinfo {year}
  {1977})}\BibitemShut {NoStop}%
\bibitem [{\citenamefont {Lefevre-Seguin}\ \emph {et~al.}(1992)\citenamefont
  {Lefevre-Seguin}, \citenamefont {Turlot}, \citenamefont {Urbina},
  \citenamefont {Esteve},\ and\ \citenamefont {Devoret}}]{LefevreSeguin1992}%
  \BibitemOpen
  \bibfield  {author} {\bibinfo {author} {\bibfnamefont {V.}~\bibnamefont
  {Lefevre-Seguin}}, \bibinfo {author} {\bibfnamefont {E.}~\bibnamefont
  {Turlot}}, \bibinfo {author} {\bibfnamefont {C.}~\bibnamefont {Urbina}},
  \bibinfo {author} {\bibfnamefont {D.}~\bibnamefont {Esteve}},\ and\ \bibinfo
  {author} {\bibfnamefont {M.~H.}\ \bibnamefont {Devoret}},\ }\href
  {https://doi.org/10.1103/physrevb.46.5507} {\bibfield  {journal} {\bibinfo
  {journal} {Physical Review B}\ }\textbf {\bibinfo {volume} {46}},\ \bibinfo
  {pages} {5507–5522} (\bibinfo {year} {1992})}\BibitemShut {NoStop}%
\bibitem [{\citenamefont {Bhupathi}\ \emph {et~al.}(2016)\citenamefont
  {Bhupathi}, \citenamefont {Groszkowski}, \citenamefont {DeFeo}, \citenamefont
  {Ware}, \citenamefont {Wilhelm},\ and\ \citenamefont
  {Plourde}}]{Bhupathi2016}%
  \BibitemOpen
  \bibfield  {author} {\bibinfo {author} {\bibfnamefont {P.}~\bibnamefont
  {Bhupathi}}, \bibinfo {author} {\bibfnamefont {P.}~\bibnamefont
  {Groszkowski}}, \bibinfo {author} {\bibfnamefont {M.}~\bibnamefont {DeFeo}},
  \bibinfo {author} {\bibfnamefont {M.}~\bibnamefont {Ware}}, \bibinfo {author}
  {\bibfnamefont {F.~K.}\ \bibnamefont {Wilhelm}},\ and\ \bibinfo {author}
  {\bibfnamefont {B.}~\bibnamefont {Plourde}},\ }\href
  {https://doi.org/10.1103/physrevapplied.5.024002} {\bibfield  {journal}
  {\bibinfo  {journal} {Physical Review Applied}\ }\textbf {\bibinfo {volume}
  {5}},\ \bibinfo {pages} {024002} (\bibinfo {year} {2016})}\BibitemShut
  {NoStop}%
\bibitem [{\citenamefont {Paradkar}\ \emph {et~al.}(2025)\citenamefont
  {Paradkar}, \citenamefont {Nicaise}, \citenamefont {Dakroury}, \citenamefont
  {Resare},\ and\ \citenamefont {Wieczorek}}]{paradkar_2025}%
  \BibitemOpen
  \bibfield  {author} {\bibinfo {author} {\bibfnamefont {A.}~\bibnamefont
  {Paradkar}}, \bibinfo {author} {\bibfnamefont {P.}~\bibnamefont {Nicaise}},
  \bibinfo {author} {\bibfnamefont {K.}~\bibnamefont {Dakroury}}, \bibinfo
  {author} {\bibfnamefont {F.}~\bibnamefont {Resare}},\ and\ \bibinfo {author}
  {\bibfnamefont {W.}~\bibnamefont {Wieczorek}},\ }\href
  {https://doi.org/10.1063/5.0235266} {\bibfield  {journal} {\bibinfo
  {journal} {Applied Physics Letters}\ }\textbf {\bibinfo {volume} {126}},\
  \bibinfo {pages} {022601} (\bibinfo {year} {2025})}\BibitemShut {NoStop}%
\bibitem [{\citenamefont {Osman}\ \emph {et~al.}(2021)\citenamefont {Osman},
  \citenamefont {Simon}, \citenamefont {Bengtsson}, \citenamefont {Kosen},
  \citenamefont {Krantz}, \citenamefont {P.~Lozano}, \citenamefont
  {Scigliuzzo}, \citenamefont {Delsing}, \citenamefont {Bylander},\ and\
  \citenamefont {Fadavi~Roudsari}}]{Osman_2021}%
  \BibitemOpen
  \bibfield  {author} {\bibinfo {author} {\bibfnamefont {A.}~\bibnamefont
  {Osman}}, \bibinfo {author} {\bibfnamefont {J.}~\bibnamefont {Simon}},
  \bibinfo {author} {\bibfnamefont {A.}~\bibnamefont {Bengtsson}}, \bibinfo
  {author} {\bibfnamefont {S.}~\bibnamefont {Kosen}}, \bibinfo {author}
  {\bibfnamefont {P.}~\bibnamefont {Krantz}}, \bibinfo {author} {\bibfnamefont
  {D.}~\bibnamefont {P.~Lozano}}, \bibinfo {author} {\bibfnamefont
  {M.}~\bibnamefont {Scigliuzzo}}, \bibinfo {author} {\bibfnamefont
  {P.}~\bibnamefont {Delsing}}, \bibinfo {author} {\bibfnamefont
  {J.}~\bibnamefont {Bylander}},\ and\ \bibinfo {author} {\bibfnamefont
  {A.}~\bibnamefont {Fadavi~Roudsari}},\ }\href
  {https://doi.org/10.1063/5.0037093} {\bibfield  {journal} {\bibinfo
  {journal} {Applied Physics Letters}\ }\textbf {\bibinfo {volume} {118}},\
  \bibinfo {pages} {064002} (\bibinfo {year} {2021})}\BibitemShut {NoStop}%
\bibitem [{\citenamefont {Osman}(2024)}]{Osman_2024}%
  \BibitemOpen
  \bibfield  {author} {\bibinfo {author} {\bibfnamefont {A.}~\bibnamefont
  {Osman}},\ }\emph {\bibinfo {title} {Scaling superconducting quantum
  processors: Coherence, frequency targeting and Crosstalk}},\ \href
  {https://research.chalmers.se/publication/543784} {Ph.D. thesis},\ \bibinfo
  {school} {Chalmers University of Technology} (\bibinfo {year}
  {2024})\BibitemShut {NoStop}%
\bibitem [{\citenamefont {Shevchuk}\ \emph {et~al.}(2017)\citenamefont
  {Shevchuk}, \citenamefont {Steele},\ and\ \citenamefont
  {Blanter}}]{Shevchuk_2017}%
  \BibitemOpen
  \bibfield  {author} {\bibinfo {author} {\bibfnamefont {O.}~\bibnamefont
  {Shevchuk}}, \bibinfo {author} {\bibfnamefont {G.~A.}\ \bibnamefont
  {Steele}},\ and\ \bibinfo {author} {\bibfnamefont {Y.~M.}\ \bibnamefont
  {Blanter}},\ }\href {https://doi.org/10.1103/PhysRevB.96.014508} {\bibfield
  {journal} {\bibinfo  {journal} {Phys. Rev. B}\ }\textbf {\bibinfo {volume}
  {96}},\ \bibinfo {pages} {014508} (\bibinfo {year} {2017})}\BibitemShut
  {NoStop}%
\bibitem [{\citenamefont {Frattini}\ \emph {et~al.}(2017)\citenamefont
  {Frattini}, \citenamefont {Vool}, \citenamefont {Shankar}, \citenamefont
  {Narla}, \citenamefont {Sliwa},\ and\ \citenamefont
  {Devoret}}]{Frattini_2017}%
  \BibitemOpen
  \bibfield  {author} {\bibinfo {author} {\bibfnamefont {N.~E.}\ \bibnamefont
  {Frattini}}, \bibinfo {author} {\bibfnamefont {U.}~\bibnamefont {Vool}},
  \bibinfo {author} {\bibfnamefont {S.}~\bibnamefont {Shankar}}, \bibinfo
  {author} {\bibfnamefont {A.}~\bibnamefont {Narla}}, \bibinfo {author}
  {\bibfnamefont {K.~M.}\ \bibnamefont {Sliwa}},\ and\ \bibinfo {author}
  {\bibfnamefont {M.~H.}\ \bibnamefont {Devoret}},\ }\href
  {https://doi.org/10.1063/1.4984142} {\bibfield  {journal} {\bibinfo
  {journal} {Applied Physics Letters}\ }\textbf {\bibinfo {volume} {110}},\
  \bibinfo {pages} {222603} (\bibinfo {year} {2017})}\BibitemShut {NoStop}%
\bibitem [{\citenamefont {Frattini}\ \emph {et~al.}(2018)\citenamefont
  {Frattini}, \citenamefont {Sivak}, \citenamefont {Lingenfelter},
  \citenamefont {Shankar},\ and\ \citenamefont {Devoret}}]{Frattini_2018}%
  \BibitemOpen
  \bibfield  {author} {\bibinfo {author} {\bibfnamefont {N.~E.}\ \bibnamefont
  {Frattini}}, \bibinfo {author} {\bibfnamefont {V.~V.}\ \bibnamefont {Sivak}},
  \bibinfo {author} {\bibfnamefont {A.}~\bibnamefont {Lingenfelter}}, \bibinfo
  {author} {\bibfnamefont {S.}~\bibnamefont {Shankar}},\ and\ \bibinfo {author}
  {\bibfnamefont {M.~H.}\ \bibnamefont {Devoret}},\ }\href
  {https://doi.org/10.1103/PhysRevApplied.10.054020} {\bibfield  {journal}
  {\bibinfo  {journal} {Phys. Rev. Appl.}\ }\textbf {\bibinfo {volume} {10}},\
  \bibinfo {pages} {054020} (\bibinfo {year} {2018})}\BibitemShut {NoStop}%
\bibitem [{\citenamefont {Lescanne}\ \emph {et~al.}(2020)\citenamefont
  {Lescanne}, \citenamefont {Villiers}, \citenamefont {Peronnin}, \citenamefont
  {Sarlette}, \citenamefont {Delbecq}, \citenamefont {Huard}, \citenamefont
  {Kontos}, \citenamefont {Mirrahimi},\ and\ \citenamefont
  {Leghtas}}]{Lescanne_2020}%
  \BibitemOpen
  \bibfield  {author} {\bibinfo {author} {\bibfnamefont {R.}~\bibnamefont
  {Lescanne}}, \bibinfo {author} {\bibfnamefont {M.}~\bibnamefont {Villiers}},
  \bibinfo {author} {\bibfnamefont {T.}~\bibnamefont {Peronnin}}, \bibinfo
  {author} {\bibfnamefont {A.}~\bibnamefont {Sarlette}}, \bibinfo {author}
  {\bibfnamefont {M.}~\bibnamefont {Delbecq}}, \bibinfo {author} {\bibfnamefont
  {B.}~\bibnamefont {Huard}}, \bibinfo {author} {\bibfnamefont
  {T.}~\bibnamefont {Kontos}}, \bibinfo {author} {\bibfnamefont
  {M.}~\bibnamefont {Mirrahimi}},\ and\ \bibinfo {author} {\bibfnamefont
  {Z.}~\bibnamefont {Leghtas}},\ }\href
  {https://doi.org/10.1038/s41567-020-0824-x} {\bibfield  {journal} {\bibinfo
  {journal} {Nature Physics}\ }\textbf {\bibinfo {volume} {16}},\ \bibinfo
  {pages} {509–513} (\bibinfo {year} {2020})}\BibitemShut {NoStop}%
\bibitem [{\citenamefont {Hillmann}\ and\ \citenamefont
  {Quijandr\'{\i}a}(2022)}]{Hillmann_2022}%
  \BibitemOpen
  \bibfield  {author} {\bibinfo {author} {\bibfnamefont {T.}~\bibnamefont
  {Hillmann}}\ and\ \bibinfo {author} {\bibfnamefont {F.}~\bibnamefont
  {Quijandr\'{\i}a}},\ }\href
  {https://doi.org/10.1103/PhysRevApplied.17.064018} {\bibfield  {journal}
  {\bibinfo  {journal} {Phys. Rev. Appl.}\ }\textbf {\bibinfo {volume} {17}},\
  \bibinfo {pages} {064018} (\bibinfo {year} {2022})}\BibitemShut {NoStop}%
\bibitem [{\citenamefont {Lu}\ \emph {et~al.}(2023)\citenamefont {Lu},
  \citenamefont {Kudra}, \citenamefont {Hillmann}, \citenamefont {Yang},
  \citenamefont {Li}, \citenamefont {Quijandría},\ and\ \citenamefont
  {Delsing}}]{Lu_2023}%
  \BibitemOpen
  \bibfield  {author} {\bibinfo {author} {\bibfnamefont {Y.}~\bibnamefont
  {Lu}}, \bibinfo {author} {\bibfnamefont {M.}~\bibnamefont {Kudra}}, \bibinfo
  {author} {\bibfnamefont {T.}~\bibnamefont {Hillmann}}, \bibinfo {author}
  {\bibfnamefont {J.}~\bibnamefont {Yang}}, \bibinfo {author} {\bibfnamefont
  {H.-X.}\ \bibnamefont {Li}}, \bibinfo {author} {\bibfnamefont
  {F.}~\bibnamefont {Quijandría}},\ and\ \bibinfo {author} {\bibfnamefont
  {P.}~\bibnamefont {Delsing}},\ }\href
  {https://doi.org/10.1038/s41534-023-00782-w} {\bibfield  {journal} {\bibinfo
  {journal} {npj Quantum Information}\ }\textbf {\bibinfo {volume} {9}},\
  \bibinfo {pages} {1} (\bibinfo {year} {2023})}\BibitemShut {NoStop}%
\bibitem [{\citenamefont {Eriksson}\ \emph {et~al.}(2024)\citenamefont
  {Eriksson}, \citenamefont {Sépulcre}, \citenamefont {Kervinen},
  \citenamefont {Hillmann}, \citenamefont {Kudra}, \citenamefont {Dupouy},
  \citenamefont {Lu}, \citenamefont {Khanahmadi}, \citenamefont {Yang},
  \citenamefont {Castillo-Moreno},\ and\ \citenamefont
  {et~al.}}]{Eriksson_2024}%
  \BibitemOpen
  \bibfield  {author} {\bibinfo {author} {\bibfnamefont {A.~M.}\ \bibnamefont
  {Eriksson}}, \bibinfo {author} {\bibfnamefont {T.}~\bibnamefont {Sépulcre}},
  \bibinfo {author} {\bibfnamefont {M.}~\bibnamefont {Kervinen}}, \bibinfo
  {author} {\bibfnamefont {T.}~\bibnamefont {Hillmann}}, \bibinfo {author}
  {\bibfnamefont {M.}~\bibnamefont {Kudra}}, \bibinfo {author} {\bibfnamefont
  {S.}~\bibnamefont {Dupouy}}, \bibinfo {author} {\bibfnamefont
  {Y.}~\bibnamefont {Lu}}, \bibinfo {author} {\bibfnamefont {M.}~\bibnamefont
  {Khanahmadi}}, \bibinfo {author} {\bibfnamefont {J.}~\bibnamefont {Yang}},
  \bibinfo {author} {\bibfnamefont {C.}~\bibnamefont {Castillo-Moreno}},\ and\
  \bibinfo {author} {\bibnamefont {et~al.}},\ }\href
  {https://doi.org/10.1038/s41467-024-46507-1} {\bibfield  {journal} {\bibinfo
  {journal} {Nature Communications}\ }\textbf {\bibinfo {volume} {15}},\
  \bibinfo {pages} {1} (\bibinfo {year} {2024})}\BibitemShut {NoStop}%
\bibitem [{\citenamefont {Wallquist}\ \emph {et~al.}(2006)\citenamefont
  {Wallquist}, \citenamefont {Shumeiko},\ and\ \citenamefont
  {Wendin}}]{Wallquist_2006}%
  \BibitemOpen
  \bibfield  {author} {\bibinfo {author} {\bibfnamefont {M.}~\bibnamefont
  {Wallquist}}, \bibinfo {author} {\bibfnamefont {V.~S.}\ \bibnamefont
  {Shumeiko}},\ and\ \bibinfo {author} {\bibfnamefont {G.}~\bibnamefont
  {Wendin}},\ }\href {https://doi.org/10.1103/PhysRevB.74.224506} {\bibfield
  {journal} {\bibinfo  {journal} {Phys. Rev. B}\ }\textbf {\bibinfo {volume}
  {74}},\ \bibinfo {pages} {224506} (\bibinfo {year} {2006})}\BibitemShut
  {NoStop}%
\bibitem [{\citenamefont {Wustmann}\ and\ \citenamefont
  {Shumeiko}(2013)}]{Wustmann2013}%
  \BibitemOpen
  \bibfield  {author} {\bibinfo {author} {\bibfnamefont {W.}~\bibnamefont
  {Wustmann}}\ and\ \bibinfo {author} {\bibfnamefont {V.}~\bibnamefont
  {Shumeiko}},\ }\href {https://doi.org/10.1103/PhysRevB.87.184501} {\bibfield
  {journal} {\bibinfo  {journal} {Phys. Rev. B}\ }\textbf {\bibinfo {volume}
  {87}},\ \bibinfo {pages} {184501} (\bibinfo {year} {2013})}\BibitemShut
  {NoStop}%
\bibitem [{\citenamefont {Potts}\ \emph {et~al.}(2001)\citenamefont {Potts},
  \citenamefont {Routley}, \citenamefont {Parker}, \citenamefont {Baumberg},\
  and\ \citenamefont {de~Groot}}]{Potts_2001}%
  \BibitemOpen
  \bibfield  {author} {\bibinfo {author} {\bibfnamefont {A.}~\bibnamefont
  {Potts}}, \bibinfo {author} {\bibfnamefont {P.~R.}\ \bibnamefont {Routley}},
  \bibinfo {author} {\bibfnamefont {G.~J.}\ \bibnamefont {Parker}}, \bibinfo
  {author} {\bibfnamefont {J.~J.}\ \bibnamefont {Baumberg}},\ and\ \bibinfo
  {author} {\bibfnamefont {P.~A.}\ \bibnamefont {de~Groot}},\ }\href
  {https://doi.org/10.1023/a:1011279908265} {\bibfield  {journal} {\bibinfo
  {journal} {Journal of Materials Science: Materials in Electronics}\ }\textbf
  {\bibinfo {volume} {12}},\ \bibinfo {pages} {289–293} (\bibinfo {year}
  {2001})}\BibitemShut {NoStop}%
\bibitem [{\citenamefont {Tinkham}(2004)}]{Tinkham_2004a}%
  \BibitemOpen
  \bibfield  {author} {\bibinfo {author} {\bibfnamefont {M.}~\bibnamefont
  {Tinkham}},\ }\href@noop {} {\emph {\bibinfo {title} {Introduction to
  Superconductivity}}},\ \bibinfo {edition} {2nd}\ ed.\ (\bibinfo  {publisher}
  {Dover Publications},\ \bibinfo {address} {Mineola, New York},\ \bibinfo
  {year} {2004})\BibitemShut {NoStop}%
\bibitem [{\citenamefont {Zheng}\ \emph {et~al.}(2023)\citenamefont {Zheng},
  \citenamefont {Li}, \citenamefont {Ding}, \citenamefont {Xiong},
  \citenamefont {Feng},\ and\ \citenamefont {Yang}}]{Zheng_2023}%
  \BibitemOpen
  \bibfield  {author} {\bibinfo {author} {\bibfnamefont {Y.}~\bibnamefont
  {Zheng}}, \bibinfo {author} {\bibfnamefont {S.}~\bibnamefont {Li}}, \bibinfo
  {author} {\bibfnamefont {Z.}~\bibnamefont {Ding}}, \bibinfo {author}
  {\bibfnamefont {K.}~\bibnamefont {Xiong}}, \bibinfo {author} {\bibfnamefont
  {J.}~\bibnamefont {Feng}},\ and\ \bibinfo {author} {\bibfnamefont
  {H.}~\bibnamefont {Yang}},\ }\href
  {https://doi.org/10.1038/s41598-023-39052-2} {\bibfield  {journal} {\bibinfo
  {journal} {Scientific Reports}\ }\textbf {\bibinfo {volume} {13}},\ \bibinfo
  {pages} {11874} (\bibinfo {year} {2023})}\BibitemShut {NoStop}%
\bibitem [{\citenamefont {Sivak}\ \emph {et~al.}(2019)\citenamefont {Sivak},
  \citenamefont {Frattini}, \citenamefont {Joshi}, \citenamefont
  {Lingenfelter}, \citenamefont {Shankar},\ and\ \citenamefont
  {Devoret}}]{Sivak_2019}%
  \BibitemOpen
  \bibfield  {author} {\bibinfo {author} {\bibfnamefont {V.}~\bibnamefont
  {Sivak}}, \bibinfo {author} {\bibfnamefont {N.}~\bibnamefont {Frattini}},
  \bibinfo {author} {\bibfnamefont {V.}~\bibnamefont {Joshi}}, \bibinfo
  {author} {\bibfnamefont {A.}~\bibnamefont {Lingenfelter}}, \bibinfo {author}
  {\bibfnamefont {S.}~\bibnamefont {Shankar}},\ and\ \bibinfo {author}
  {\bibfnamefont {M.}~\bibnamefont {Devoret}},\ }\href
  {https://doi.org/10.1103/PhysRevApplied.11.054060} {\bibfield  {journal}
  {\bibinfo  {journal} {Phys. Rev. Appl.}\ }\textbf {\bibinfo {volume} {11}},\
  \bibinfo {pages} {054060} (\bibinfo {year} {2019})}\BibitemShut {NoStop}%
\bibitem [{\citenamefont {Probst}\ \emph {et~al.}(2015)\citenamefont {Probst},
  \citenamefont {Song}, \citenamefont {Bushev}, \citenamefont {Ustinov},\ and\
  \citenamefont {Weides}}]{Probst_2015}%
  \BibitemOpen
  \bibfield  {author} {\bibinfo {author} {\bibfnamefont {S.}~\bibnamefont
  {Probst}}, \bibinfo {author} {\bibfnamefont {F.~B.}\ \bibnamefont {Song}},
  \bibinfo {author} {\bibfnamefont {P.~A.}\ \bibnamefont {Bushev}}, \bibinfo
  {author} {\bibfnamefont {A.~V.}\ \bibnamefont {Ustinov}},\ and\ \bibinfo
  {author} {\bibfnamefont {M.}~\bibnamefont {Weides}},\ }\href
  {https://doi.org/10.1063/1.4907935} {\bibfield  {journal} {\bibinfo
  {journal} {Review of Scientific Instruments}\ }\textbf {\bibinfo {volume}
  {86}},\ \bibinfo {pages} {024706} (\bibinfo {year} {2015})}\BibitemShut
  {NoStop}%
\bibitem [{\citenamefont {Rieger}\ \emph {et~al.}(2023)\citenamefont {Rieger},
  \citenamefont {G\"unzler}, \citenamefont {Spiecker}, \citenamefont
  {Nambisan}, \citenamefont {Wernsdorfer},\ and\ \citenamefont
  {Pop}}]{Rieger_2023}%
  \BibitemOpen
  \bibfield  {author} {\bibinfo {author} {\bibfnamefont {D.}~\bibnamefont
  {Rieger}}, \bibinfo {author} {\bibfnamefont {S.}~\bibnamefont {G\"unzler}},
  \bibinfo {author} {\bibfnamefont {M.}~\bibnamefont {Spiecker}}, \bibinfo
  {author} {\bibfnamefont {A.}~\bibnamefont {Nambisan}}, \bibinfo {author}
  {\bibfnamefont {W.}~\bibnamefont {Wernsdorfer}},\ and\ \bibinfo {author}
  {\bibfnamefont {I.}~\bibnamefont {Pop}},\ }\href
  {https://doi.org/10.1103/PhysRevApplied.20.014059} {\bibfield  {journal}
  {\bibinfo  {journal} {Phys. Rev. Appl.}\ }\textbf {\bibinfo {volume} {20}},\
  \bibinfo {pages} {014059} (\bibinfo {year} {2023})}\BibitemShut {NoStop}%
\bibitem [{\citenamefont {Baity}\ \emph {et~al.}(2024)\citenamefont {Baity},
  \citenamefont {Maclean}, \citenamefont {Seferai}, \citenamefont {Bronstein},
  \citenamefont {Shu}, \citenamefont {Hemakumara},\ and\ \citenamefont
  {Weides}}]{Baitly_2024}%
  \BibitemOpen
  \bibfield  {author} {\bibinfo {author} {\bibfnamefont {P.~G.}\ \bibnamefont
  {Baity}}, \bibinfo {author} {\bibfnamefont {C.}~\bibnamefont {Maclean}},
  \bibinfo {author} {\bibfnamefont {V.}~\bibnamefont {Seferai}}, \bibinfo
  {author} {\bibfnamefont {J.}~\bibnamefont {Bronstein}}, \bibinfo {author}
  {\bibfnamefont {Y.}~\bibnamefont {Shu}}, \bibinfo {author} {\bibfnamefont
  {T.}~\bibnamefont {Hemakumara}},\ and\ \bibinfo {author} {\bibfnamefont
  {M.}~\bibnamefont {Weides}},\ }\href
  {https://doi.org/10.1103/PhysRevResearch.6.013329} {\bibfield  {journal}
  {\bibinfo  {journal} {Phys. Rev. Res.}\ }\textbf {\bibinfo {volume} {6}},\
  \bibinfo {pages} {013329} (\bibinfo {year} {2024})}\BibitemShut {NoStop}%
\bibitem [{\citenamefont {Diaz-Naufal}\ \emph {et~al.}(2025)\citenamefont
  {Diaz-Naufal}, \citenamefont {Deeg}, \citenamefont {Zoepfl}, \citenamefont
  {Schneider}, \citenamefont {Juan}, \citenamefont {Kirchmair},\ and\
  \citenamefont {Metelmann}}]{Diaz-Naufal_2025}%
  \BibitemOpen
  \bibfield  {author} {\bibinfo {author} {\bibfnamefont {N.}~\bibnamefont
  {Diaz-Naufal}}, \bibinfo {author} {\bibfnamefont {L.}~\bibnamefont {Deeg}},
  \bibinfo {author} {\bibfnamefont {D.}~\bibnamefont {Zoepfl}}, \bibinfo
  {author} {\bibfnamefont {C.~M.~F.}\ \bibnamefont {Schneider}}, \bibinfo
  {author} {\bibfnamefont {M.~L.}\ \bibnamefont {Juan}}, \bibinfo {author}
  {\bibfnamefont {G.}~\bibnamefont {Kirchmair}},\ and\ \bibinfo {author}
  {\bibfnamefont {A.}~\bibnamefont {Metelmann}},\ }\href
  {https://doi.org/10.1103/PhysRevA.111.053505} {\bibfield  {journal} {\bibinfo
   {journal} {Phys. Rev. A}\ }\textbf {\bibinfo {volume} {111}},\ \bibinfo
  {pages} {053505} (\bibinfo {year} {2025})}\BibitemShut {NoStop}%
\bibitem [{\citenamefont {Watanabe}\ \emph {et~al.}(2009)\citenamefont
  {Watanabe}, \citenamefont {Inomata}, \citenamefont {Yamamoto},\ and\
  \citenamefont {Tsai}}]{Watanabe_2009}%
  \BibitemOpen
  \bibfield  {author} {\bibinfo {author} {\bibfnamefont {M.}~\bibnamefont
  {Watanabe}}, \bibinfo {author} {\bibfnamefont {K.}~\bibnamefont {Inomata}},
  \bibinfo {author} {\bibfnamefont {T.}~\bibnamefont {Yamamoto}},\ and\
  \bibinfo {author} {\bibfnamefont {J.-S.}\ \bibnamefont {Tsai}},\ }\href
  {https://doi.org/10.1103/physrevb.80.174502} {\bibfield  {journal} {\bibinfo
  {journal} {Physical Review B}\ }\textbf {\bibinfo {volume} {80}},\ \bibinfo
  {pages} {174502} (\bibinfo {year} {2009})}\BibitemShut {NoStop}%
\bibitem [{\citenamefont {Latorre}\ \emph {et~al.}(2022)\citenamefont
  {Latorre}, \citenamefont {Paradkar}, \citenamefont {Hambraeus}, \citenamefont
  {Higgins},\ and\ \citenamefont {Wieczorek}}]{martiIEEE2022}%
  \BibitemOpen
  \bibfield  {author} {\bibinfo {author} {\bibfnamefont {M.~G.}\ \bibnamefont
  {Latorre}}, \bibinfo {author} {\bibfnamefont {A.}~\bibnamefont {Paradkar}},
  \bibinfo {author} {\bibfnamefont {D.}~\bibnamefont {Hambraeus}}, \bibinfo
  {author} {\bibfnamefont {G.}~\bibnamefont {Higgins}},\ and\ \bibinfo {author}
  {\bibfnamefont {W.}~\bibnamefont {Wieczorek}},\ }\href
  {https://doi.org/10.1109/TASC.2022.3147730} {\bibfield  {journal} {\bibinfo
  {journal} {IEEE Transactions on Applied Superconductivity}\ }\textbf
  {\bibinfo {volume} {32}},\ \bibinfo {pages} {1800305} (\bibinfo {year}
  {2022})}\BibitemShut {NoStop}%
\bibitem [{\citenamefont {Latorre}\ \emph {et~al.}(2023)\citenamefont
  {Latorre}, \citenamefont {Higgins}, \citenamefont {Paradkar}, \citenamefont
  {Bauch},\ and\ \citenamefont {Wieczorek}}]{martiPRA2023}%
  \BibitemOpen
  \bibfield  {author} {\bibinfo {author} {\bibfnamefont {M.~G.}\ \bibnamefont
  {Latorre}}, \bibinfo {author} {\bibfnamefont {G.}~\bibnamefont {Higgins}},
  \bibinfo {author} {\bibfnamefont {A.}~\bibnamefont {Paradkar}}, \bibinfo
  {author} {\bibfnamefont {T.}~\bibnamefont {Bauch}},\ and\ \bibinfo {author}
  {\bibfnamefont {W.}~\bibnamefont {Wieczorek}},\ }\href
  {https://doi.org/10.1103/PhysRevApplied.19.054047} {\bibfield  {journal}
  {\bibinfo  {journal} {Phys. Rev. Appl.}\ }\textbf {\bibinfo {volume} {19}},\
  \bibinfo {pages} {054047} (\bibinfo {year} {2023})}\BibitemShut {NoStop}%
\bibitem [{\citenamefont {Jaycox}\ and\ \citenamefont
  {Ketchen}(1981)}]{Ketchen_1981}%
  \BibitemOpen
  \bibfield  {author} {\bibinfo {author} {\bibfnamefont {J.}~\bibnamefont
  {Jaycox}}\ and\ \bibinfo {author} {\bibfnamefont {M.}~\bibnamefont
  {Ketchen}},\ }\href {https://doi.org/10.1109/TMAG.1981.1060902} {\bibfield
  {journal} {\bibinfo  {journal} {IEEE Transactions on Magnetics}\ }\textbf
  {\bibinfo {volume} {17}},\ \bibinfo {pages} {400} (\bibinfo {year}
  {1981})}\BibitemShut {NoStop}%
\bibitem [{\citenamefont {Ketchen}\ and\ \citenamefont
  {Jaycox}(1982)}]{Ketchen_1982}%
  \BibitemOpen
  \bibfield  {author} {\bibinfo {author} {\bibfnamefont {M.~B.}\ \bibnamefont
  {Ketchen}}\ and\ \bibinfo {author} {\bibfnamefont {J.~M.}\ \bibnamefont
  {Jaycox}},\ }\href {https://doi.org/10.1063/1.93210} {\bibfield  {journal}
  {\bibinfo  {journal} {Applied Physics Letters}\ }\textbf {\bibinfo {volume}
  {40}},\ \bibinfo {pages} {736–738} (\bibinfo {year} {1982})}\BibitemShut
  {NoStop}%
\bibitem [{\citenamefont {Gross}\ \emph {et~al.}(1990)\citenamefont {Gross},
  \citenamefont {Chaudhari}, \citenamefont {Kawasaki}, \citenamefont
  {Ketchen},\ and\ \citenamefont {Gupta}}]{Gross_1990}%
  \BibitemOpen
  \bibfield  {author} {\bibinfo {author} {\bibfnamefont {R.}~\bibnamefont
  {Gross}}, \bibinfo {author} {\bibfnamefont {P.}~\bibnamefont {Chaudhari}},
  \bibinfo {author} {\bibfnamefont {M.}~\bibnamefont {Kawasaki}}, \bibinfo
  {author} {\bibfnamefont {M.~B.}\ \bibnamefont {Ketchen}},\ and\ \bibinfo
  {author} {\bibfnamefont {A.}~\bibnamefont {Gupta}},\ }\href
  {https://doi.org/10.1063/1.103600} {\bibfield  {journal} {\bibinfo  {journal}
  {Applied Physics Letters}\ }\textbf {\bibinfo {volume} {57}},\ \bibinfo
  {pages} {727–729} (\bibinfo {year} {1990})}\BibitemShut {NoStop}%
\bibitem [{\citenamefont {Drung}\ \emph {et~al.}(2007)\citenamefont {Drung},
  \citenamefont {Abmann}, \citenamefont {Beyer}, \citenamefont {Kirste},
  \citenamefont {Peters}, \citenamefont {Ruede},\ and\ \citenamefont
  {Schurig}}]{Drung_2007}%
  \BibitemOpen
  \bibfield  {author} {\bibinfo {author} {\bibfnamefont {D.}~\bibnamefont
  {Drung}}, \bibinfo {author} {\bibfnamefont {C.}~\bibnamefont {Abmann}},
  \bibinfo {author} {\bibfnamefont {J.}~\bibnamefont {Beyer}}, \bibinfo
  {author} {\bibfnamefont {A.}~\bibnamefont {Kirste}}, \bibinfo {author}
  {\bibfnamefont {M.}~\bibnamefont {Peters}}, \bibinfo {author} {\bibfnamefont
  {F.}~\bibnamefont {Ruede}},\ and\ \bibinfo {author} {\bibfnamefont
  {T.}~\bibnamefont {Schurig}},\ }\href
  {https://doi.org/10.1109/TASC.2007.897403} {\bibfield  {journal} {\bibinfo
  {journal} {IEEE Transactions on Applied Superconductivity}\ }\textbf
  {\bibinfo {volume} {17}},\ \bibinfo {pages} {699} (\bibinfo {year}
  {2007})}\BibitemShut {NoStop}%
\bibitem [{\citenamefont {Granata}\ and\ \citenamefont
  {Vettoliere}(2016)}]{Granata_2016}%
  \BibitemOpen
  \bibfield  {author} {\bibinfo {author} {\bibfnamefont {C.}~\bibnamefont
  {Granata}}\ and\ \bibinfo {author} {\bibfnamefont {A.}~\bibnamefont
  {Vettoliere}},\ }\href {https://doi.org/10.1016/j.physrep.2015.12.001}
  {\bibfield  {journal} {\bibinfo  {journal} {Physics Reports}\ }\textbf
  {\bibinfo {volume} {614}},\ \bibinfo {pages} {1–69} (\bibinfo {year}
  {2016})}\BibitemShut {NoStop}%
\bibitem [{\citenamefont {Xie}\ \emph {et~al.}(2017)\citenamefont {Xie},
  \citenamefont {Chukharkin}, \citenamefont {Ruffieux}, \citenamefont
  {Schneiderman}, \citenamefont {Kalabukhov}, \citenamefont {Arzeo},
  \citenamefont {Bauch}, \citenamefont {Lombardi},\ and\ \citenamefont
  {Winkler}}]{Xie_2017}%
  \BibitemOpen
  \bibfield  {author} {\bibinfo {author} {\bibfnamefont {M.}~\bibnamefont
  {Xie}}, \bibinfo {author} {\bibfnamefont {M.~L.}\ \bibnamefont {Chukharkin}},
  \bibinfo {author} {\bibfnamefont {S.}~\bibnamefont {Ruffieux}}, \bibinfo
  {author} {\bibfnamefont {J.~F.}\ \bibnamefont {Schneiderman}}, \bibinfo
  {author} {\bibfnamefont {A.}~\bibnamefont {Kalabukhov}}, \bibinfo {author}
  {\bibfnamefont {M.}~\bibnamefont {Arzeo}}, \bibinfo {author} {\bibfnamefont
  {T.}~\bibnamefont {Bauch}}, \bibinfo {author} {\bibfnamefont
  {F.}~\bibnamefont {Lombardi}},\ and\ \bibinfo {author} {\bibfnamefont
  {D.}~\bibnamefont {Winkler}},\ }\href
  {https://doi.org/10.1088/1361-6668/aa8e14} {\bibfield  {journal} {\bibinfo
  {journal} {Superconductor Science and Technology}\ }\textbf {\bibinfo
  {volume} {30}},\ \bibinfo {pages} {115014} (\bibinfo {year}
  {2017})}\BibitemShut {NoStop}%
\bibitem [{\citenamefont {Dantsker}\ \emph {et~al.}(1997)\citenamefont
  {Dantsker}, \citenamefont {Tanaka},\ and\ \citenamefont
  {Clarke}}]{Dantsker_1997}%
  \BibitemOpen
  \bibfield  {author} {\bibinfo {author} {\bibfnamefont {E.}~\bibnamefont
  {Dantsker}}, \bibinfo {author} {\bibfnamefont {S.}~\bibnamefont {Tanaka}},\
  and\ \bibinfo {author} {\bibfnamefont {J.}~\bibnamefont {Clarke}},\ }\href
  {https://doi.org/10.1063/1.118776} {\bibfield  {journal} {\bibinfo  {journal}
  {Applied Physics Letters}\ }\textbf {\bibinfo {volume} {70}},\ \bibinfo
  {pages} {2037–2039} (\bibinfo {year} {1997})}\BibitemShut {NoStop}%
\bibitem [{\citenamefont {Rosa}(1908)}]{rosa1908self}%
  \BibitemOpen
  \bibfield  {author} {\bibinfo {author} {\bibfnamefont {E.~B.}\ \bibnamefont
  {Rosa}},\ }\href@noop {} {\emph {\bibinfo {title} {The self and mutual
  inductances of linear conductors}}},\ \bibinfo {number} {80}\ (\bibinfo
  {publisher} {US Department of Commerce and Labor, Bureau of Standards},\
  \bibinfo {year} {1908})\BibitemShut {NoStop}%
\bibitem [{\citenamefont {Wade}\ and\ \citenamefont
  {Banister}(1973)}]{Wade1973}%
  \BibitemOpen
  \bibfield  {author} {\bibinfo {author} {\bibfnamefont {K.}~\bibnamefont
  {Wade}}\ and\ \bibinfo {author} {\bibfnamefont {A.}~\bibnamefont
  {Banister}},\ }\href {https://books.google.se/books?id=QwNPDAAAQBAJ} {\emph
  {\bibinfo {title} {The Chemistry of Aluminium, Gallium, Indium and
  Thallium}}},\ \bibinfo {edition} {1st}\ ed.,\ \bibinfo {series} {Pergamon
  Texts in Inorganic Chemistry}, Vol.~\bibinfo {volume} {12}\ (\bibinfo
  {publisher} {Pergamon Press},\ \bibinfo {address} {Oxford, UK},\ \bibinfo
  {year} {1973})\ p.\ \bibinfo {pages} {107}\BibitemShut {NoStop}%
\bibitem [{\citenamefont {Deeg}\ \emph {et~al.}(2025)\citenamefont {Deeg},
  \citenamefont {Zoepfl}, \citenamefont {Diaz-Naufal}, \citenamefont {Juan},
  \citenamefont {Metelmann},\ and\ \citenamefont {Kirchmair}}]{Deeg_2025}%
  \BibitemOpen
  \bibfield  {author} {\bibinfo {author} {\bibfnamefont {L.~F.}\ \bibnamefont
  {Deeg}}, \bibinfo {author} {\bibfnamefont {D.}~\bibnamefont {Zoepfl}},
  \bibinfo {author} {\bibfnamefont {N.}~\bibnamefont {Diaz-Naufal}}, \bibinfo
  {author} {\bibfnamefont {M.~L.}\ \bibnamefont {Juan}}, \bibinfo {author}
  {\bibfnamefont {A.}~\bibnamefont {Metelmann}},\ and\ \bibinfo {author}
  {\bibfnamefont {G.}~\bibnamefont {Kirchmair}},\ }\href
  {https://doi.org/10.1103/PhysRevApplied.23.014082} {\bibfield  {journal}
  {\bibinfo  {journal} {Phys. Rev. Appl.}\ }\textbf {\bibinfo {volume} {23}},\
  \bibinfo {pages} {014082} (\bibinfo {year} {2025})}\BibitemShut {NoStop}%
\bibitem [{\citenamefont {Lee}\ \emph {et~al.}(1995)\citenamefont {Lee},
  \citenamefont {Liu},\ and\ \citenamefont {Itoh}}]{Chung-Yi_1995}%
  \BibitemOpen
  \bibfield  {author} {\bibinfo {author} {\bibfnamefont {C.-Y.}\ \bibnamefont
  {Lee}}, \bibinfo {author} {\bibfnamefont {Y.}~\bibnamefont {Liu}},\ and\
  \bibinfo {author} {\bibfnamefont {T.}~\bibnamefont {Itoh}},\ }\href
  {https://doi.org/10.1109/22.475632} {\bibfield  {journal} {\bibinfo
  {journal} {IEEE Transactions on Microwave Theory and Techniques}\ }\textbf
  {\bibinfo {volume} {43}},\ \bibinfo {pages} {2759} (\bibinfo {year}
  {1995})}\BibitemShut {NoStop}%
\bibitem [{\citenamefont {Chen}\ \emph {et~al.}(2014)\citenamefont {Chen},
  \citenamefont {Megrant}, \citenamefont {Kelly}, \citenamefont {Barends},
  \citenamefont {Bochmann}, \citenamefont {Chen}, \citenamefont {Chiaro},
  \citenamefont {Dunsworth}, \citenamefont {Jeffrey}, \citenamefont {Mutus},\
  and\ \citenamefont {et~al.}}]{Chen_2014}%
  \BibitemOpen
  \bibfield  {author} {\bibinfo {author} {\bibfnamefont {Z.}~\bibnamefont
  {Chen}}, \bibinfo {author} {\bibfnamefont {A.}~\bibnamefont {Megrant}},
  \bibinfo {author} {\bibfnamefont {J.}~\bibnamefont {Kelly}}, \bibinfo
  {author} {\bibfnamefont {R.}~\bibnamefont {Barends}}, \bibinfo {author}
  {\bibfnamefont {J.}~\bibnamefont {Bochmann}}, \bibinfo {author}
  {\bibfnamefont {Y.}~\bibnamefont {Chen}}, \bibinfo {author} {\bibfnamefont
  {B.}~\bibnamefont {Chiaro}}, \bibinfo {author} {\bibfnamefont
  {A.}~\bibnamefont {Dunsworth}}, \bibinfo {author} {\bibfnamefont
  {E.}~\bibnamefont {Jeffrey}}, \bibinfo {author} {\bibfnamefont {J.~Y.}\
  \bibnamefont {Mutus}},\ and\ \bibinfo {author} {\bibnamefont {et~al.}},\
  }\href {https://doi.org/10.1063/1.4863745} {\bibfield  {journal} {\bibinfo
  {journal} {Applied Physics Letters}\ }\textbf {\bibinfo {volume} {104}},\
  \bibinfo {pages} {052602} (\bibinfo {year} {2014})}\BibitemShut {NoStop}%
\bibitem [{\citenamefont {Burnett}\ \emph {et~al.}(2018)\citenamefont
  {Burnett}, \citenamefont {Bengtsson}, \citenamefont {Niepce},\ and\
  \citenamefont {Bylander}}]{Burnett_2018}%
  \BibitemOpen
  \bibfield  {author} {\bibinfo {author} {\bibfnamefont {J.}~\bibnamefont
  {Burnett}}, \bibinfo {author} {\bibfnamefont {A.}~\bibnamefont {Bengtsson}},
  \bibinfo {author} {\bibfnamefont {D.}~\bibnamefont {Niepce}},\ and\ \bibinfo
  {author} {\bibfnamefont {J.}~\bibnamefont {Bylander}},\ }\href
  {https://doi.org/10.1088/1742-6596/969/1/012131} {\bibfield  {journal}
  {\bibinfo  {journal} {Journal of Physics: Conference Series}\ }\textbf
  {\bibinfo {volume} {969}},\ \bibinfo {pages} {012131} (\bibinfo {year}
  {2018})}\BibitemShut {NoStop}%
\bibitem [{\citenamefont {Bizn\'arov\'a}\ \emph {et~al.}(2024)\citenamefont
  {Bizn\'arov\'a}, \citenamefont {Rivera~Hern\'andez}, \citenamefont
  {Forchheimer}, \citenamefont {Bylander}, \citenamefont {Haviland},\ and\
  \citenamefont {Andersson}}]{Biznarova_2024}%
  \BibitemOpen
  \bibfield  {author} {\bibinfo {author} {\bibfnamefont {J.}~\bibnamefont
  {Bizn\'arov\'a}}, \bibinfo {author} {\bibfnamefont {J.}~\bibnamefont
  {Rivera~Hern\'andez}}, \bibinfo {author} {\bibfnamefont {D.}~\bibnamefont
  {Forchheimer}}, \bibinfo {author} {\bibfnamefont {J.}~\bibnamefont
  {Bylander}}, \bibinfo {author} {\bibfnamefont {D.~B.}\ \bibnamefont
  {Haviland}},\ and\ \bibinfo {author} {\bibfnamefont {G.}~\bibnamefont
  {Andersson}},\ }\href {https://doi.org/10.1103/PhysRevApplied.22.014063}
  {\bibfield  {journal} {\bibinfo  {journal} {Phys. Rev. Appl.}\ }\textbf
  {\bibinfo {volume} {22}},\ \bibinfo {pages} {014063} (\bibinfo {year}
  {2024})}\BibitemShut {NoStop}%
\bibitem [{\citenamefont {Svensson}\ \emph {et~al.}(2018)\citenamefont
  {Svensson}, \citenamefont {Pierre}, \citenamefont {Simoen}, \citenamefont
  {Wustmann}, \citenamefont {Krantz}, \citenamefont {Bengtsson}, \citenamefont
  {Johansson}, \citenamefont {Bylander}, \citenamefont {Shumeiko},\ and\
  \citenamefont {Delsing}}]{Svensson_2018}%
  \BibitemOpen
  \bibfield  {author} {\bibinfo {author} {\bibfnamefont {I.-M.}\ \bibnamefont
  {Svensson}}, \bibinfo {author} {\bibfnamefont {M.}~\bibnamefont {Pierre}},
  \bibinfo {author} {\bibfnamefont {M.}~\bibnamefont {Simoen}}, \bibinfo
  {author} {\bibfnamefont {W.}~\bibnamefont {Wustmann}}, \bibinfo {author}
  {\bibfnamefont {P.}~\bibnamefont {Krantz}}, \bibinfo {author} {\bibfnamefont
  {A.}~\bibnamefont {Bengtsson}}, \bibinfo {author} {\bibfnamefont
  {G.}~\bibnamefont {Johansson}}, \bibinfo {author} {\bibfnamefont
  {J.}~\bibnamefont {Bylander}}, \bibinfo {author} {\bibfnamefont
  {V.}~\bibnamefont {Shumeiko}},\ and\ \bibinfo {author} {\bibfnamefont
  {P.}~\bibnamefont {Delsing}},\ }\href
  {https://doi.org/10.1088/1742-6596/969/1/012146} {\bibfield  {journal}
  {\bibinfo  {journal} {Journal of Physics: Conference Series}\ }\textbf
  {\bibinfo {volume} {969}},\ \bibinfo {pages} {012146} (\bibinfo {year}
  {2018})}\BibitemShut {NoStop}%
\bibitem [{\citenamefont {Paul}(2011)}]{Paul_2011}%
  \BibitemOpen
  \bibfield  {author} {\bibinfo {author} {\bibfnamefont {C.~R.}\ \bibnamefont
  {Paul}},\ }\href@noop {} {\emph {\bibinfo {title} {Inductance loop and
  partial Clayton R. Paul}}}\ (\bibinfo  {publisher} {John Wiley \& Sons},\
  \bibinfo {year} {2011})\ p.\ \bibinfo {pages} {126}\BibitemShut {NoStop}%
\bibitem [{\citenamefont {Qi}(2001)}]{Qi_2001}%
  \BibitemOpen
  \bibfield  {author} {\bibinfo {author} {\bibfnamefont {X.}~\bibnamefont
  {Qi}},\ }\emph {\bibinfo {title} {High-frequency characterization and
  modeling of on-chip interconnects and RF IC Wire bonds}},\ \href@noop {}
  {Ph.D. thesis},\ \bibinfo  {school} {Stanford University} (\bibinfo {year}
  {2001})\BibitemShut {NoStop}%
\end{thebibliography}%
\end{document}